\documentclass[submitting]{nst}
\usepackage{ulem}
\usepackage{hyperref}
\usepackage[all]{hypcap}
\usepackage{mathrsfs}
\usepackage{graphicx}
\usepackage{epstopdf}
\usepackage{amsmath,amsthm,amssymb}

\usepackage{color}

\begin{document}

\title{The medium temperature dependence of jet transport coefficient in high-energy nucleus-nucleus collisions}
\thanks{This work is supported in part by the Guangdong Major Project of Basic and Applied Basic Research No.
2020B0301030008 and the Science and Technology Program of Guangzhou No. 2019050001, by the National Science Foundation of China under Grant Nos. 12347130 and 11935007.}

\author{Man Xie}
\email{manxie@m.scnu.edu.cn;}
\affiliation{Key Laboratory of Atomic and Subatomic Structure and Quantum Control (MOE), Guangdong Basic Research Center of Excellence for Structure and Fundamental Interactions of Matter, Institute of Quantum Matter
    \\ South China Normal University, Guangzhou 510006, China}
\affiliation{Guangdong-Hong Kong Joint Laboratory of Quantum Matter, Guangdong Provincial Key Laboratory of Nuclear Science, Southern Nuclear Science Computing Center,
\\South China Normal University, Guangzhou 510006, China}
\affiliation{Key Laboratory of Quark and Lepton Physics (MOE) and Institute of Particle Physics,
	\\ Central China Normal University, Wuhan 430079, China}

\author{Qing-Fei Han}
\email{hanqingfei@mails.ccnu.edu.cn;}
\affiliation{Key Laboratory of Quark and Lepton Physics (MOE) and Institute of Particle Physics,
	\\ Central China Normal University, Wuhan 430079, China}

\author{Enke Wang}
\email{wangek@scnu.edu.cn;}
\affiliation{Key Laboratory of Atomic and Subatomic Structure and Quantum Control (MOE), Guangdong Basic Research Center of Excellence for Structure and Fundamental Interactions of Matter, Institute of Quantum Matter
    \\ South China Normal University, Guangzhou 510006, China}
\affiliation{Guangdong-Hong Kong Joint Laboratory of Quantum Matter, Guangdong Provincial Key Laboratory of Nuclear Science, Southern Nuclear Science Computing Center,
\\South China Normal University, Guangzhou 510006, China}
\affiliation{Key Laboratory of Quark and Lepton Physics (MOE) and Institute of Particle Physics,
	\\ Central China Normal University, Wuhan 430079, China}

\author{Ben-Wei Zhang}
\email{bwzhang@mail.ccnu.edu.cn;}
\affiliation{Key Laboratory of Quark and Lepton Physics (MOE) and Institute of Particle Physics,
	\\ Central China Normal University, Wuhan 430079, China}
\affiliation{Key Laboratory of Atomic and Subatomic Structure and Quantum Control (MOE), Guangdong Basic Research Center of Excellence for Structure and Fundamental Interactions of Matter, Institute of Quantum Matter
    \\ South China Normal University, Guangzhou 510006, China}

\author{Han-Zhong Zhang}
\email[Corresponding author, ]{zhanghz@mail.ccnu.edu.cn.}
\affiliation{Key Laboratory of Quark and Lepton Physics (MOE) and Institute of Particle Physics,
	\\ Central China Normal University, Wuhan 430079, China}
\affiliation{Key Laboratory of Atomic and Subatomic Structure and Quantum Control (MOE), Guangdong Basic Research Center of Excellence for Structure and Fundamental Interactions of Matter, Institute of Quantum Matter
    \\ South China Normal University, Guangzhou 510006, China}

\begin{abstract}
The medium-temperature $T$ dependence of the jet transport coefficient $\hat q$ was studied via the nuclear modification factor $R_{AA}(p_{\rm T})$ and elliptical flow parameter $v_2(p_{\rm T})$ for large transverse momentum $p_{\rm T}$ hadrons in high-energy nucleus-nucleus collisions.
Within a next-to-leading-order perturbative QCD parton model for hard scatterings with modified fragmentation functions due to jet quenching controlled by $\hat q$, we check the suppression and azimuthal anisotropy for large $p_{\rm T}$ hadrons, and extract $\hat q$ by global fits to $R_{AA}(p_{\rm T})$ and $v_2(p_{\rm T})$ data in A + A collisions at RHIC and LHC, respectively.
The numerical results from the best fits show that $\hat q/T^3$ goes down with local medium temperature $T$ in the parton jet trajectory.
Compared with the case of a constant $\hat{q}/T^3$, the going-down $T$ dependence of $\hat{q}/T^3$ makes a hard parton jet to lose more energy near $T_{\rm c}$ and therefore strengthens the azimuthal anisotropy for large $p_{\rm T}$ hadrons.
As a result, $v_2(p_{\rm T})$ for large $p_{\rm T}$ hadrons was enhanced by approximately 10\% to better fit the data at RHIC/LHC. Considering the first-order phase transition from QGP to the hadron phase and the additional energy loss in the hadron phase, $v_2(p_{\rm T})$ is again enhanced by 5\%-10\% at RHIC/LHC.

\end{abstract}


\maketitle

\section{Introduction}

The suppression and azimuthal anisotropy of the high transverse momentum ($p_{\rm T}$) hadrons are two valuable pieces of evidence for the existence of the quark-gluon plasma (QGP) that might be created in high-energy nucleus-nucleus collisions performed at both the Relativistic Heavy-Ion Collider (RHIC) \cite{PHENIX:2008saf,PHENIX:2012jha,PHENIX:2010nlr} and the Large Hadron Collider (LHC) \cite{ALICE:2012aqc,CMS:2012aa,ALICE:2018vuu,CMS:2016xef,ALICE:2018hza,CMS:2012tqw,ALICE:2012vgf,ALICE:2016ccg,CMS:2017xgk}.
When high-energy partons propagate through the color-deconfined QGP medium, they encounter multiple scatterings and lose energy through medium-induced gluon radiation. Then, the final state hadrons observed in the nucleus-nucleus (A + A) collisions are suppressed compared to those observed in proton-proton (p + p) collisions. In general, the suppression strength is given by the nuclear modification factor $R_{AA}(p_{\rm T})$ defined as the ratio of single hadron spectrum in A + A collisions to that in p + p collisions. In a typical noncentral A + A collision, the initial geometric anisotropy can be converted into the azimuthal anisotropy in the gluon density distribution of the produced QGP medium, which leads to azimuthal anisotropy of the total energy loss for energetic jets owing to the path length and gluon density dependence of the jet energy loss.
To characterize this anisotropy, one can introduce the elliptic flow coefficient $v_2(p_{\rm T})$, which is defined as the second-order Fourier coefficient in the azimuthal angular distribution of the final-state high $p_{\rm T}$ hadrons. Both of the two observables $R_{AA}(p_{\rm T})$ and $v_2(p_{\rm T})$ for large $p_{\rm T}$ hadrons are the consequence of jet quenching or energy loss \cite{Gyulassy:1990ye,Wang:1991xy,Qin:2015srf,Lin:2021mdn,Tang:2020ame,Ma:2017ybx,Song:2017wtw,Luo:2017faz}, which are expected to give a consistent jet quenching description.

The strength of the jet energy loss is controlled by the jet transport coefficient $\hat q$, which is proportional to the medium gluon number density $\rho$ and is defined as the average transverse momentum broadening $q_{\rm T}$ squared per unit length for a jet propagating inside the medium \cite{Baier:1996sk}:
\begin{eqnarray}
\hat{q} = \rho\int dq^2_{\rm T} \frac{d\sigma}{dq^2_{\rm T}} q^2_{\rm T} .
\label{eq:qhat}
\end{eqnarray}
Quantitative extraction of the energy loss parameter was first performed by the JET Collaboration, utilizing different theoretical models and different approximations compared to experimental data for single hadron production at RHIC and LHC \cite{Burke:2013yra}. For simplicity, $\hat{q}/T^3$ is generally assumed to be constant when studying bulk matter evolution \cite{Chen:2010te} and the suppression of large $p_{\rm T}$ single hadron and dihadron production \cite{Chen:2011vt,Xie:2019oxg}. The jet transport coefficient and mean free path at the initial time were simultaneously extracted for the jet energy loss \cite{Liu:2015vna}. A comparison of the extracted $\hat{q}/T^3$ for different initial temperatures in the center of the QGP between the RHIC and LHC cases indicated that $\hat{q}/T^3$ decreased slightly with increasing medium temperature \cite{Burke:2013yra,Chen:2010te,Chen:2011vt,Xie:2019oxg,Liu:2015vna}. MARTINI \cite{Schenke:2009gb}, MCGILL-AMY \cite{Qin:2007rn} and other theoretical studies \cite{Das:2015ana,Cao:2017umt} yielded similar conclusions. In fact, perturbative studies with resummed hard thermal loops in finite-temperature QCD have given rise to an additional temperature dependence of $\hat{q}/T^3$ for a fixed strong coupling constant \cite{Wang:2000uj, CasalderreySolana:2007sw}. CUJET model \cite{Liao:2008dk,Xu:2014tda} considered there might exist a strong dependence of $\hat{q}/T^3$ on temperature and tried to give a systemic description on $R_{AA}(p_{\rm T})$ and $v_2(p_{\rm T})$ simultaneously with a Gaussian-like temperature dependence form of $\hat{q}/T^3$ \cite{Xu:2015bbz,Shi:2019nyp} within the opacity expansion energy loss formalism \cite{Gyulassy:2000er}. There were also many other descriptions and developments for the jet quenching parameter, such as the radiative corrections to $\hat{q}$ \cite{Wu:2014nca,Mueller:2016xoc,Iancu:2018trm} and nonperturbative calculations for it using the AdS/CFT correspondence at strong coupling in string theory \cite{Liu:2006ug,Zhang:2015hkz,Ghiglieri:2018ltw} and lattice approaches \cite{Panero:2013pla,Panero:2014qxa,Panero:2014sua,Kumar:2018cgf}.
Recently, with the newly developed Bayesian analysis, JETSCAPE studied the medium temperature, virtuality, and jet energy dependence of $\hat{q}$ via single hadron suppression at RHIC and LHC energies \cite{JETSCAPE:2021ehl}. Meanwhile, the LIDO model \cite{Ke:2020clc} and JETSCAPE \cite{JETSCAPE:2022jer} also extracted the $\hat{q}$ value using two types of observables, single inclusive hadron and jet suppression. In Ref. \cite{Xie:2022ght}, with non-parametric prior distribution of $\hat{q}$, using single hadron production, dihadron, and $\gamma$-hadron correlation data calibrated the temperature-dependent $\hat{q}$. All these studies indicate that $\hat{q}/T^3$ should have a larger value at critical temperature $T_{\rm c}$.

In this study, we investigated the additional temperature dependence of $\hat{q}/T^3$ by comparing theoretical calculations with experimental data for both $R_{AA}(p_{\rm T})$ and $v_2(p_{\rm T})$ at large $p_{\rm T}$ at RHIC and LHC. To reveal a clear tendency for the additional temperature dependence, we assume a linear or Gaussian distribution form for the temperature dependence of $\hat{q}/T^3$ within a high-twist energy-loss formalism \cite{Wang:2009qb,Wang:2001cs,Wang:2002ri}. A (3+1)d ideal hydrodynamic description of the bulk matter evolution is used for the medium expansion, which was outputted in references \cite{Hirano:2001eu,Hirano:2002ds} for Au + Au collisions at 200 GeV and Pb + Pb collisions at 2.76 TeV. The initial conditions for the ideal hydrodynamic equations were fixed, such that the final bulk hadron spectra from the experiments were reproduced. To fit both $R_{AA}(p_{\rm T})$ and $v_2(p_{\rm T})$ simultaneously, we first consider only the QGP phase for the jet energy loss, and then the hadron phase contribution \cite{Chen:2010te} is also included.
Our calculations provide a good description of $R_{AA}(p_{\rm T})$ for different temperature-dependent schemes of $\hat{q}/T^3$, whereas the theoretical results for $v_2(p_{\rm T})$ underestimate the experimental data.
However, compared to the case with a constant $\hat{q}/T^3$, we find that the going-down temperature dependence of $\hat{q}/T^3$ gives an approximately 10\% rise to $v_2(p_{\rm T})$ in the QGP phase, and an additional 10\% rise at RHIC and a 5\% rise at the LHC when hadron phase contribution is included.

The remainder of this paper is organized as follows. We first review the next-to-leading-order (NLO) perturbative QCD (pQCD) parton model with medium-modified fragmentation functions in Sec. ~\ref{sec:pQCD}. Then shown in Sec.~\ref{sec:linear-QGP} and Sec. ~\ref{sec:gauss-QGP} are our numerical results fitted to $R_{AA}(p_{\rm T})$ and $v_2(p_{\rm T})$ data for the linear and Gaussian temperature dependence of $\hat{q}/T^3$ in QGP phase, respectively. In Sec.~\ref{sec:QGP-hadron} the hadron phase contribution is included in the linear temperature dependence of $\hat{q}/T^3$. Finally, we conclude this paper in Sec. \ref{sec:summary} with a summary.

\section{NLO pQCD parton model with modified fragmentation functions}\label{sec:pQCD}

Within the NLO pQCD parton model, the collinear factorized differential cross section of single hadron production in p + p collisions can be factorized into the convolution of parton distribution functions (PDFs), short-distance partonic cross sections, and fragmentation functions (FFs) \cite{Owens:1986mp,CTEQ:1993hwr},
\begin{eqnarray}
	\frac{d\sigma_{pp}^h}{dyd^2p_{\rm T}}&=&\sum_{abcd}\int dx_a dx_b f_{a/p}(x_a,\mu^2) f_{b/p}(x_b,\mu^2)\nonumber \\
&&\times
	\frac{1}{\pi}\frac{d\sigma_{ab\rightarrow cd}}{d\hat{t}}\frac{D_{c}^{h}(z_c,\mu^2)}{z_c}+\mathcal {O}(\alpha_s^3),
\label{eq:pp-sin-spec}
\end{eqnarray}
where $f_{a/p}(x_a,\mu^2)$ is the parton distribution function for parton $a$ with momentum fraction $x_a$ from a free nucleon and CT14 parameterization is used \cite{Hou:2016nqm}. The fragmentation functions $D_{c}^{h}(z_c,\mu^2)$ for a parton in vacuum is given by the AKK parameterization \cite{Albino:2008fy}, in which $z_c$ is the momentum fraction carried by the outgoing hadrons from the parent parton $c$. $d\sigma(ab\rightarrow cd)/d\hat{t}$ is the parton-parton hard-scattering cross section at LO $\alpha_s^2$. The partonic scattering cross sections in our numerical simulations were computed up to the NLO implied in $\mathcal{O}(\alpha_s^3)$. The NLO corrections include 1-loop contributions to $2 \rightarrow 2$ tree level and $2 \rightarrow 3$ tree level contributions. More detailed discussions on the NLO calculations can be found in \cite{Harris:2001sx}.

In A + A collisions, the cross section for single hadron production at high transverse momentum is given by \cite{Zhang:2007ja,Zhang:2009rn}
\begin{eqnarray}
\frac{dN_{AA}^h}{dyd^2p_{\rm T}}&&=\sum_{abcd} \int d^2r
t_A(\vec{r}) t_B(\vec{r}+\vec{b}) \int dx_adx_b \nonumber \\
&&\times\
f_{a/A}(x_a,\mu^2,\vec{r}) f_{b/B}(x_b,\mu^2,\vec{r}+\vec{b}) \nonumber \\
&&\times \frac{1}{\pi}
\frac{d\sigma_{ab\rightarrow cd}}{d\hat{t}}
	\frac{\tilde{D}_{c}^{h}(z_{c},\mu^2,\Delta E_c)}{z_c} +\mathcal {O}(\alpha_s^3), \ \
\label{eq:AA-sin-spec}
\end{eqnarray}
{where $t_A(\vec{r})=\int \rho_A(\vec{r})dz$} is the nuclear thickness function given by the Woods–Saxon distribution and is normalized as $\int d^2r t_A(\vec{r}) = A$.
$f_{a/A}(x_a,\mu^2,\vec{r})$ is the nucleus-modified parton distribution function, which is assumed to be factorized into parton distributions in a free nucleon $f_{a/N}(x_a,\mu^2)$ and the nuclear shadowing factor $S_{a/A}(x_a,\mu^2,\vec{r})$ \cite{Wang:1996yf,Li:2001xa},
\begin{eqnarray}
f_{a/A}(x_a,\mu^2,\vec{r}) &&= S_{a/A}(x_a,\mu^2,\vec{r})\left[\frac{Z}{A}f_{a/p}(x_a,\mu^2)\right. \nonumber\\
&&+\left.\left(1-\frac{Z}{A}\right)f_{a/n}(x_a,\mu^2)\right],
\end{eqnarray}
where $Z$ is the proton number of the nucleus and $A$ is the nuclear mass number. Assuming that the shadowing is proportional to the local nuclear density, the shadowing factor $S_{a/A}(x_a,\mu^2,\vec{r})$ can be obtained using the following form \cite{Emelyanov:1999pkc, Hirano:2003pw}:
\begin{eqnarray}
S_{a/A}(x_a,\mu^2,\vec{r})=1+[S_{a/A}(x_a,\mu^2)-1] \frac{At_A(\vec{r})}{\int{d^2}r [t_A(\vec{r})]^2}, \ \ \
\end{eqnarray}
where $S_{a/A}(x_a,\mu^2)$ is obtained from the EPPS16 \cite{Eskola:2016oht}.

The medium-modified fragmentation function $\tilde{D}_{c}^{h}$ can be calculated as follows \cite{Wang:1996yh,Wang:1996pe,Wang:2004yv,Zhang:2007ja,Zhang:2009rn}:
\begin{eqnarray}
	&& \tilde{D}_{c}^{h}(z_c,\mu^2,\Delta{E_c}) = (1-e^{-\langle{N_g}\rangle})\left[\frac{z'_c}{z_c}D_{c}^{h}(z'_c,\mu^2)\right. \nonumber\\
&& \phantom{XX}+\left.{\langle{N_g}\rangle}\frac{{z_g}'}{z_c}D_{g}^{h}({z_g}',\mu^2)\right]+e^{-\langle{N_g}\rangle}D_{c}^{h}({z_c},\mu^2),
\label{eq:mffs}
\end{eqnarray}
where $z_c=p_{\rm T}/{p_{\rm T}}_c$ is the momentum fraction for a parton fragmenting into a hadron in vacuum. ${z_c}'=p_{\rm T}/({p_{\rm T}}_c-\Delta{E_c})$ is the rescaled momentum fraction and denotes that a parton with ${p_{\rm T}}_c$ propagating through the medium loses energy $\Delta{E}_c$ and fragments into a hadron with $p_{\rm T}$. ${z_g}'= p_{\rm T}/(\Delta{E_c}/\langle{N_g}\rangle)$ is the momentum fraction of the radiated gluon fragmenting into a hadron. $\langle{N_g}\rangle$ is the number of radiated gluons.

The parton energy loss caused by the medium-induced gluon radiation can be calculated using a higher-twist (HT) approach  \cite{Wang:2009qb,Wang:2001cs,Wang:2002ri}.
For a light quark $c$ with an initial energy $E$, the radiative energy loss $\Delta E_c$ can be calculated as
\begin{eqnarray}
\frac{\Delta{E}_c}{E} &=& \frac{2C_A\alpha_s}{\pi} \int d\tau \int \frac{dl_{\rm T}^2}{l_{\rm T}^4}\int dz \nonumber\\
	&&\times \left[1+(1-z)^2\right] \hat{q} \sin^2(\frac{l_{\rm T}^2\tau}{4z(1-z)E}),
\label{eq:De}
\end{eqnarray}
where $C_A=3$, $\alpha_s$ is the strong coupling constant, and $l_{\rm T}$ is the transverse momentum of the radiated gluons. We assume that the energy loss of a gluon is $9/4$ times that of a quark owing to the different color factors for the quark-gluon vertex and gluon-gluon vertex \cite{Wang:2009qb}.
The average number of radiated gluons from the propagating hard parton is calculated as \cite{Chang:2014fba}:
\begin{eqnarray}
	\langle N_g \rangle &=& \frac{2C_A \alpha_{s}}{\pi} \int d\tau \int \frac{dl_{\rm T}^2}{l_{\rm T}^4}\int \frac{dz}{z} \nonumber\\
	&&\times \left[1+(1-z)^2\right] \hat{q} \sin^2(\frac{l_{\rm T}^2\tau}{4z(1-z)E}).
\label{eq:Ng}
\end{eqnarray}
{The HT formalism contains the transverse momentum $l_{\rm T}$ of the radiated gluon, which also indicates the changes in the transverse momenta of the partons \cite{Zhang:2003yn,Zhang:2004qm,Qin:2015srf}. In our numerical simulations, we adopt the small angle approximation within the collinear factorization theorem, according to Eq. (\ref{eq:AA-sin-spec}). Consequently, we focused solely on the effect of energy loss and assumed that the parton direction remains unchanged in the fragmentation functions. Such an approximation has been used in many current jet energy-loss formalisms and has successfully explained experimental data \cite{Armesto:2011ht, Cao:2018ews, Shi:2018izg, Dai:2018mhw, Ke:2020clc, JETSCAPE:2022jer}.}

The parton energy loss and number of radiated gluons are both controlled by the jet transport parameter $\hat{q}$ \citep{Baier:1996sk}.
According to Eq. ($\ref{eq:qhat}$) for $\hat{q}$ proportional to the medium gluon density $\rho$, one can simply assume a constant value for the scaled jet transport parameter \cite{Xie:2019oxg,Liu:2015vna}:
\begin{eqnarray}
\frac{\hat q}{T^3} = \frac{\hat q_0}{T_0^3}\frac{p^{\mu}\cdot u_{\mu}}{p_0},
\label{eq:qhat-cons}
\end{eqnarray}
where $p^{\mu}$ is the four momentum of the parton, $u^{\mu}$ is the local four flow velocity of the fluid, $T$ is the local temperature of the medium and $T_0$ is a reference temperature taken as the highest temperature at the center of the medium at the initial time $\tau_0$.

For an additional $T$-dependence of ${\hat q}/{T^3}$, one can simply assume a linear form such as ${\hat q}/{T^3} \sim aT+b$.
In the following actual calculations, we write the linear form as
\begin{eqnarray}
\frac{\hat q}{T^3} = [ (\frac{\hat q_0}{T_0^3} - \frac{\hat q_{\rm c}}{{T_{\rm c}}^3})\frac{T-T_{\rm c}}{T_0-T_{\rm c}}+\frac{\hat q_{\rm c}}{{T_{\rm c}}^3} ]
\frac{p^{\mu}\cdot u_{\mu}}{p_0}.
\label{eq:qhat-linear}
\end{eqnarray}
We also check the additional $T$ dependence of ${\hat q}/{T^3}$ in the Gaussian form:
\begin{eqnarray}
\frac{\hat q}{T^3} = \frac{\hat{q}_0}{T_0^3}\frac{e^{-(T/T_{\rm c}-1)^2/(2\sigma_T^2/T_{\rm c}^2)}}{e^{-(T_0/T_{\rm c}-1)^2/(2\sigma_T^2/T_{\rm c}^2)}}
\frac{p^{\mu}\cdot u_{\mu}}{p_0}.
\label{eq:qhat-gaus}
\end{eqnarray}
Parameters $\hat {q}_0$, $\hat {q}_{\rm c}$, and $\sigma_T$ were introduced to adjust the strength of the additional temperature dependence. $T_{\rm c} = 170$ MeV is the critical temperature. When $T = T_0$ in Eq. ($\ref{eq:qhat-linear}$) and $\sigma_T^2 = \infty$ for Eq.~ ($\ref{eq:qhat-gaus}$), both equations return to Eq.~ ($\ref{eq:qhat-cons}$).

To describe jet quenching in high-energy nucleus-nucleus collisions, it is necessary to provide the space-time evolution of the jet transport coefficient in Eq. ($\ref{eq:qhat-cons}$, $\ref{eq:qhat-linear}$, $\ref{eq:qhat-gaus}$) along the parton propagation. In our studies, the dynamic evolution of the medium that governs the space-time evolution of the local temperature $T$ and flow velocity $u$ was obtained using a (3+1)-dimensional hydrodynamic model \cite{Hirano:2001eu,Hirano:2002ds}. This model provides results on the transverse dynamics of the bulk medium in A + A collisions under the initial conditions. Furthermore, the model includes the first-order phase transition between the QGP and hadron phases at $T_{\rm c}$ = 170 MeV and provides the hadron phase fraction $f(\vec{r})$, which is defined as
\begin{equation}
f(\vec{r}) = \left\{
\begin{array}{l}
0 \\
0 \sim 1 \\
1
\end{array}
 \;\; \;\;
\begin{array}{l}
 \text{if T$>$170 MeV},\\
 \text{if T=170 MeV},\\
 \text{if T$<$170 MeV},
\end{array}
\right.
\label{eq:hadr-f}
\end{equation}
where $\vec{r}$ denotes the local position of jet. As the geometric position moves closer to the periphery of the medium or the medium evolution time increases, the hadronic phase fraction gradually increases from 0 to 1 at $T_{\rm c} = 170$ MeV, which is determined by the proportion of the hadron and parton number density \cite{Nonaka:2000ek,Hirano:2001eu,Hirano:2007ei}. The energy loss of the jet propagating through both the QGP and hadronic phases can be simultaneously described using the higher-twist approach, except that $\hat{q}$ in the separate phase is different. To include the contributions to $\hat q$ from both QGP and hadron phases, we changed
Eq. (\ref{eq:qhat-cons}), (\ref{eq:qhat-linear}), and (\ref{eq:qhat-gaus}) as follows:
\begin{eqnarray}
\frac{\hat q}{T^3} \rightarrow \frac{\hat q}{T^3}(1-f)+\frac{\hat q_h}{T^3}f,
\label{eq:qhat-f}
\end{eqnarray}
where $\hat{q}_h$ is the jet transport parameter of the hadronic phase. By combining Eq. (\ref{eq:hadr-f}), one can see that, for studies exclusively concerning the QGP phase, we consider a pure partonic medium at $T>170$ MeV, along with the QGP fraction $(1-f)$ in the mixed phase at $T_{\rm c}=170$ MeV. For studies on the hadronic phase, we only need to consider the $\frac{\hat{q}_h}{T^3}f$ term, which accounts for the contributions from the hadronic phase during the mixed phase, as well as the entire hadronic medium when the temperature is below 170 MeV, until the system reaches dynamic freeze-out.
When the hadron phase was considered for the jet energy loss, the extracted jet transport parameter for the QGP phase was reduced owing to the long evolution time of the mixed phase, as shown in Ref. \cite{Chen:2010te}.

The jet transport parameter in the hadron phase can be expressed as follows \cite{Chen:2010te}:
\begin{eqnarray}
\hat{q}_{h}=\frac{\hat{q}_N}{\rho_N}[\frac{2}{3}\sum\limits_{M}\rho_M(T)+\sum\limits_{B}\rho_B(T)],
\label{eq:qhat-had}
\end{eqnarray}
where $\hat{q}_N \approx 0.02$ GeV$^2$/fm is the extracted jet transport parameter at the center of the cold nucleonic matter of a large nucleus and $\rho_N \approx 0.17$ fm$^{-3}$ is the nucleon density at the center of the large nucleus \cite{Wang:2009qb}. $\rho_M$ and $\rho_B$ are the meson and baryon density in the hadronic resonance gas at a given temperature, respectively. Factor 2/3 represents the ratio of the constituent quark numbers of the meson and the baryon.
The hadron density at a given temperature $T$ and zero chemical potential is expressed as \cite{Chen:2010te}
\begin{eqnarray}
\sum\limits_{h}\rho_h(T)=\frac{T^3}{2\pi^2}\sum\limits_{h}(\frac{m_h}{T})^2\sum\limits_{n=1}^{\infty}\frac{\eta_h^{n+1}}{n}K_2(n\frac{m_h}{T}),
\end{eqnarray}
where $\eta_h=\pm $ for mesons (M)/baryons (B).
In the following calculations, hadron resonances with masses below 1 GeV were included: 17 types of mesons: $\pi^+, \pi^-, \pi^0, K^+, K^-, K^0, \overline{K^{0}}, \eta, \eta^{'}, \rho^+, \rho^-, \rho^0, K^{*+},$ $K^{*-}, K^{*0}, \overline{K^{*0}}, \omega$; and 2 kinds of baryons, $p, n$. Here, we ignore the contribution of antinucleons to $\hat{q}_h$, which is less than 3\%.


\section{Linear temperature dependence of $\hat{q}/T^3$ in QGP phase} \label{sec:linear-QGP}

With the spectrum in p + p collisions as a baseline, the nuclear suppression factor $R_{AA}(p_{\rm T})$ for single hadron production in A + A collisions can be expressed as \cite{Wang:2004yv,Wang:2021xpv},
\begin{eqnarray}
R_{AA}(p_{\rm T})=\frac{dN_{AA}^h/dyd^2p_{\rm T}}{T_{AA}(\vec{b}) d{\sigma}_{pp}^h/dyd^2p_{\rm T}},
\label{eq:Raa}
\end{eqnarray}
where $T_{AA}(\vec{b}) =\int d^2r t_A(\vec{r})t_B(\vec{r}+\vec{b})$ is the overlap function of the two colliding nuclei.

The anisotropy of the final-state hadrons in the transverse momentum can be quantified using the Fourier expansion of the hadrons distribution in the azimuthal angle.
We focus on the second Fourier coefficient, namely elliptic anisotropy coefficient $v_2(p_{\rm T})$, which can be written as \cite{Poskanzer:1998yz,Wang:2000fq, Tang:2023wcd,Wang:2022det,Lan:2022rrc,Wang:2022fwq},
\begin{eqnarray}
v_2(p_{\rm T})= \frac{\int^{\pi}_{-\pi} d\phi cos(2\phi) dN_{AA}^h / dy d^2p_{\rm T} d\phi}{\int^{\pi}_{-\pi} d\phi dN_{AA}^h / dy d^2p_{\rm T} d\phi},
\label{eq:v2}
\end{eqnarray}
where $\phi$ is the jet azimuthal angle between the jet propagation direction and the impact parameter.

In this section, we will use Eq. ($\ref{eq:qhat-linear}$) and ($\ref{eq:qhat-f}$) with $\hat{q}_h = 0$ to consider the linear temperature dependence of $\hat{q}/T^3$ in QGP phase. $\chi^2$ fitting to both $R_{AA}(p_{\rm T})$ and $v_2(p_{\rm T})$ for hadrons in the middle rapidity region will be performed for different introduced parameters, which is given by
\begin{eqnarray}
\chi^2 = \sum_{i=1}^N \left[{(V_{\rm th}-V_{\rm exp})^2}/{(\sigma_{\rm sys}^2+\sigma_{\rm stat}^2)}\right],
\label{eq:chi2}
\end{eqnarray}
where $V_{\rm th}$ and $V_{\rm exp}$ denote the theoretical and experimental results, respectively, and $\sigma_{\rm sys}$ and $\sigma_{\rm stat}$ provide the systematic and statistical errors for the data, respectively. For a global fit of both $R_{AA}(p_{\rm T})$ and $v_2(p_{\rm T})$, the data number $N$ for the degrees of freedom (${\rm d.o.f}.$) is the sum of $R_{AA}(p_{\rm T})$ and $v_2(p_{\rm T})$ data numbers. The $\chi^2/{\rm d.o.f}$ value was minimized to near unity to determine the best-fit temperature dependence of $\hat{q}/T^3$ \cite{Shi:2019nyp}. In the $\chi^2/{\rm d.o.f}$ calculations, we selected only the experimental data points with $p_{\rm T}>7.5$ GeV$/c$ at both RHIC and the LHC energies to ensure the validity of pQCD parton model.


\subsection{Fit $R_{AA}$ and $v_2$ at RHIC}

\begin{figure*}[tbh]

\includegraphics[width = 0.37\textwidth, height=0.28\textwidth]{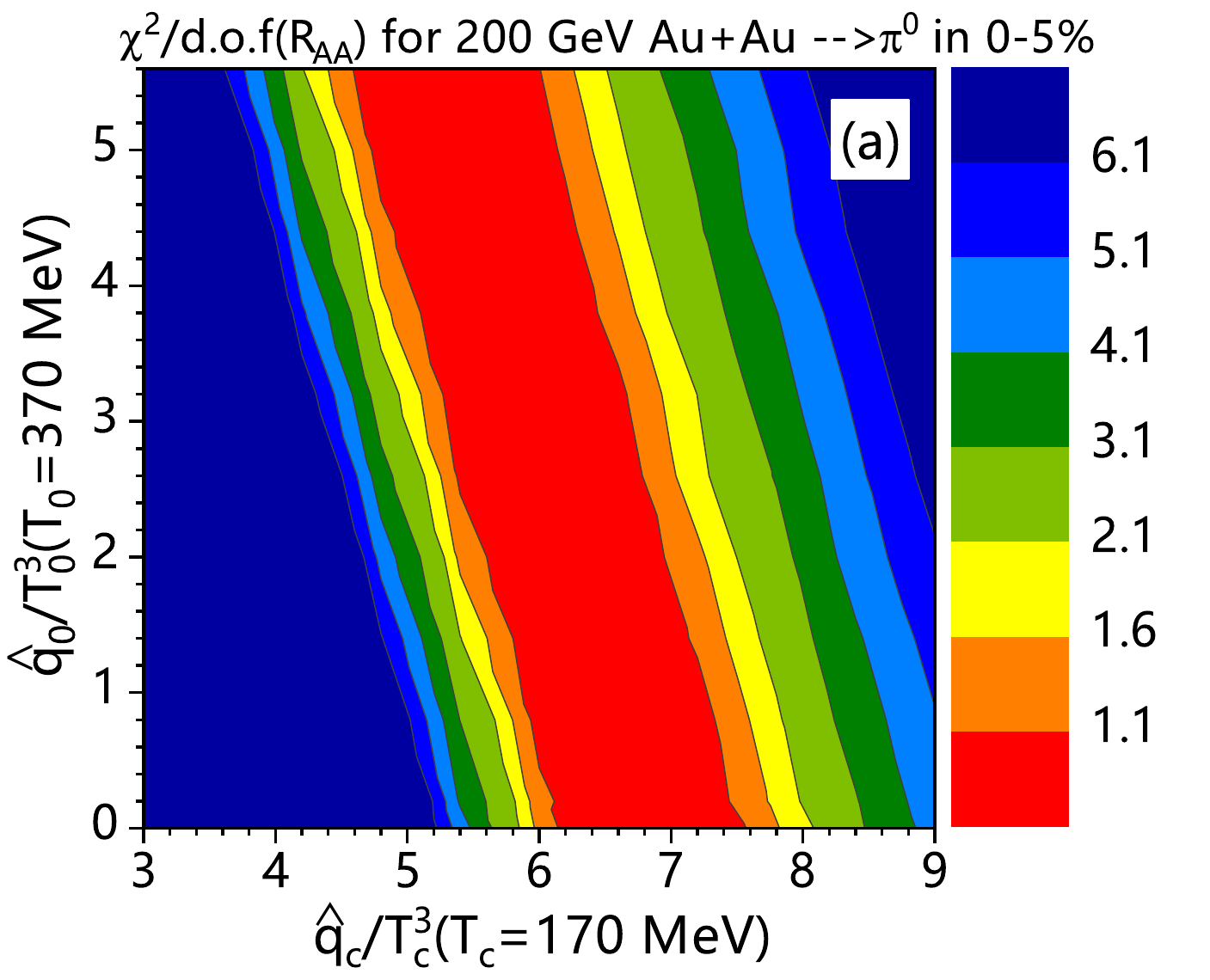}
\hspace{-5mm}
\includegraphics[width = 0.32\textwidth, height=0.27\textwidth]{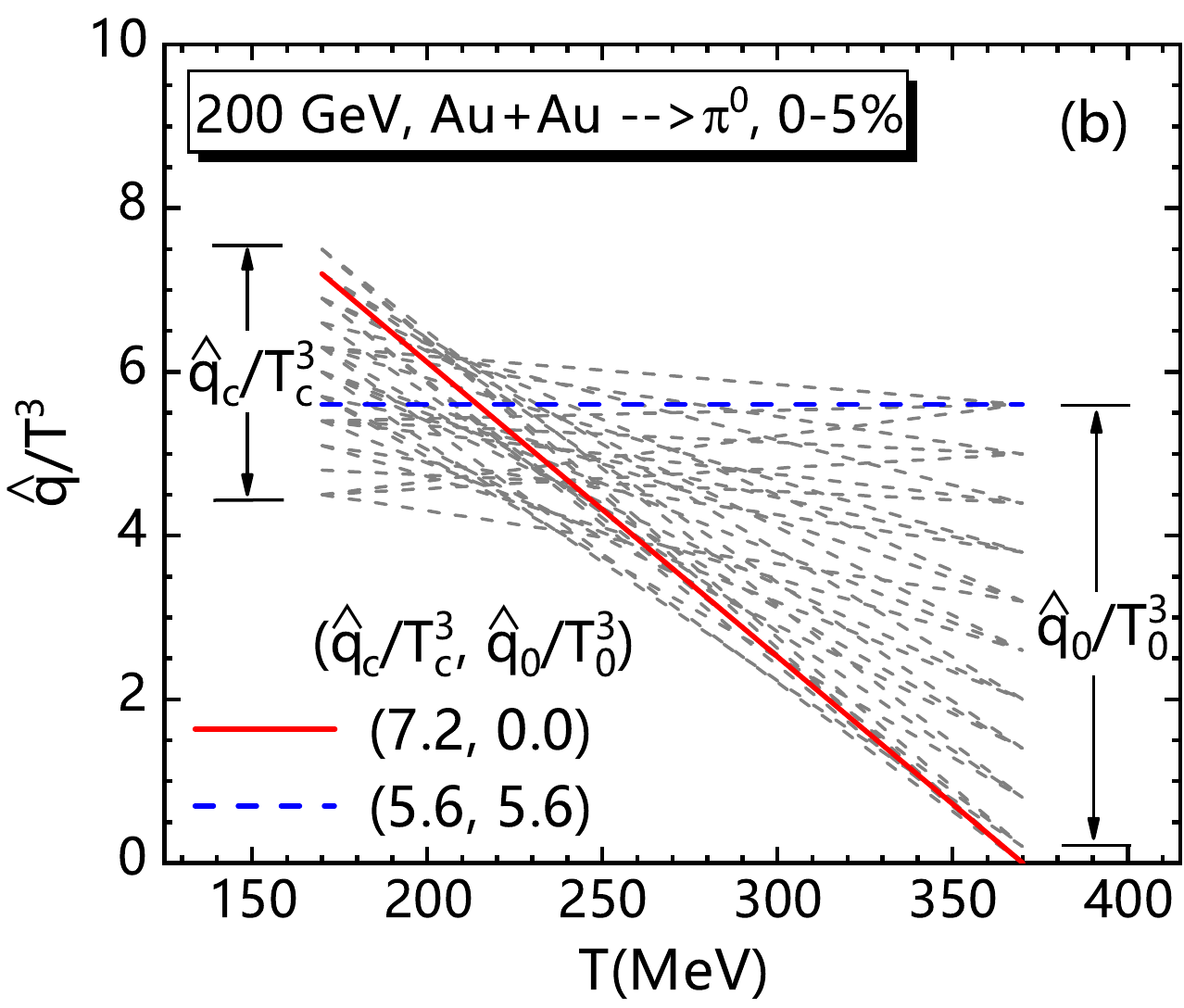}
\hspace{-1mm}
\includegraphics[width = 0.32\textwidth, height=0.27\textwidth]{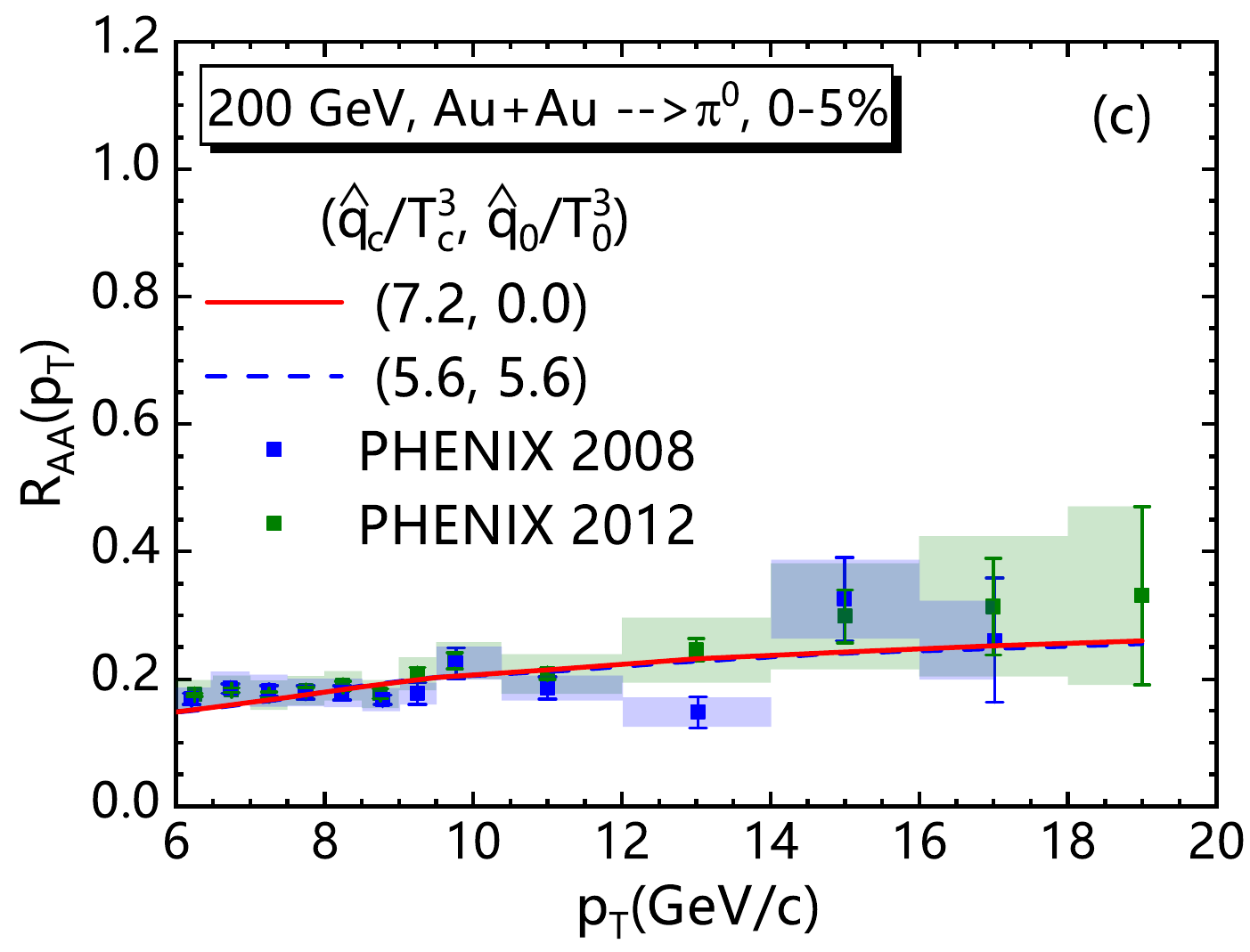}

\caption{(Linear-dependence) Panel (a): The $\chi^2/{\rm d.o.f}$ analyses for single hadron $R_{AA}(p_{\rm T})$ as a function of $\hat{q}_{\rm c}/T_{\rm c}^3$ and $\hat{q}_0/T_0^3$ from fitting to PHENIX data \cite{PHENIX:2008saf,PHENIX:2012jha} in the most central 0-5\% Au + Au collisions at $\sqrt{s_{\rm NN}}=200$~GeV. Panel (b): The scaled dimensionless jet transport parameters $\hat{q}/T^3$ as a function of medium temperature $T$ from the best fitting region of the panel (a).  Panel (c): The single hadron suppression factors $R_{AA}(p_{\rm T})$ with couples of $(\hat{q}_{\rm c}/T_{\rm c}^3,\hat{q}_0/T_0^3)=(7.2,0.0)$ (red solid curve) and $(5.6,5.6)$ (blue dashed curve) compared with PHENIX \cite{PHENIX:2008saf,PHENIX:2012jha} data.}

\label{fig:L-RHIC-0-5}
\end{figure*}

Current studies indicate that $\hat{q}_{\rm c}/T_{\rm c}^3$ at the critical temperature has a larger value, and that the value of $\hat{q}_0/T_0^3$ at the highest temperature is smaller. Therefore, we first choose $\hat{q}_{\rm c}/T_{\rm c}^3 \in [3.0, 9.0]$ and $\hat{q}_0/T_0^3 \in [0.2, 5.6]$ with a bin size of 0.3 and get 399 pairs of $(\hat{q}_{\rm c}/T_{\rm c}^3,\hat{q}_0/T_0^3)$ for Eq. ($\ref{eq:qhat-linear}$) and ($\ref{eq:qhat-f}$), with $\hat{q}_h = 0$. To determine the limit value of $\hat{q}_0/T_0^3$, we can expand it to zero. Therefore, we make 420 times of calculations for the suppression factor $R_{AA}(p_{\rm T})$ as a function of $p_{\rm T}$ for single hadrons produced in the most central 0–5\% Au + Au collisions at $\sqrt{s_{\rm NN}}=200$ GeV. Each result for $R_{AA}(p_{\rm T})$ with a given pair of $(\hat{q}_{\rm c}/T_{\rm c}^3,\hat{q}_0/T_0^3)$ provides a value of $\chi^2/{\rm d.o.f}$ to fit the experimental data for $R_{AA}(p_{\rm T})$ \cite{PHENIX:2008saf,PHENIX:2012jha}. As shown in Fig. \ref{fig:L-RHIC-0-5} (a) is such a 2-dimensional figure for $\chi^2/{\rm d.o.f}$ as functions of ($\hat{q}_{\rm c}/T_{\rm c}^3,\hat{q}_0/T_0^3$). Different colors represent different fitting values. The $\chi^2/{\rm d.o.f}$ value was minimized to near unity to determine the best-fitting couples of $(\hat{q}_{\rm c}/T_{\rm c}^3,\hat{q}_0/T_0^3)$.

\begin{figure*}[tbh]
\begin{center}
\includegraphics[width = 0.32\textwidth]{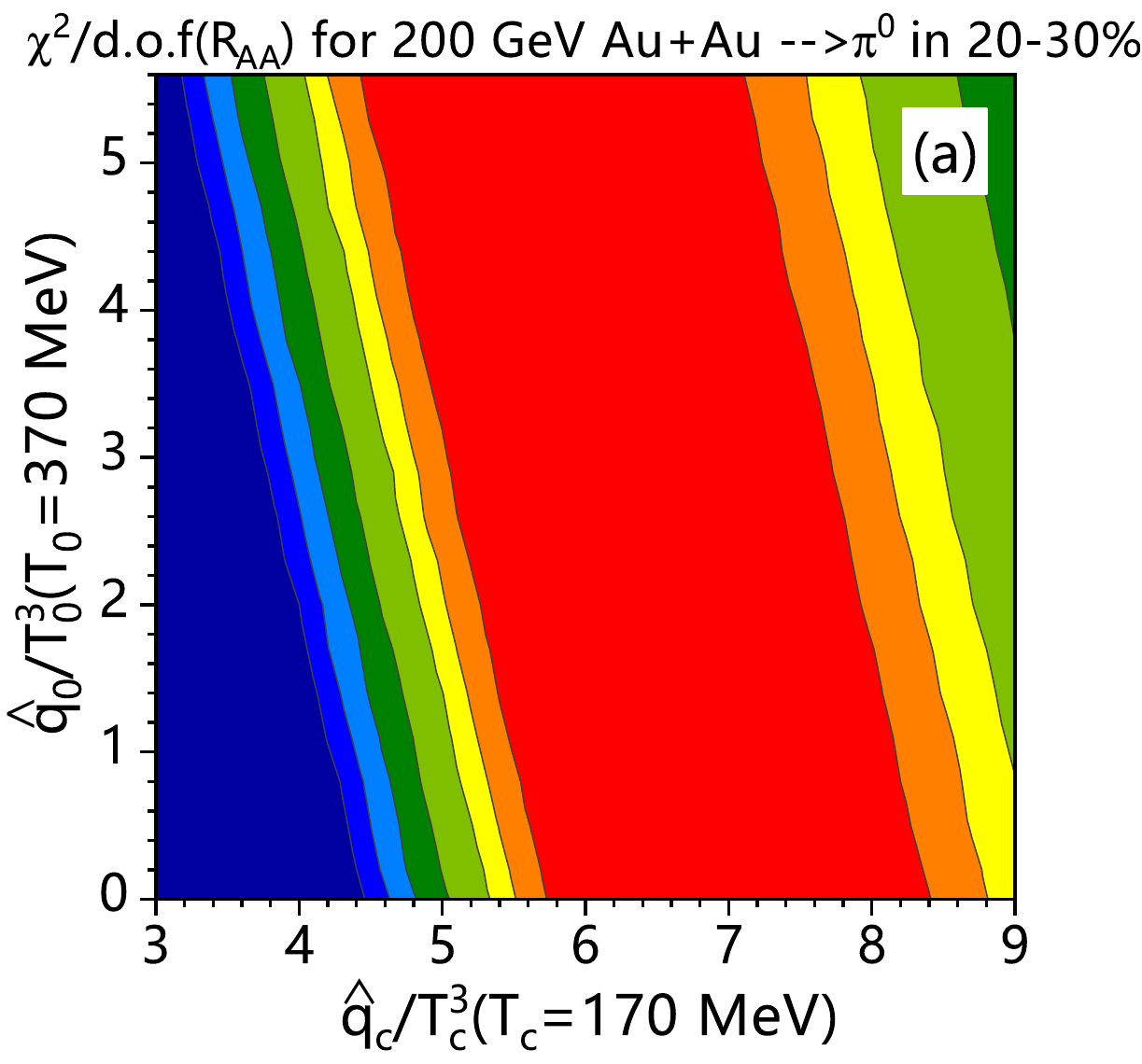}
\hspace{-3mm}
\includegraphics[width = 0.312\textwidth]{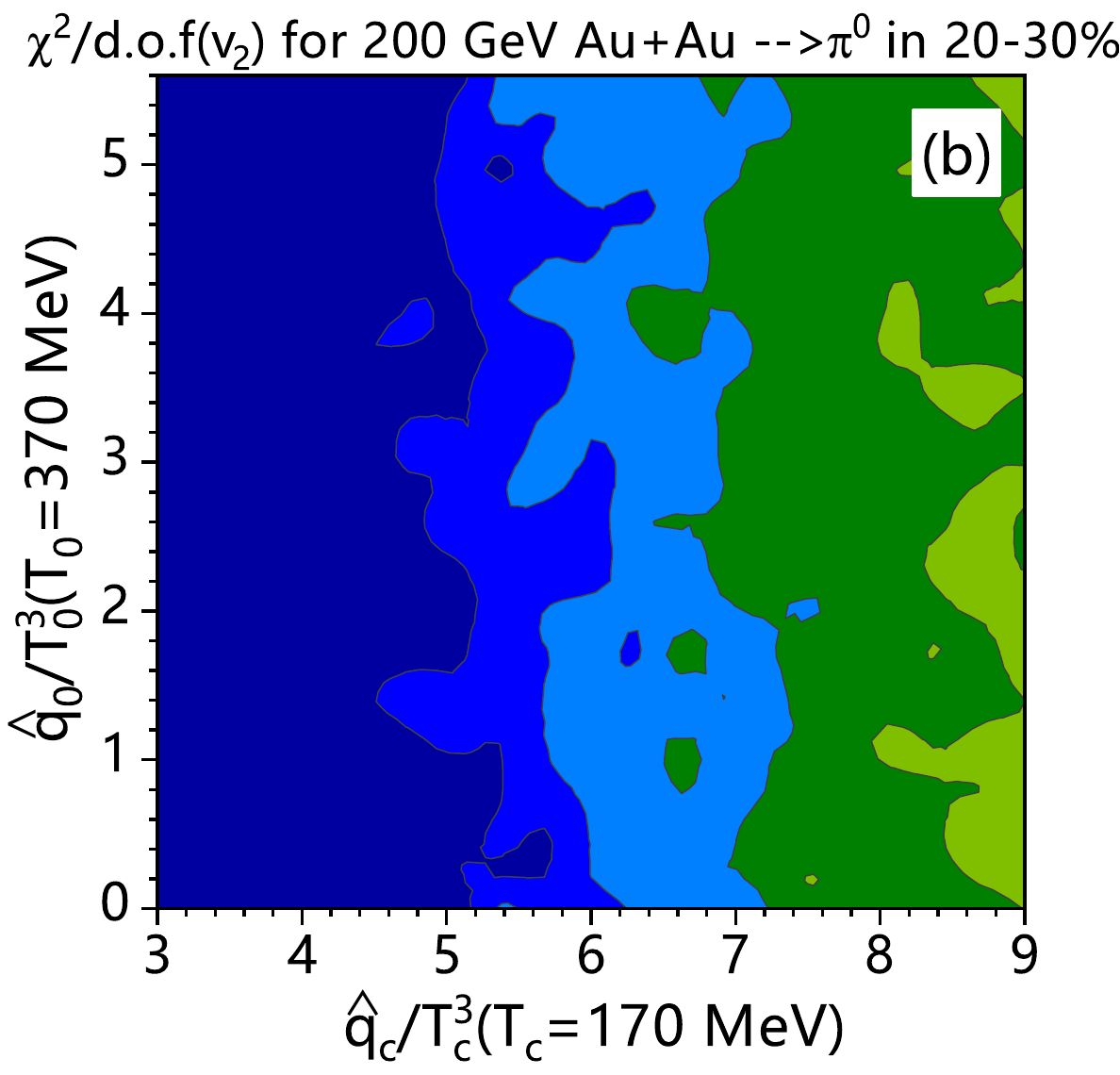}
\hspace{-3mm}
\includegraphics[width = 0.37\textwidth]{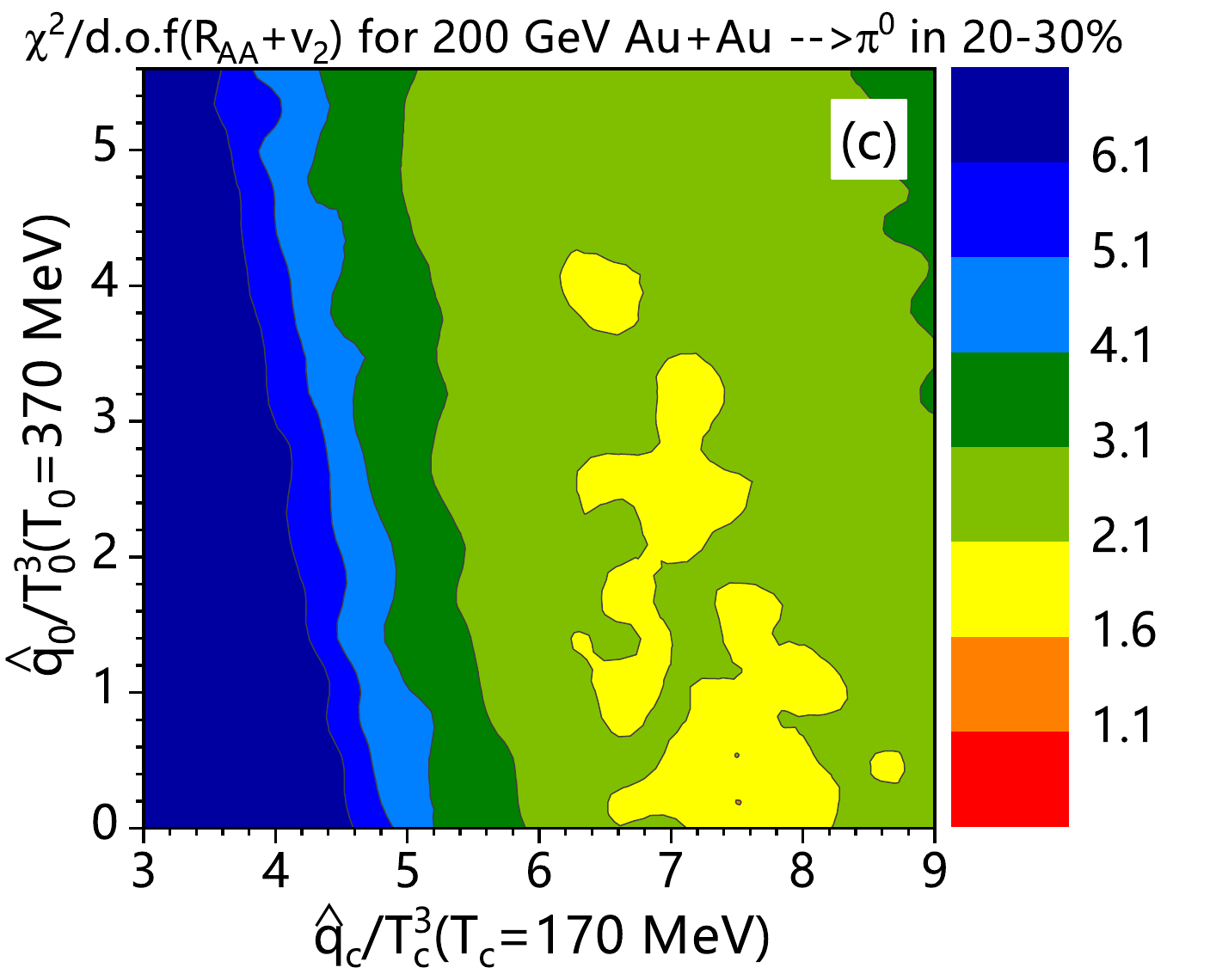}

\caption{(Linear-dependence) The $\chi^2/{\rm d.o.f}$ analyses for single hadron $R_{AA}(p_{\rm T})$ (panel (a)) and elliptic flow $v_{2}(p_{\rm T})$ (panel (b)) as a function of $\hat{q}_{\rm c}/T_{\rm c}^3$ and $\hat{q}_0/T_0^3$ from fitting to experimental data \cite{PHENIX:2008saf,PHENIX:2012jha,PHENIX:2010nlr} in 20-30\% Au + Au collisions at $\sqrt{s_{\rm NN}}=200$ GeV. The global $\chi^2/{\rm d.o.f}$ fitting results for both  $R_{AA}(p_{\rm T})$ and $v_{2}(p_{\rm T})$ are shown in the panel (c).}
\label{fig:L-RHIC-20-30-x2}
\end{center}
\end{figure*}

\begin{figure*}[tbh]
\begin{center}
\includegraphics[width = 0.30\textwidth,height=0.27\textwidth]{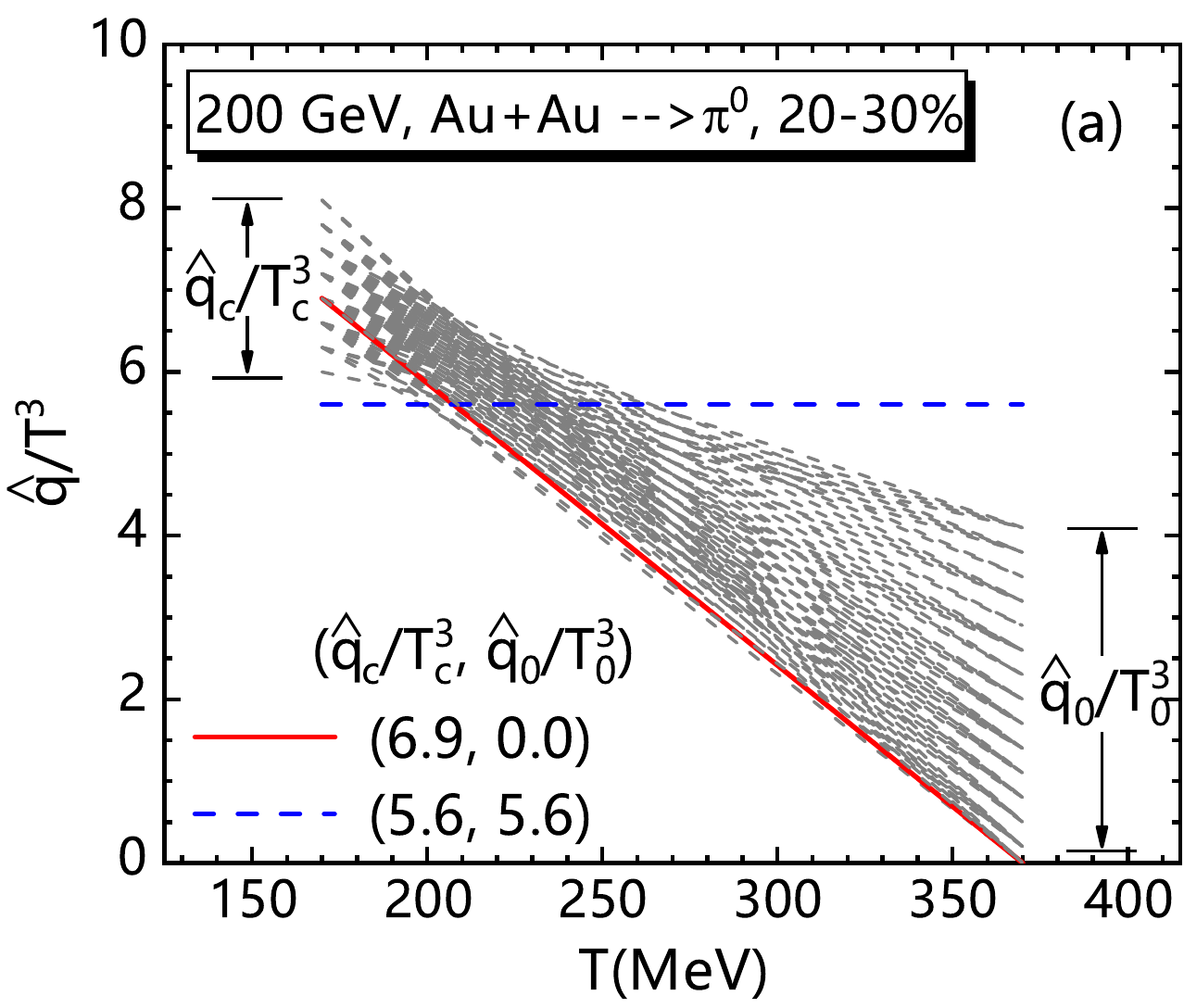}
\includegraphics[width = 0.32\textwidth,height=0.27\textwidth]{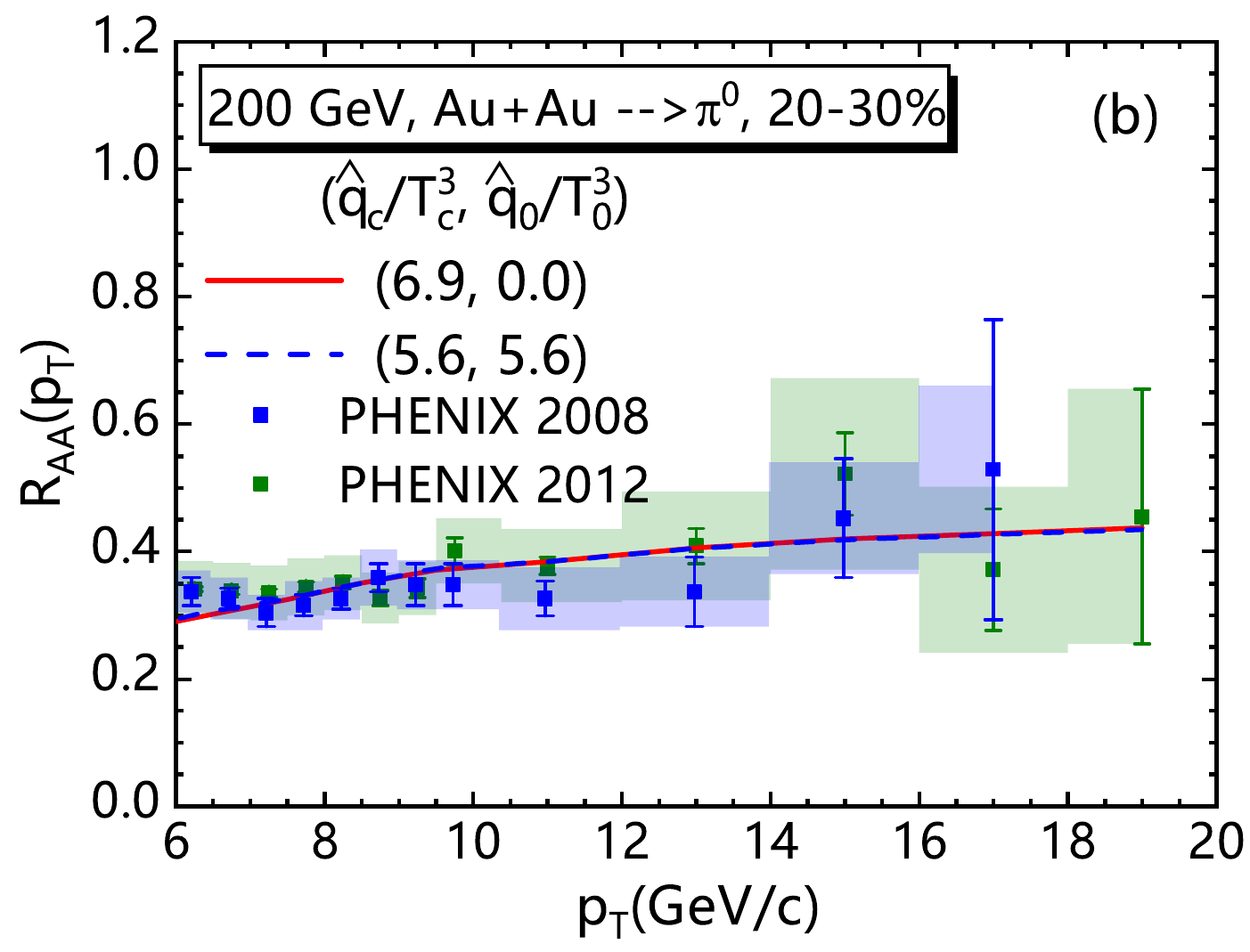}
\includegraphics[width = 0.32\textwidth,height=0.27\textwidth]{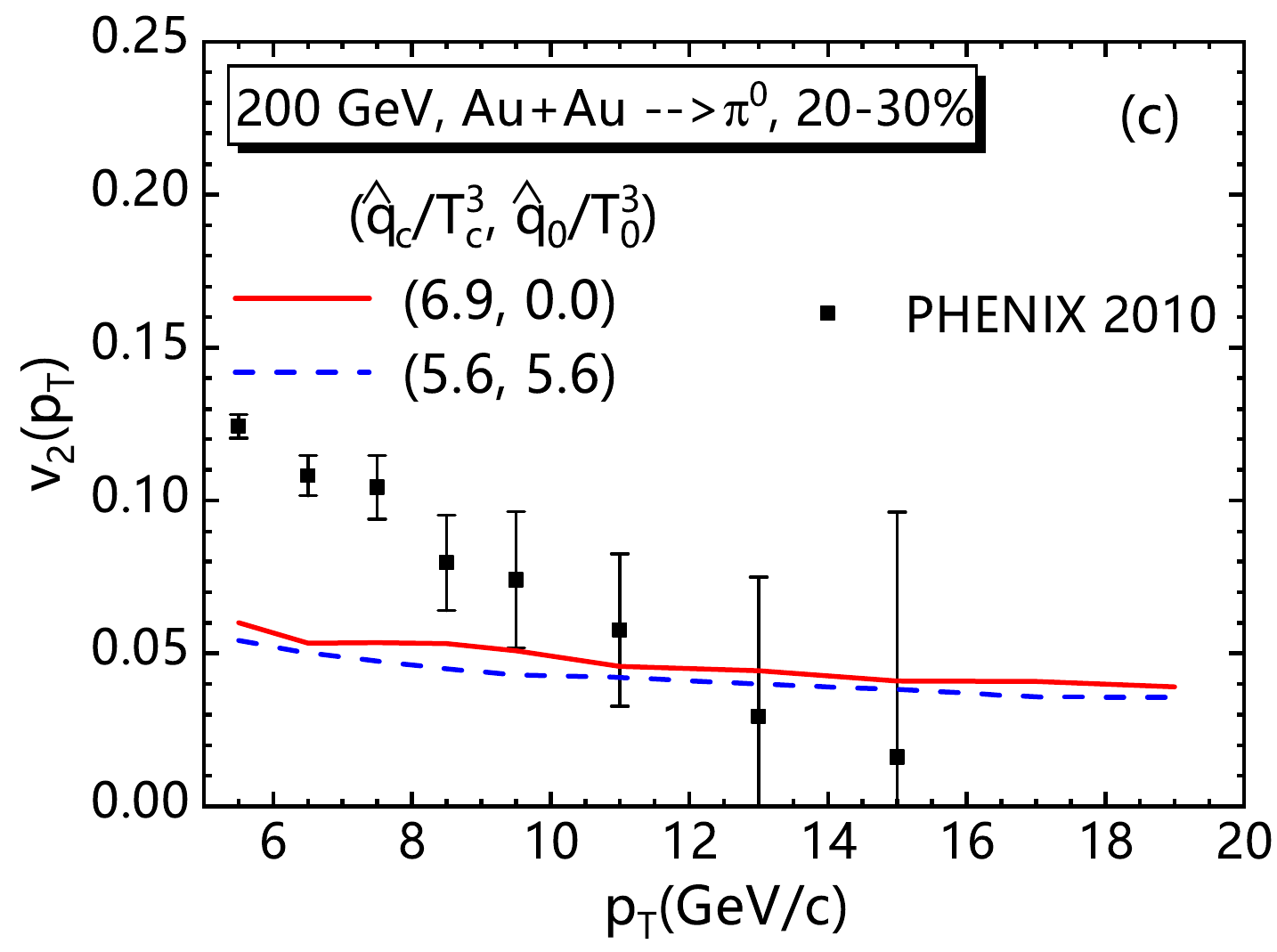}

\caption{(Linear-dependence) Panel (a): the scaled dimensionless jet transport parameters $\hat{q}/T^3$ as a function of medium temperature $T$ from the best fitting region of global $\chi^2$ fits of Fig. \ref{fig:L-RHIC-20-30-x2} (c) in 20-30\% Au + Au collisions at $\sqrt{s_{\rm NN}}=200$ GeV. The single hadron suppression factors $R_{AA}(p_{\rm T})$ and elliptic flow $v_{2}(p_{\rm T})$ are shown in panel (b) and (c), respectively, with couples of $(\hat{q}_{\rm c}/T_{\rm c}^3,\hat{q}_0/T_0^3)=(6.9,0.0)$ (red solid curve) and $(5.6,5.6)$ (blue dashed curve) compared with PHENIX \cite{PHENIX:2008saf,PHENIX:2012jha,PHENIX:2010nlr} data.}
\end{center}
\label{fig:L-RHIC-RAA-v2-20-30}
\end{figure*}

\begin{figure*}[tbh]
\begin{center}

\includegraphics[width = 0.37\textwidth]{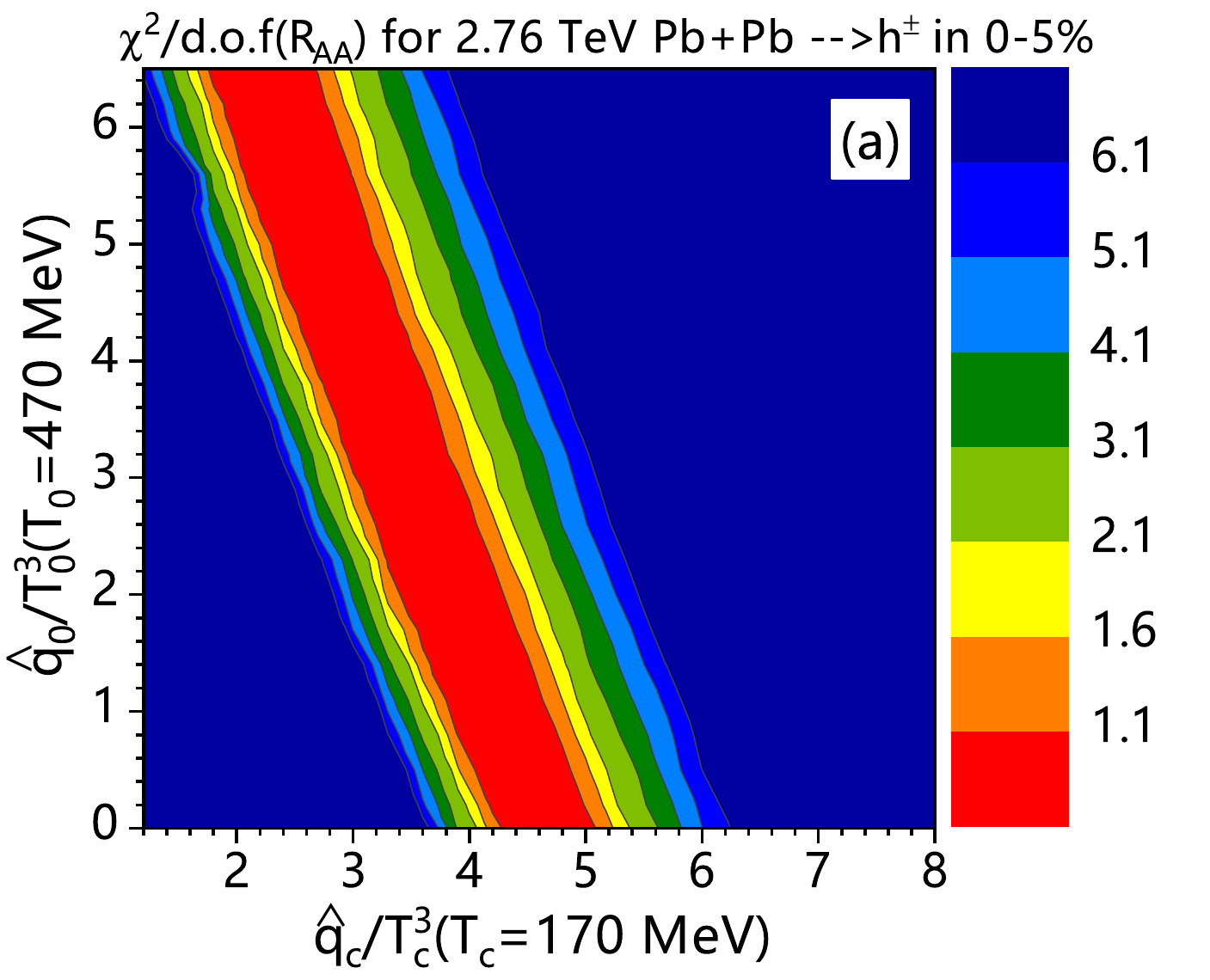}
\hspace{5mm}
\includegraphics[width = 0.324\textwidth,height=0.28\textwidth]{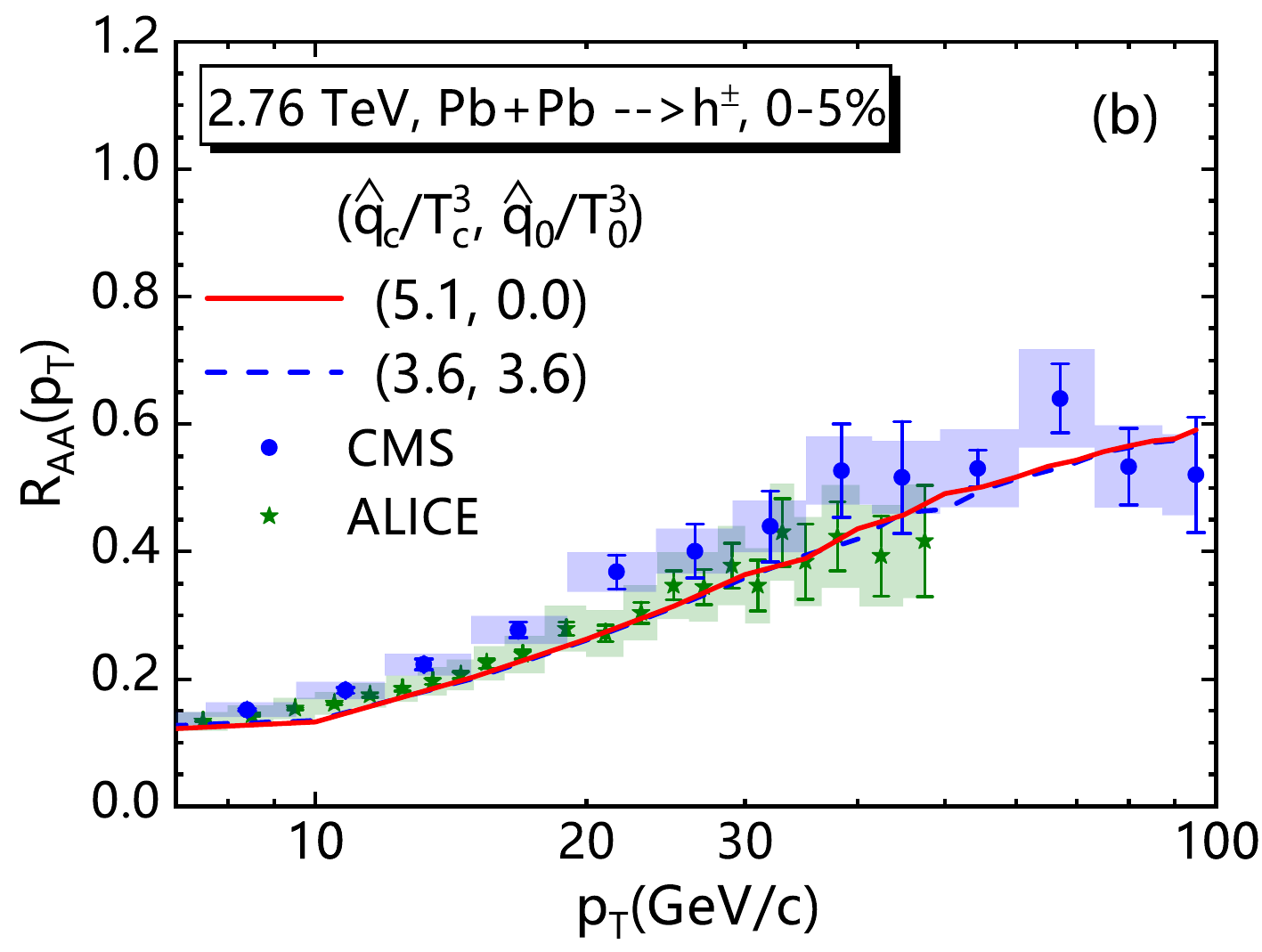}

\caption{(Linear-dependence) Panel (a): The $\chi^2/{\rm d.o.f}$ analyses for single hadron $R_{AA}(p_{\rm T})$ as a function of $\hat{q}_{\rm c}/T_{\rm c}^3$ and $\hat{q}_0/T_0^3$ from fitting to experimental data \cite{ALICE:2012aqc,CMS:2012aa} in the most central 0-5\% Pb + Pb collisions at $\sqrt{s_{\rm NN}}=2.76$~TeV. Panel (b): The single hadron suppression factors $R_{AA}(p_{\rm T})$ with couples of $(\hat{q}_{\rm c}/T_{\rm c}^3, \hat{q}_0/T_0^3)=(5.1,0.0)$ (red solid curve) and $(3.6,3.6)$ (blue dashed curve) compared with experimental data.}
\end{center}
\label{fig:L-LHC-0-5}
\end{figure*}

\begin{figure*}[tbh]
\begin{center}
\includegraphics[width = 0.32\linewidth]{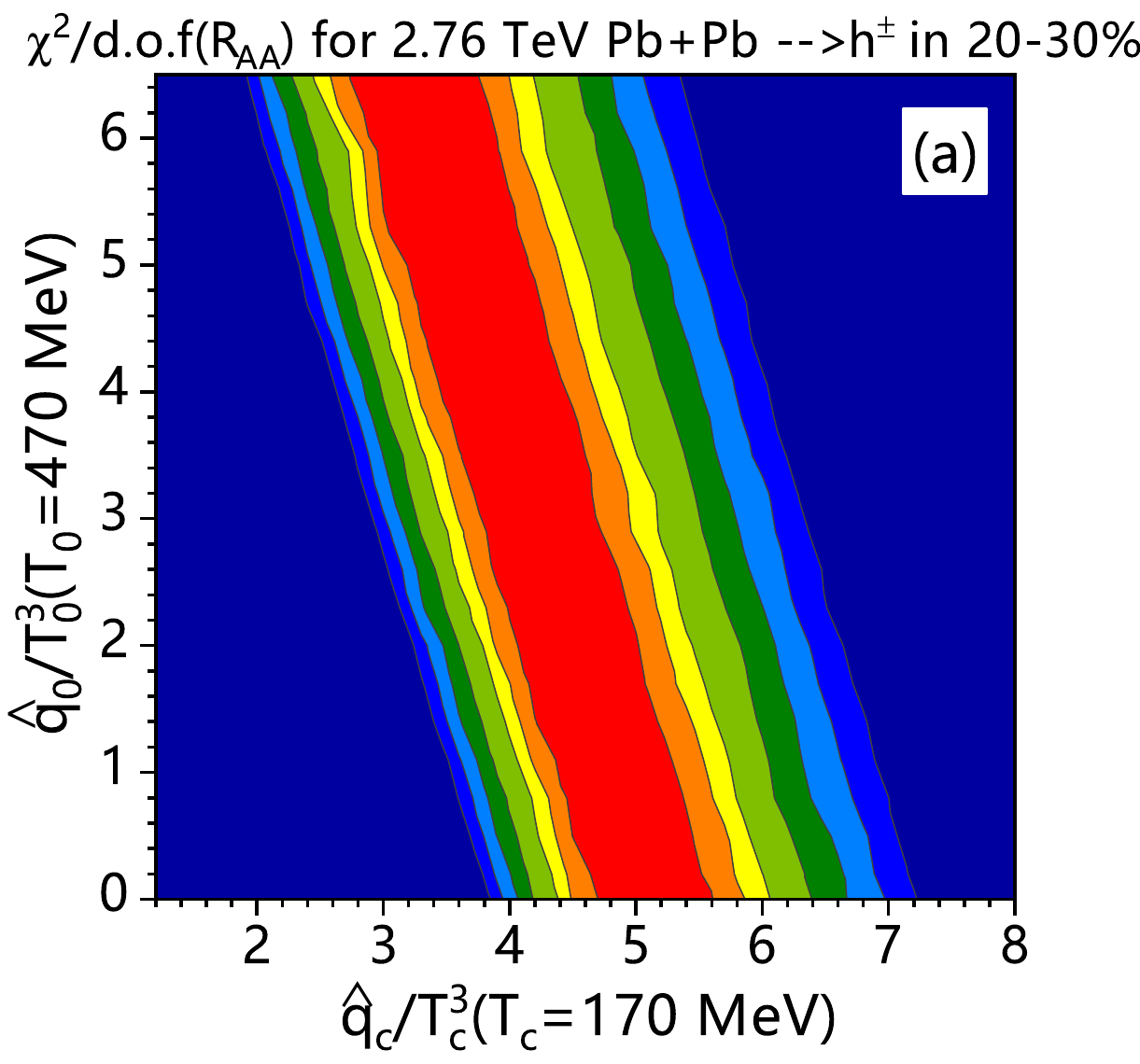}
\hspace{-3mm}
\includegraphics[width = 0.312\linewidth]{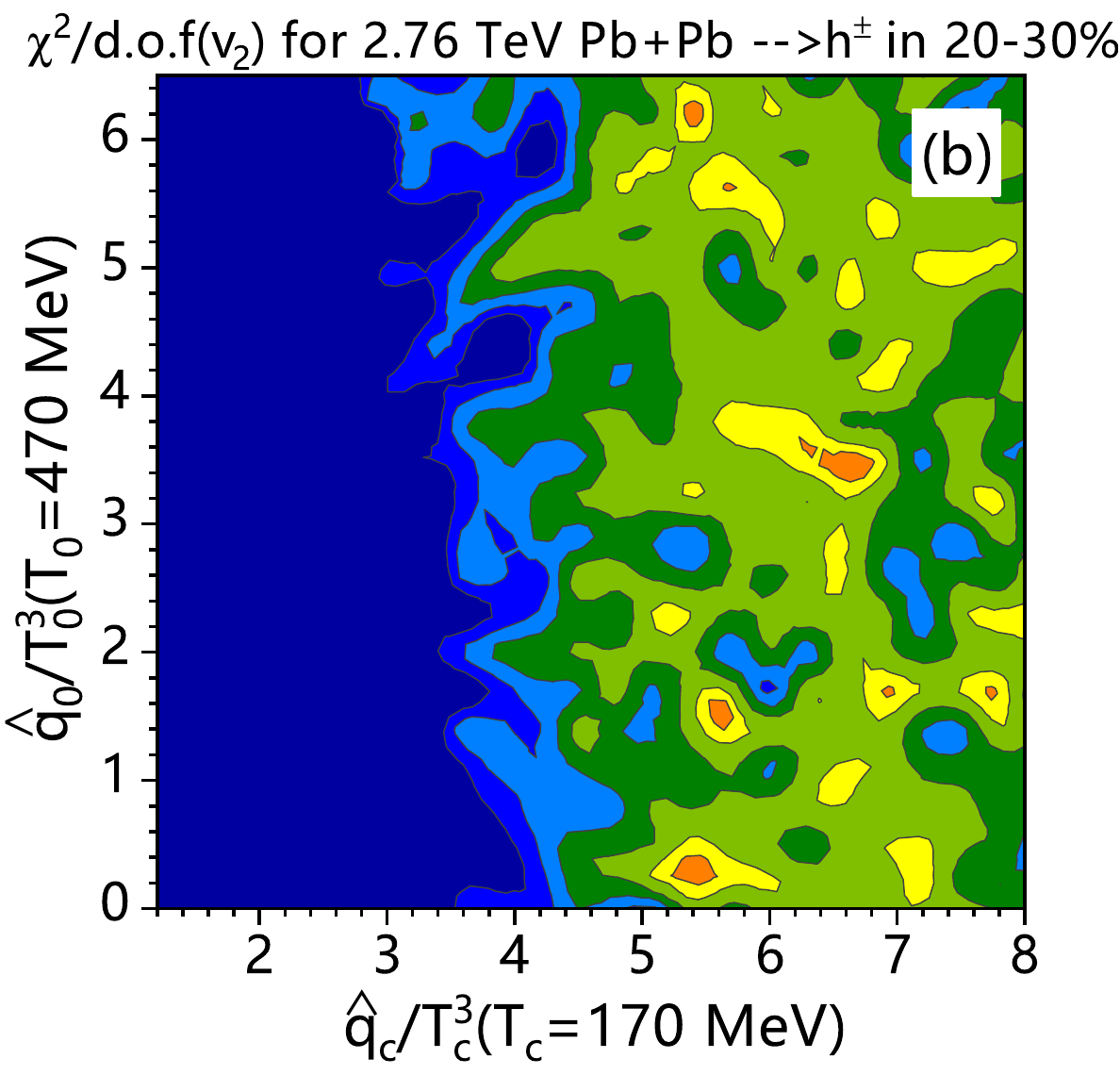}
\hspace{-3mm}
\includegraphics[width = 0.37\linewidth]{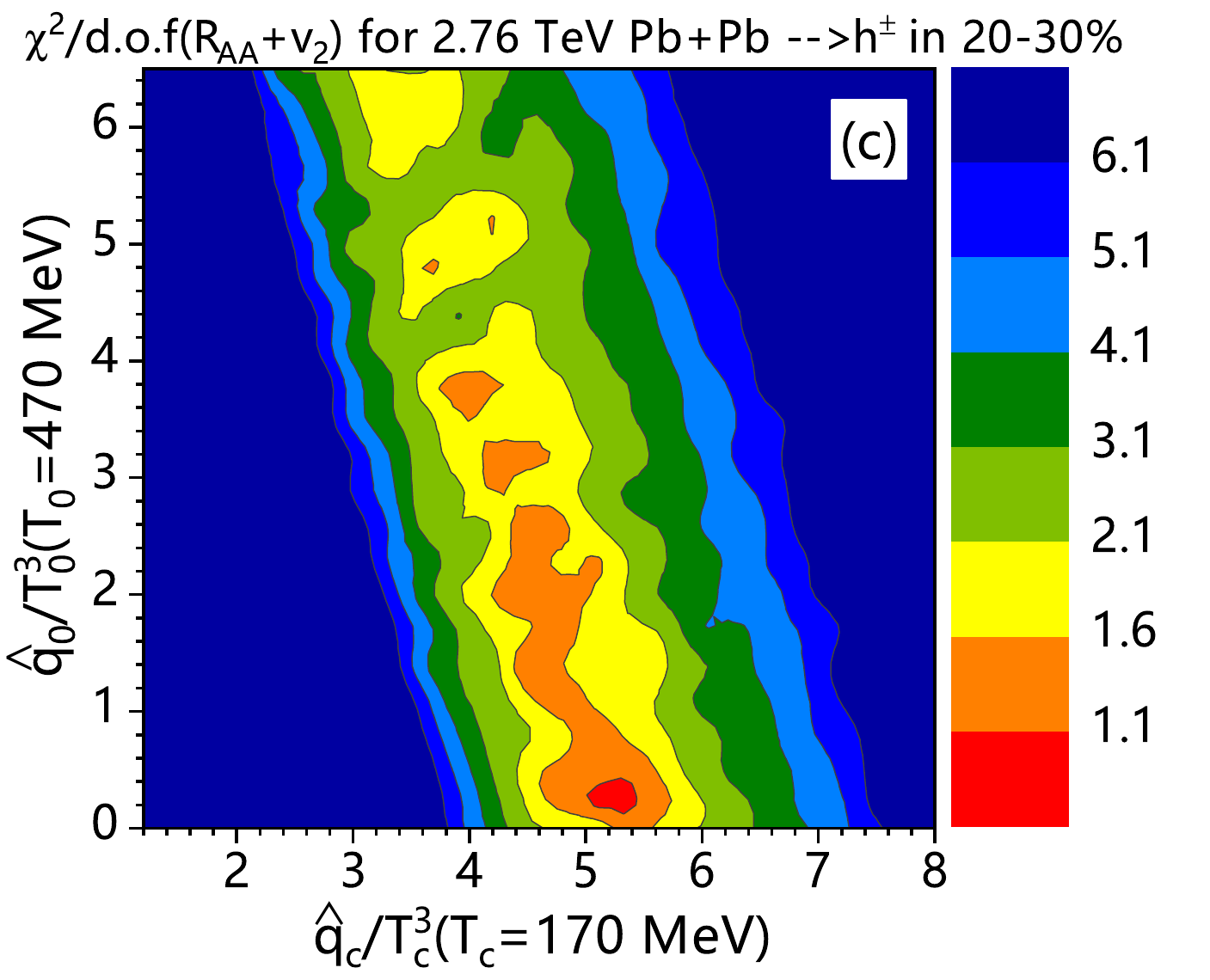}

\caption{(Linear-dependence) The $\chi^2/{\rm d.o.f}$ analyses of single hadron $R_{AA}(p_{\rm T})$ (panel (a)) and elliptic flow $v_{2}(p_{\rm T})$ (panel (b)) as a function of $\hat{q}_{\rm c}/T_{\rm c}^3$ and $\hat{q}_0/T_0^3$ from fitting to experimental data \cite{ALICE:2012aqc,CMS:2012aa,CMS:2012tqw,ALICE:2012vgf} in 20-30\% Pb + Pb collisions at $\sqrt{s_{\rm NN}}=2.76$~TeV.  The global $\chi^2/{\rm d.o.f}$ fitting results for both  $R_{AA}(p_{\rm T})$ and $v_{2}(p_{\rm T})$ are shown in the panel (c).}
\end{center}
\label{fig:L-LHC-20-30-x2}
\end{figure*}

\begin{figure*}[tbh]
\begin{center}
\includegraphics[width = 0.30\textwidth,height=0.27\textwidth]{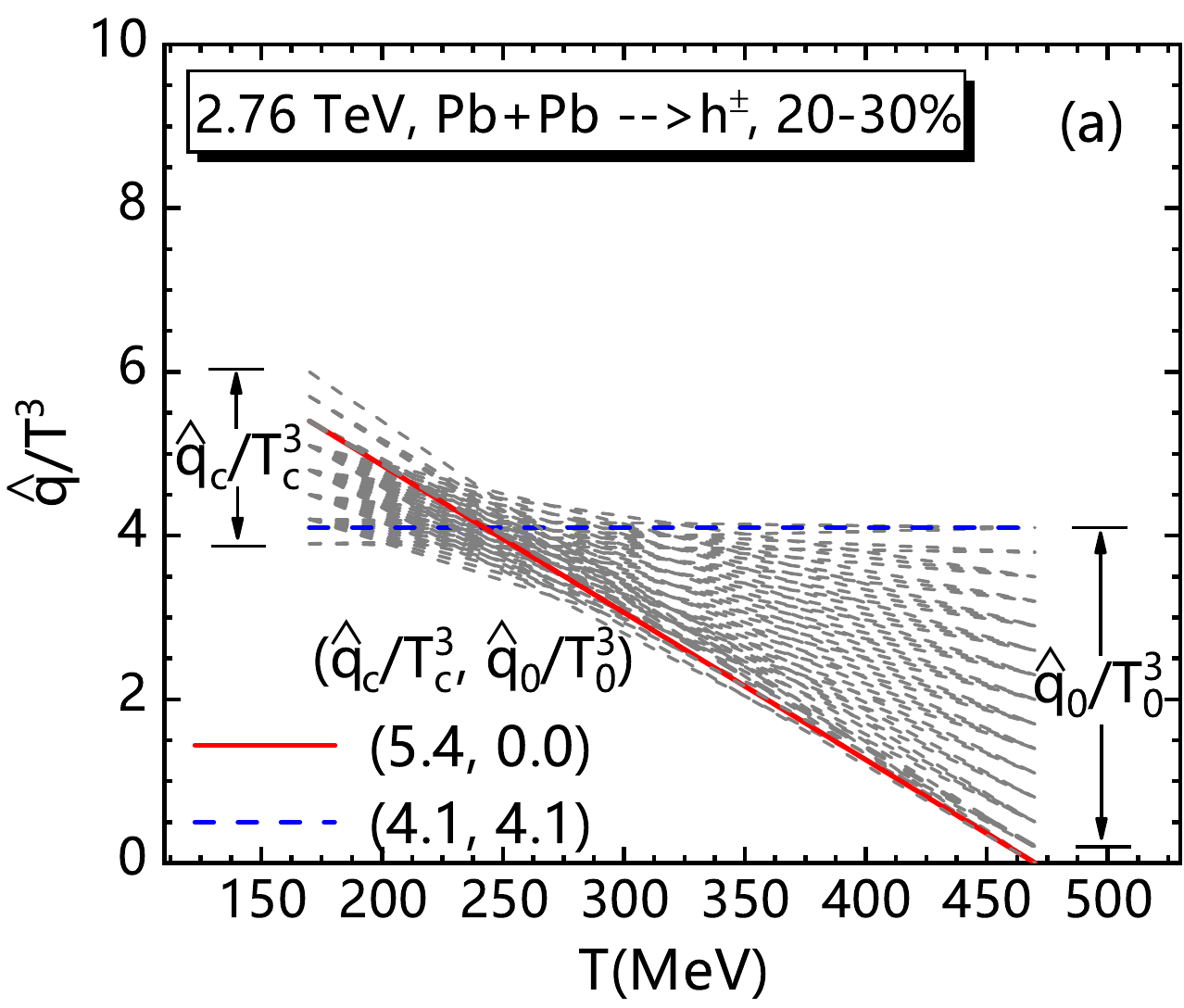}
\includegraphics[width = 0.32\textwidth,height=0.27\textwidth]{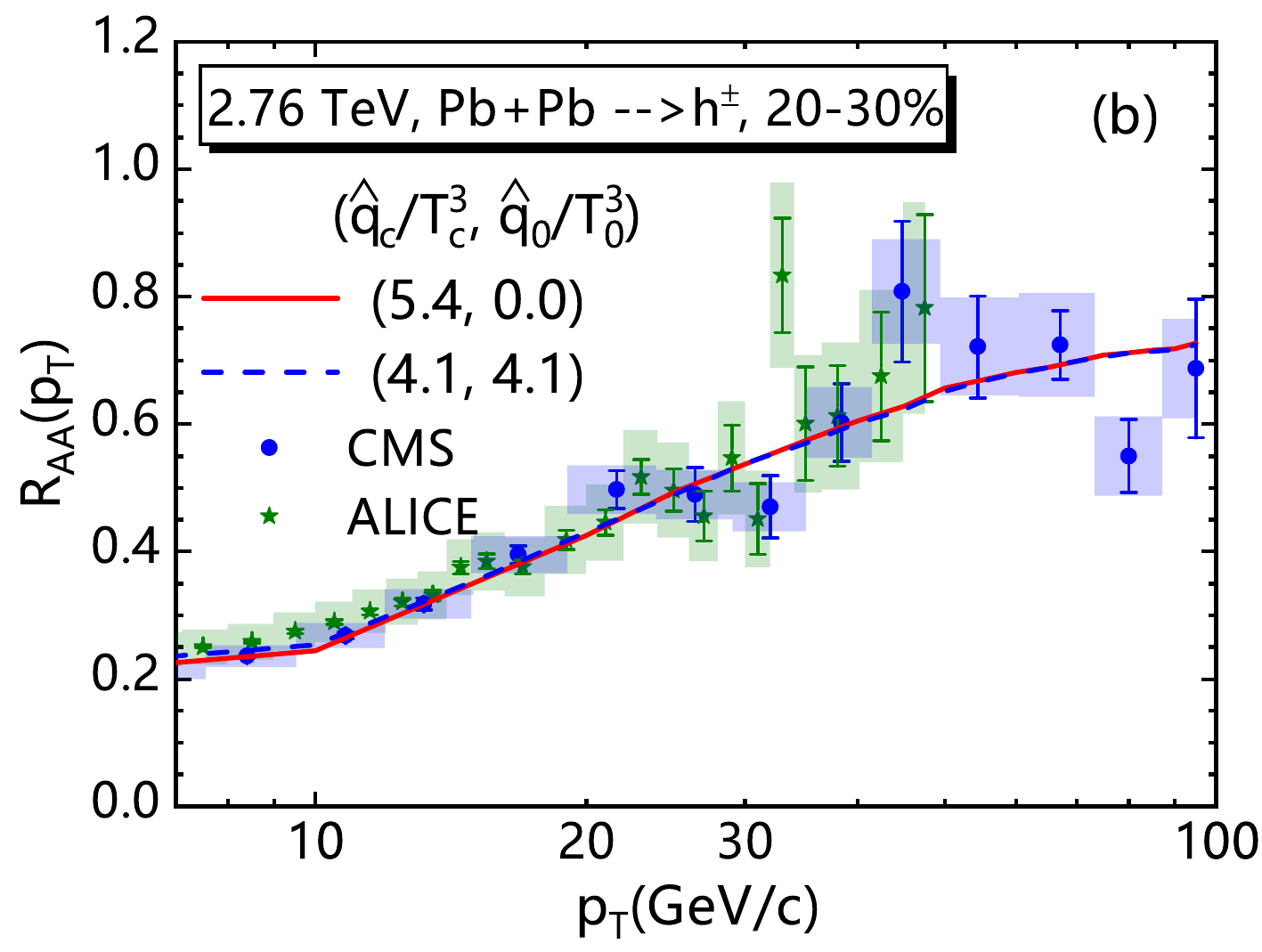}
\includegraphics[width = 0.32\textwidth,height=0.27\textwidth]{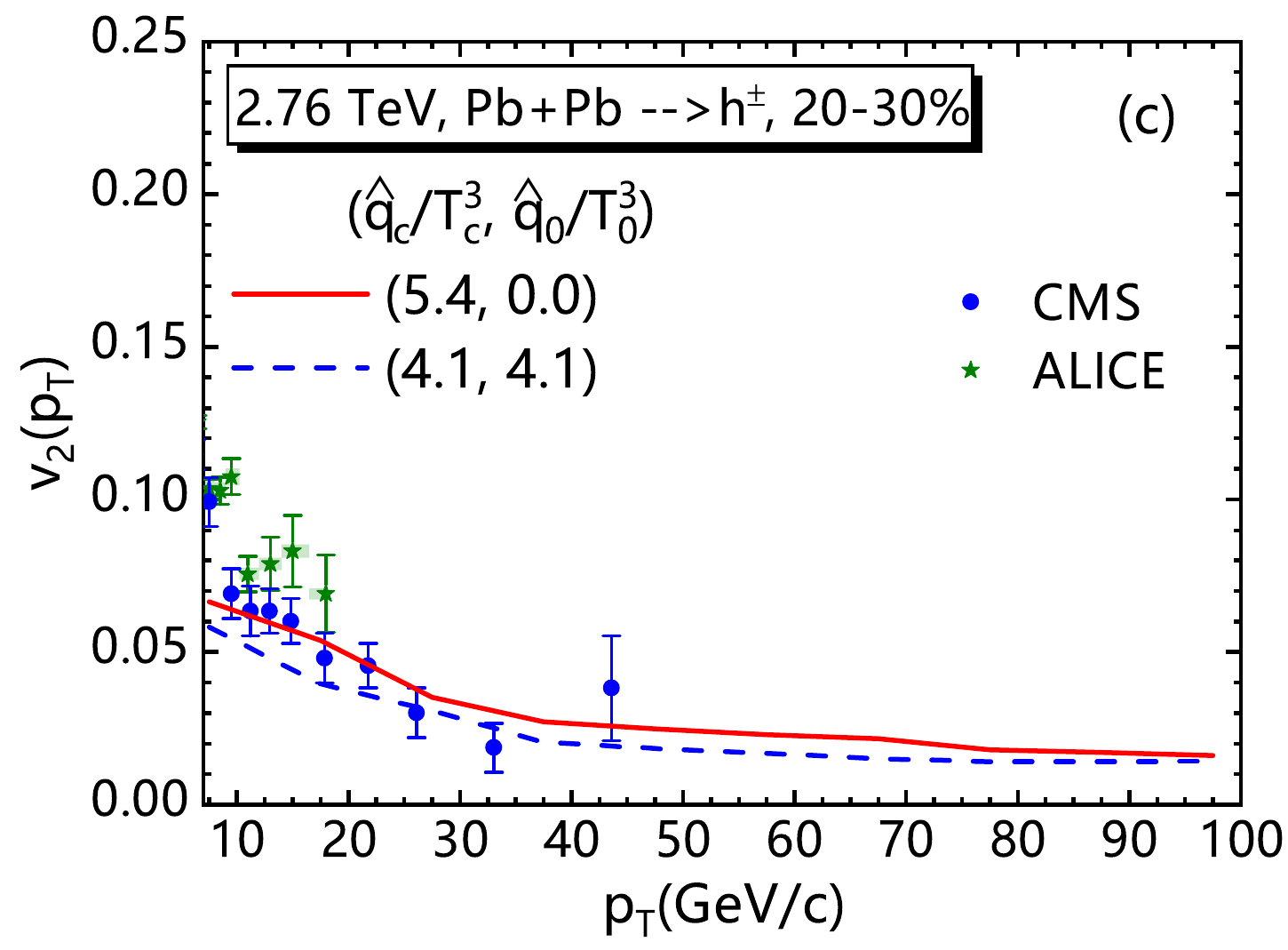}

\caption{(Linear-dependence) In 20-30\% Pb + Pb collisions at $\sqrt{s_{\rm NN}}=2.76$ TeV, the scaled dimensionless jet transport parameters $\hat{q}/T^3$ as a function of medium temperature $T$ from the best fitting region of global $\chi^2$ fits of Fig. \ref{fig:L-LHC-20-30-x2} (c) are shown in the panel (a). The single hadron suppression factors $R_{AA}(p_{\rm T})$ and elliptic flow $v_{2}(p_{\rm T})$ are shown in the (b) and (c) panels, respectively, with couples of $(\hat{q}_{\rm c}/T_{\rm c}^3,\hat{q}_0/T_0^3)=(5.4,0.0)$ (red soild curve) and $(4.1,4.1)$ (blue dashed curve) compared with experimental data \cite{ALICE:2012aqc,CMS:2012aa,CMS:2012tqw,ALICE:2012vgf}.}
\end{center}
\label{fig:L-LHC-RAA-v2-20-30}
\end{figure*}

The best fit given by the red region indicates that the single hadron $R_{AA}(p_{\rm T})$ is more sensitive to the value of $\hat{q}_{\rm c}/T_{\rm c}^3 $ than $\hat{q}_0/T_0^3$.
However, the fitting fails to obtain a unique pair of $(\hat{q}_{\rm c}/T_{\rm c}^3,\hat{q}_0/T_0^3)$ for an explicit dependence form of $\hat{q}/T^3$ on $T$ only through the constraint of single hadron $R_{AA}(p_{\rm T})$.
To demonstrate the different linear temperature dependencies of $\hat{q}/T^3$ for the same suppression of single hadrons, in Fig. \ref{fig:L-RHIC-0-5} (b) we draw the gray dashed curves for $\hat{q}/T^3$ as a function of $T$, which are constrained by the best-fitting region of $\chi^2/{\rm d.o.f}$.

Among the gray dashed curves, we choose one horizontal line (blue) for a constant $\hat{q}/T^3$ with $(\hat{q}_{\rm c}/T_{\rm c}^3,\hat{q}_0/T_0^3)=(5.6,5.6)$ and one leaning line (red) for a linear $T$ dependence of $\hat{q}/T^3$ with ${(\hat{q}_{\rm c}/T_{\rm c}^3,\hat{q}_0/T_0^3)=(7.2,0.0)}$. Using these two pairs of $(\hat{q}_{\rm c}/T_{\rm c}^3,\hat{q}_0/T_0^3)$, we obtain almost the same $R_{AA}(p_{\rm T})$ as in Fig. \ref{fig:L-RHIC-0-5} (c) which shows that the single hadron suppression is a consequence of total jet energy loss and is not sensitive to the $T$ dependence of $\hat{q}/T^3$ in central Au + Au collisions.

Due to the jet path length and medium density dependence in the jet trajectory inside the hot medium, the jet energy loss in noncentral Au + Au collisions exhibits azimuthal anisotropy. The hadron suppression
depends on the azimuthal angle concerning the reaction plane, thus leading to azimuthal anisotropy in the high-$p_{\rm T}$ hadron spectra. The same energy loss mechanism permits to perform a global fit to constrain $(\hat{q}_{\rm c}/T_{\rm c}^3,\hat{q}_0/T_0^3)$ with both the suppression factor $R_{AA}(p_{\rm T})$ and elliptic flow parameter $v_2(p_{\rm T})$ for large $p_{\rm T}$ hadrons in noncentral A + A collisions.
With the same couples of $(\hat{q}_{\rm c}/T_{\rm c}^3,\hat{q}_0/T_0^3)$ as in 0-5\% centrality, we simultaneously make {420} times of calculations for $R_{AA}(p_{\rm T})$ and $v_2(p_{\rm T})$ as a function of $p_{\rm T}$ to fit to the experimental data \cite{PHENIX:2008saf,PHENIX:2012jha,PHENIX:2010nlr} in 20-30\% Au + Au collisions, and get the $\chi^2/{\rm d.o.f}$ results for $R_{AA}(p_{\rm T})$ as shown in Fig. \ref{fig:L-RHIC-20-30-x2} (a) and $v_{2}(p_{\rm T})$ in Fig. \ref{fig:L-RHIC-20-30-x2} (b), respectively.
The $\chi^2/{\rm d.o.f}$ fitting for $R_{AA}$ in noncentral collisions is similar to that in central collisions. This implies that only $R_{AA}(p_{\rm T})$ constraint does not give an explicit dependence form of $\hat{q}/T^3$ on $T$.
The $\chi^2/{\rm d.o.f}$ fitting of $v_2$ in Fig. \ref{fig:L-RHIC-20-30-x2} (b) shows that the data of elliptic flow $v_2(p_{\rm T})$ favor larger $\hat{q}_{\rm c}/T_{\rm c}^3$ and are almost insensitive to $\hat{q}_0/T_0^3$.
Global $\chi^2/{\rm d.o.f}$ fitting was performed for both $R_{AA}(p_{\rm T})$ and $v_2(p_{\rm T})$ in Fig. \ref{fig:L-RHIC-20-30-x2} (c) in which the limited yellow region is found to constrain $(\hat{q}_{\rm c}/T_{\rm c}^3,\hat{q}_0/T_0^3)$.

Shown in Fig. \ref{fig:L-RHIC-RAA-v2-20-30} (a) is the scaled dimensionless jet transport parameters $\hat{q}/T^3$ as a function of medium temperature $T$ from the best fitting region of global $\chi^2$ fits of Fig. \ref{fig:L-RHIC-20-30-x2} (c). The blue dashed curve is also for the constant-dependence case with $(\hat{q}_{\rm c}/T_{\rm c}^3,\hat{q}_0/T_0^3)=(5.6,5.6)$, and the red solid curve for a linear $T$ dependence of $\hat{q}/T^3$ with ${(\hat{q}_{\rm c}/T_{\rm c}^3,\hat{q}_0/T_0^3)=(6.9,0.0)}$. These two dependence forms yielded almost the same $R_{AA}(p_{\rm T})$ as shown in Fig. \ref{fig:L-RHIC-RAA-v2-20-30} (b), which is similar to the situation in central collisions. However, these two dependencies of $\hat{q}/T^3$ provide different contributions to $v_2(p_{\rm T})$ as shown in Fig. \ref{fig:L-RHIC-RAA-v2-20-30} (c). Numerical results show that the linearly-decreasing $T$ dependence of $\hat{q}/T^3$ with ${(\hat{q}_{\rm c}/T_{\rm c}^3,\hat{q}_0/T_0^3)=(6.9,0.0)}$ makes an enhancement by 10\% for $v_2(p_{\rm T})$ comparing to the constant dependence case with $(\hat{q}_{\rm c}/T_{\rm c}^3,\hat{q}_0/T_0^3)=(5.6,5.6)$. This linearly decreasing $T$-dependence of $\hat{q}/T^3$ with ${(\hat{q}_{\rm c}/T_{\rm c}^3,\hat{q}_0/T_0^3)=(6.9,0.0)}$ indicates that more energy loss occurs near the critical temperature $T_{\rm c}$.

\subsection{Fit $R_{AA}$ and $v_2$ at the LHC}

Similarly, we present the relevant results for the Pb + Pb collisions at $\sqrt{s_{\rm NN}}=2.76$~TeV.
{Here, we choose $\hat{q}_{\rm c}/T_{\rm c}^3 \in [1.2, 8.1]$ and $\hat{q}_0/T_0^3 \in [0.2, 6.5]$ with the same bin size of 0.3, and further include $\hat{q}_0/T_0^3=0.0$ to obtain 552 couples of $(\hat{q}_{\rm c}/T_{\rm c}^3,\hat{q}_0/T_0^3)$ for Eq. ($\ref{eq:qhat-linear}$) and ($\ref{eq:qhat-f}$), with $\hat{q}_h = 0$.}
The $\chi^2/{\rm d.o.f}$ results for central 0–5\% Pb + Pb collisions were performed on single hadron suppression factors, as shown in Fig. \ref{fig:L-LHC-0-5} (a). The best-fitting contour is similar to that shown in Fig. \ref{fig:L-RHIC-0-5} (a) but with a smaller $\hat{q}_{\rm c}/T_{\rm c}^3$.
With constant and linear forms for $\hat{q}/T^3$, we again obtain the same single hadron suppression, as shown in Fig. \ref{fig:L-LHC-0-5} (b).

For 20–30\% Pb + Pb collisions, $\chi^2/{\rm d.o.f}$ fitting for only $R_{AA}(p_{\rm T})$ or $v_2(p_{\rm T})$ and the global fitting for both are shown in Fig. \ref{fig:L-LHC-20-30-x2} (a), (b) and (c), respectively. Similar to noncentral Au + Au collisions, both separated $\chi^2/{\rm d.o.f}$ fitting for $R_{AA}(p_{\rm T})$ and $v_2(p_{\rm T})$ cannot provide a clear constraint on the $T$-dependence of $\hat{q}/T^3$. However, the difference between $\chi^2/{\rm d.o.f}(R_{AA})$, $\chi^2/{\rm d.o.f}(v_2)$ shows that the data of $v_2(p_{\rm T})$ prefer a larger jet energy loss near $T_{\rm c}$ and are insensitive to changes in $\hat{q}_0/T_0^3$. Consequently, the global fits for both $R_{AA}(p_{\rm T})$ and $v_2(p_{\rm T})$ impose a constraint to some extent on the $T$-dependence of $\hat{q}/T^3$ as shown in Fig. \ref{fig:L-LHC-20-30-x2} (c), similarly to Fig. \ref{fig:L-RHIC-20-30-x2} (c).

Choosing $\chi^2/{\rm d.o.f}<1.6$ in Fig. \ref{fig:L-LHC-20-30-x2} (c) for the best fitting, one can get the curves for the $T$ dependence of $\hat{q}/T^3$ in Fig. \ref{fig:L-LHC-RAA-v2-20-30} (a). We again observed a tendency for $\hat{q}/T^3$ to decrease with an increase in the local temperature along the jet trajectory. Among the best-fitting values, selecting ${(\hat{q}_{\rm c}/T_{\rm c}^3,\hat{q}_0/T_0^3)=(5.4,0.0)}$ for the linear $T$ dependence and $(4.1,4.1)$ for the constant dependence, we calculate the $R_{AA}(p_{\rm T})$ and $v_2(p_{\rm T})$ as a function of $p_{\rm T}$ shown in Fig. \ref{fig:L-LHC-RAA-v2-20-30} (b) and (c), respectively. Two almost identical $R_{AA}(p_{\rm T})$ were obtained, whereas $v_2(p_{\rm T})$ was enhanced by 10\% because of the larger jet energy loss near $T_{\rm c}$ for the linearly decreasing $T$ dependence of $\hat{q}/T^3$ at the LHC.

Regardless of whether in Pb + Pb or Au + Au collisions, $R_{AA}(p_{\rm T})$ and $v_2(p_{\rm T})$ are both more sensitive to the jet energy loss near the critical temperature $T_{\rm c}$ than near the initial highest temperature $T_0$. The data for $v_2(p_{\rm T})$ prefer larger values of $\hat{q}_{\rm c}/T_{\rm c}^3$. Furthermore, the anisotropy of the final-state hadrons at high transverse momentum can be strengthened up to 10\% by increasing the jet energy loss near $T_{\rm c}$ with a linearly decreasing $T$ dependence of $\hat{q}/T^3$.

\subsection{Jet energy loss distribution}

Given a parton jet with any creation site and any moving direction in the initial hard scattering, we consider the jet energy loss distribution when propagating through the hot medium. The average energy loss rate in the jet trajectory is given by

\begin{eqnarray}
\langle \frac{d\Delta E }{ d\tau } \rangle = \frac{\int d\phi \int d^2r t_A(\vec{r})t_B(\vec{r}+\vec{b}) d\Delta E(\vec{r}+\vec{n}\tau) / d\tau }{\int d\phi \int d^2r t_A(\vec{r})t_B(\vec{r}+\vec{b})},\nonumber\\
\label{eq:aver-dEdt}
\end{eqnarray}
where $\Delta E(\vec{r})$ is given by Eq. ($\ref{eq:De}$), $\vec{r}$ is the initial creation point for an energy-given jet, and $\vec{n}$ is the unit vector of the jet movement direction $\phi$ which is the same as that in Eq. ($\ref{eq:v2}$). The average cumulative energy loss for the jet traversing the medium is then given as
\begin{eqnarray}
\langle \Delta E (\tau ) \rangle = \frac{\int d\phi \int d^2r t_A(\vec{r})t_B(\vec{r}+\vec{b}) \int_{\tau_0}^{\tau} d\tau \frac{d\Delta E(\vec{r}+\vec{n}\tau) }{ d\tau}}{\int d\phi \int d^2r t_A(\vec{r})t_B(\vec{r}+\vec{b})}.\nonumber\\
\label{eq:aver-E}
\end{eqnarray}

\begin{figure*}[tbh]
\begin{center}
\includegraphics[width=0.32\linewidth]{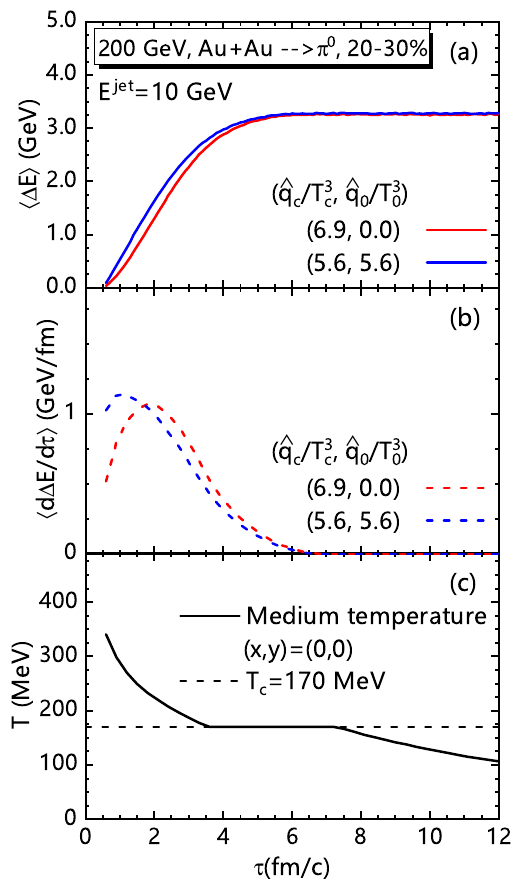}
\hspace{5mm}
\includegraphics[width=0.32\linewidth]{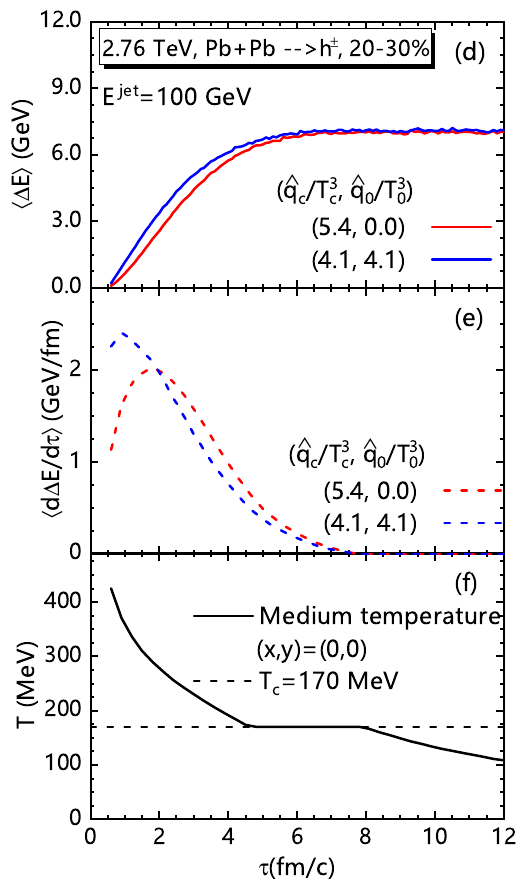}

\caption{(Linear-dependence) Panels (a) and (d): the average accumulative energy loss for one 10 GeV jet in 20-30\% Au + Au collisions at $\sqrt{s_{\rm NN}}=200$ GeV and for one 100 GeV jet in 20-30\% Pb + Pb collisions at $\sqrt{s_{\rm NN}}=2.76$ TeV, respectively. Panels (b) and (e): the corresponding differential energy loss.
Red solid or dashed curves are for the linearly-decreasing $T$ dependence of $(\hat{q}/T^3$, and blue for the constant $(\hat{q}/T^3$. Panels (c) and (f): the medium temperature as a function of time at the center point $(x,y)=(0,0)$ for the two collision systems, respectively.}
\end{center}
\label{fig:L-de-tau}
\end{figure*}

Shown in Fig. \ref{fig:L-de-tau} are the average accumulative (solid curves) and differential (dashed curves) energy loss for one 10 GeV jet with ${(\hat{q}_{\rm c}/T_{\rm c}^3,\hat{q}_0/T_0^3)=(6.9,0.0)}$ (red curves) and $(5.6,5.6)$ (blue curves) in 20-30\% Au + Au collisions at $\sqrt{s_{\rm NN}}=200$ GeV (panel (a) and (b)), and for one 100 GeV jet with ${(\hat{q}_{\rm c}/T_{\rm c}^3,\hat{q}_0/T_0^3)=(5.4,0.0)}$ (red curves) and $(4.1,4.1)$ (blue curves) in 20-30\% Pb + Pb collisions at $\sqrt{s_{\rm NN}}=2.76$ TeV (panel (d) and (e)), respectively.
The medium temperature as a function of time at the center point $(x,y)=(0,0)$ for the two collision systems is shown in the lower panels (c) and (f).

When the jet passes out of the critical region from QGP to the hadron phase, it is over for the jet to accumulate the lost energy, as shown in Eq. ($\ref{eq:qhat-f}$), with $\hat{q}_h = 0$. For the two well-chosen cases of constant dependence and linearly decreasing $T$ dependence of $\hat{q}/T^3$, the final total energy losses were similar, as shown in Fig. \ref{fig:L-de-tau} (a) and (d). In the meantime, the peak of the jet energy loss distribution $\langle{d\Delta{E}}/d\tau\rangle$ along the jet path is ``pushed" to move to critical temperature $T_{\rm c}$ nearby due to the linearly-decreasing $T$ dependence of $\hat{q}/T^3$ (red dashed curves) compared to the constant dependence (blue dashed curves), as shown in Fig. \ref{fig:L-de-tau} (b) and (e).
More energy loss occurs as the critical temperature approaches and enhances the final hadron azimuthal anisotropy. Therefore, $v_2(p_{\rm T})$ is strengthened for the linearly decreasing $T$ dependence case, as shown in Fig. \ref{fig:L-RHIC-RAA-v2-20-30} (c) and Fig. \ref{fig:L-LHC-RAA-v2-20-30} (c).

\begin{figure*}[tbh]
\begin{center}
\includegraphics[width=0.32\linewidth]{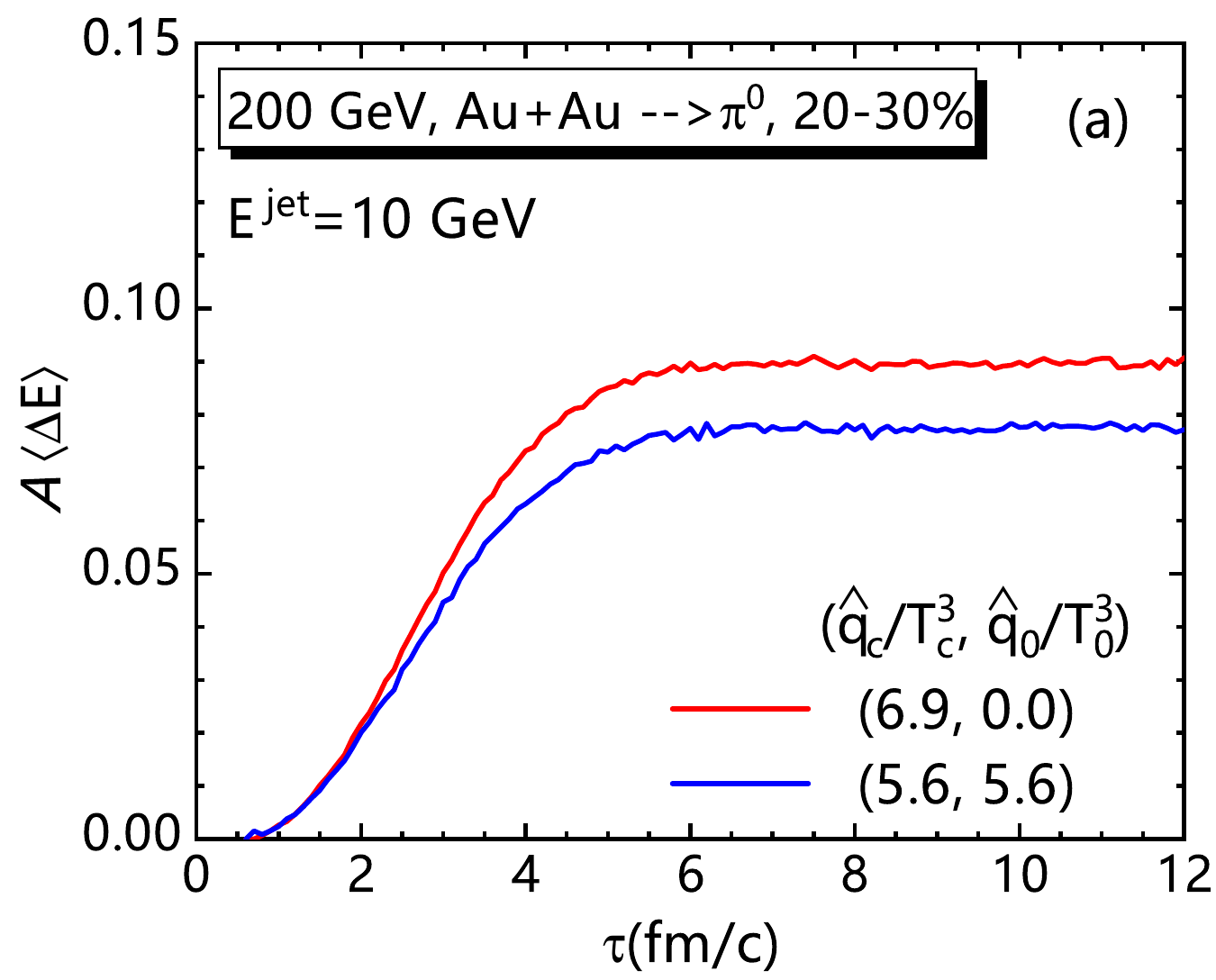}
\hspace{5mm}
\includegraphics[width=0.32\linewidth]{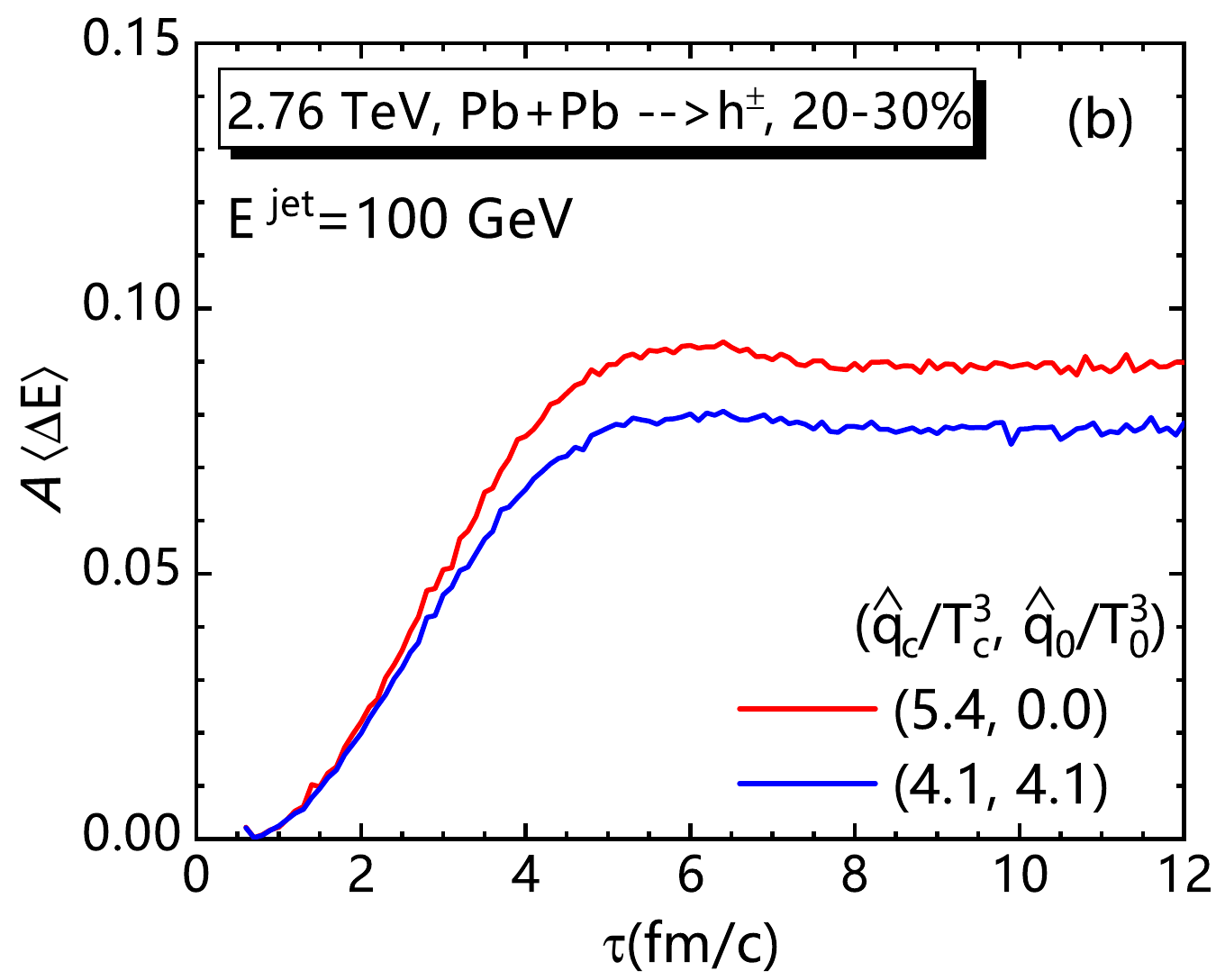}

\caption{(Linear-dependence) The energy loss asymmetry between the jet propagating direction of $\phi=\pi/2$ and $\phi=0$ for one 10 GeV jet with $(\hat{q}_{\rm c}/T_{\rm c}^3,\hat{q}_0/T_0^3)=(6.9,0.0)$ (red curve) and $(5.6,5.6)$ (blue curve) in 20-30\% Au + Au collisions at $\sqrt{s_{\rm NN}}=200$ GeV (panel (a)) and for one 100 GeV jet with $(\hat{q}_{\rm c}/T_{\rm c}^3,\hat{q}_0/T_0^3)=(5.4,0.0)$ (red curve) and $(4.1,4.1)$ (blue curve) in  20-30\% Pb + Pb collisions at $\sqrt{s_{\rm NN}}=2.76$ TeV (panel (b)).}
\end{center}
\label{fig:L-deltaE}
\end{figure*}

To clearly illustrate the enhanced azimuthal anisotropy, we define the energy loss asymmetry as follows:
\begin{eqnarray}
A\langle \Delta E(\tau)\rangle = \frac{\langle \Delta E(\tau) \rangle^{\phi=\pi/2} - \langle \Delta E(\tau) \rangle^{\phi=0}}
         {\langle \Delta E(\tau) \rangle^{\phi=\pi/2} + \langle \Delta E(\tau) \rangle^{\phi=0}},
\label{eq:asym-dE}
\end{eqnarray}
where $\langle\Delta E(\tau,\phi)\rangle$ is given by Eq. ($\ref{eq:aver-E}$), in which the $\phi$ integration for the azimuthal average was removed. On average, a parton jet encounters the greatest energy loss because it has the longest path length at $\phi=\pi/2$ and the shortest at $\phi=0$. Shown in Fig. \ref{fig:L-deltaE} (a) and (b) are such energy loss asymmetries for an energy-given parton jet in 20-30\% Au + Au collisions at 200 GeV and Pb + Pb collisions at 2.76 TeV, respectively. The red solid curves represent the linearly decreasing $T$-dependence of $\hat{q}/T^3$, whereas the blue solid curves represent the constant case. The former is 10\% larger than the latter in both panels, which is similar to the enhancement in $v_2(p_{\rm T})$ shown in Fig. \ref{fig:L-RHIC-RAA-v2-20-30} (c) and \ref{fig:L-LHC-RAA-v2-20-30} (c).

Owing to the medium-temperature evolution, different $T$ dependencies of the jet transport coefficient result in different energy-loss distributions for jet propagation. The large $p_{\rm T}$ hadron suppression $R_{AA}$ was a consequence of the total energy loss and was independent of the jet energy loss distribution. However, compared with the constant case for a given total energy loss, the linearly decreasing $T$ dependence of $\hat{q}/T^3$ causes an energy loss to redistribute and leads to more energy loss near the critical temperature, and therefore, a stronger azimuthal anisotropy for hadron production.

\begin{figure*}[tbh]
\begin{center}
\includegraphics[width = 0.36\textwidth, height=0.28\textwidth]{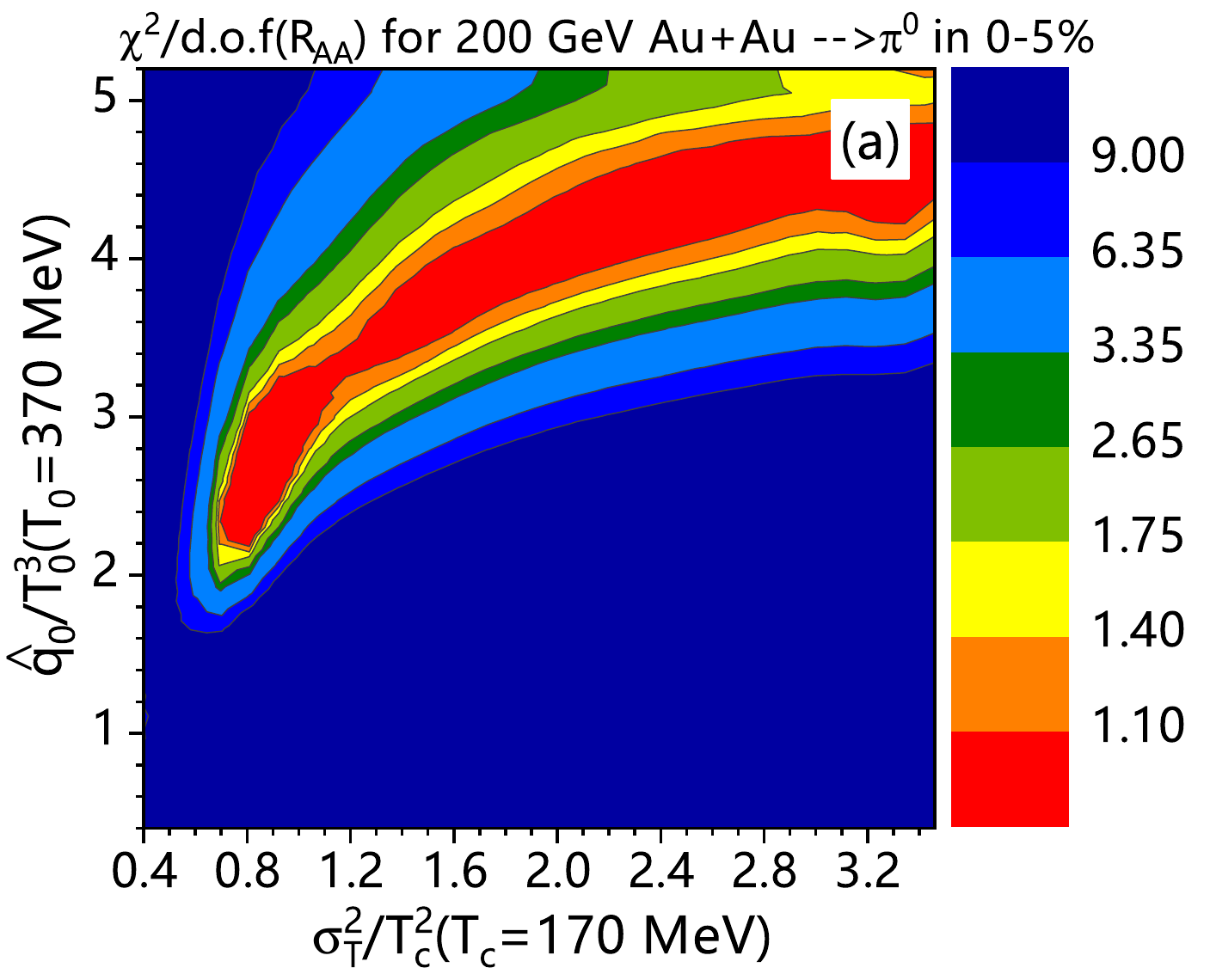}
\hspace{-4mm}
\includegraphics[width = 0.31\textwidth, height=0.27\textwidth]{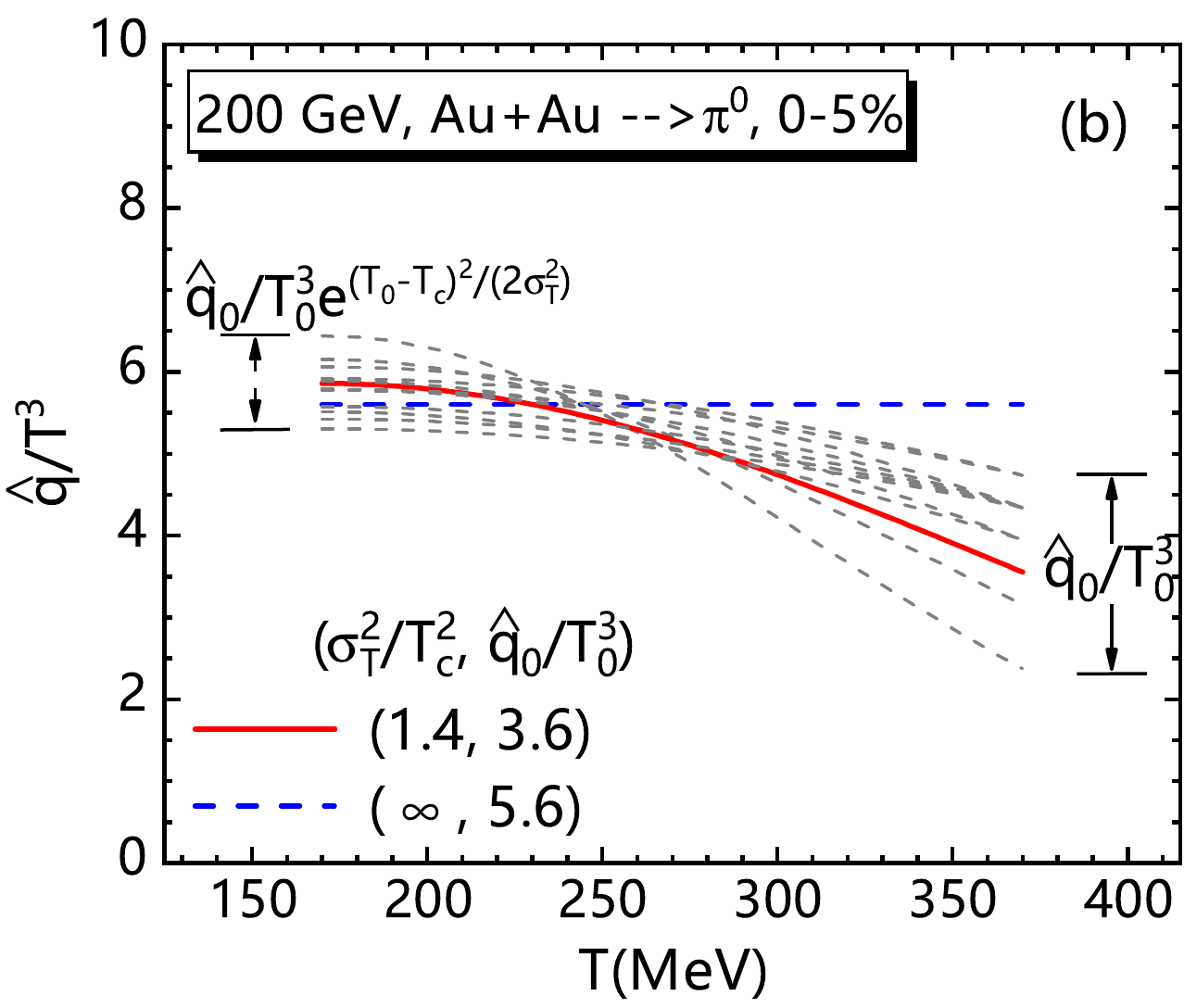}
\hspace{-1mm}
\includegraphics[width = 0.31\textwidth, height=0.27\textwidth]{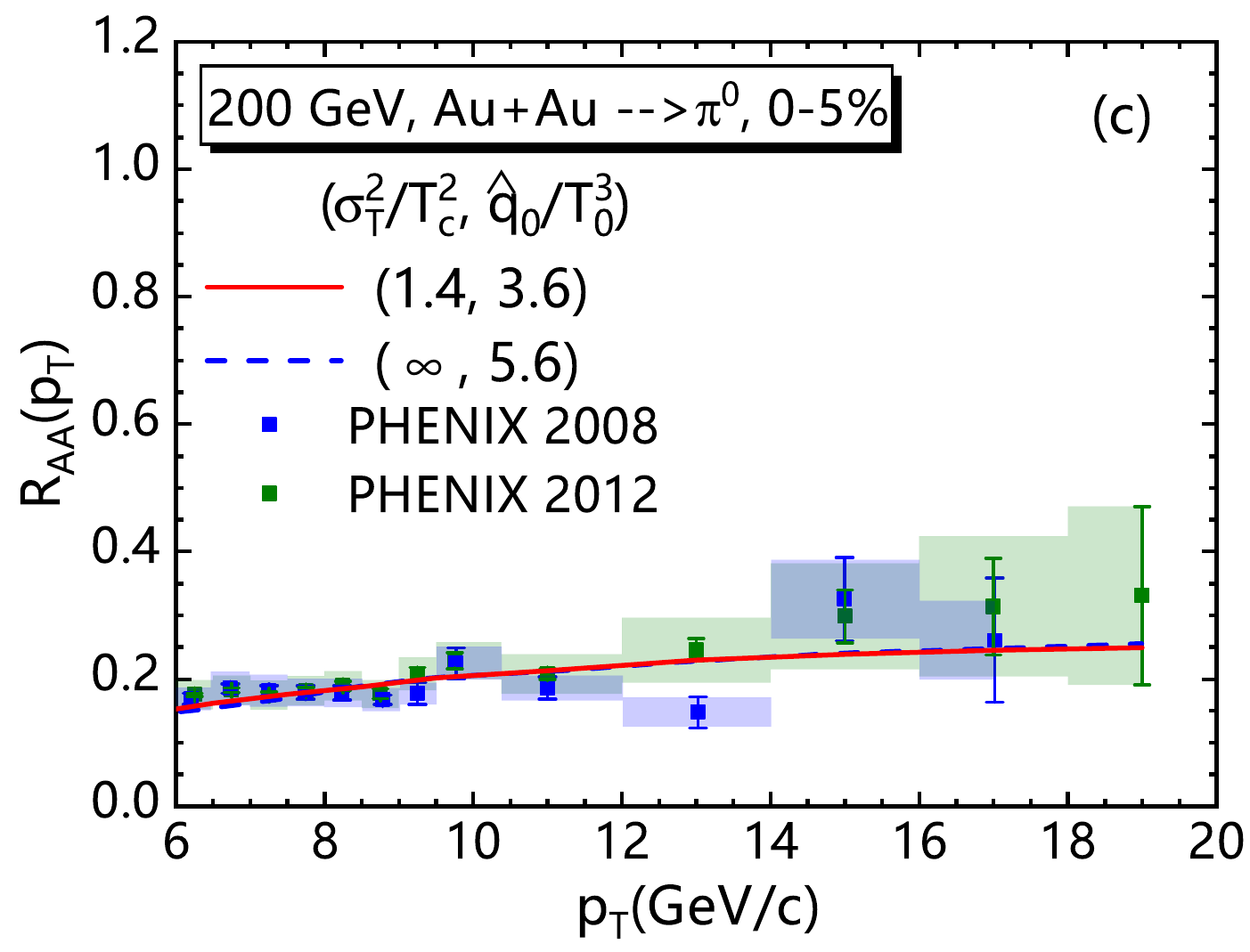}
\caption{(Gaussian-dependence) Panel (a): The $\chi^2/{\rm d.o.f}$ analyses for single hadron $R_{AA}(p_{\rm T})$ as a function of $\sigma_{T}^2/T_{\rm c}^2$ and $\hat{q}_0/T_0^3$ from fitting to experimental data \cite{PHENIX:2008saf,PHENIX:2012jha} in the most central 0-5\% Au + Au collisions at $\sqrt{s_{\rm NN}}=200$~GeV. Panel (b): The scaled dimensionless jet transport parameters $\hat{q}/T^3$ as a function of medium temperature $T$ from the best fitting region of the panel (a). Panel (c): The single hadron suppression factors $R_{AA}(p_{\rm T})$ with couples of $(\sigma_T^2/T_{\rm c}^2,\hat{q}_0/T_0^3)=(1.4,3.6)$ (red solid curve) and $(\sigma_T^2/T_{\rm c}^2,\hat{q}_0/T_0^3)=(\infty,5.6)$ (blue dashed curve) compared with PHENIX \cite{PHENIX:2008saf,PHENIX:2012jha} data.}
\end{center}
\label{fig:G-RHIC-0-5}
\end{figure*}

\begin{figure*}[tbh]
\begin{center}
\includegraphics[width = 0.32\textwidth]{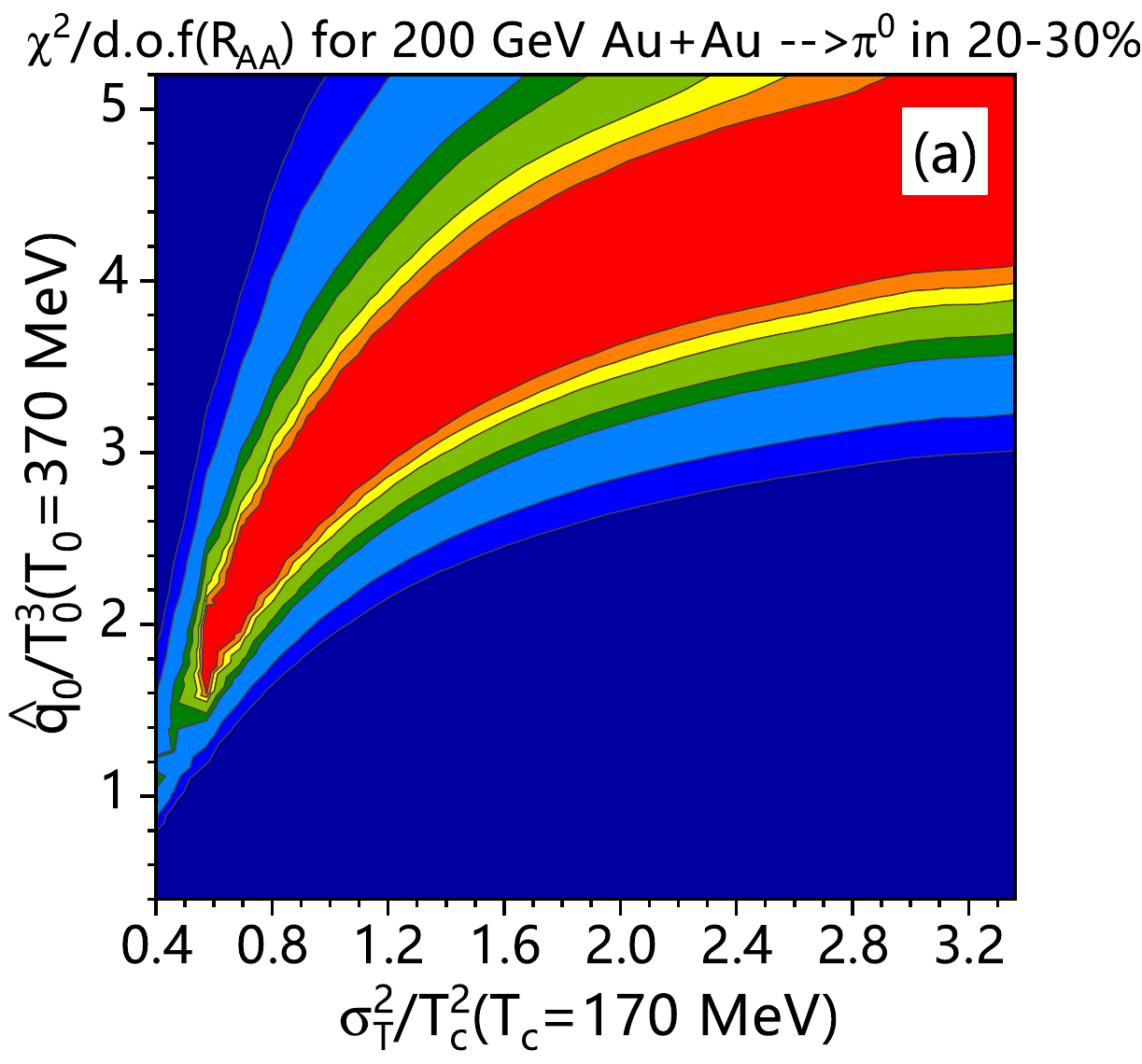}
\hspace{-3mm}
\includegraphics[width = 0.312\textwidth]{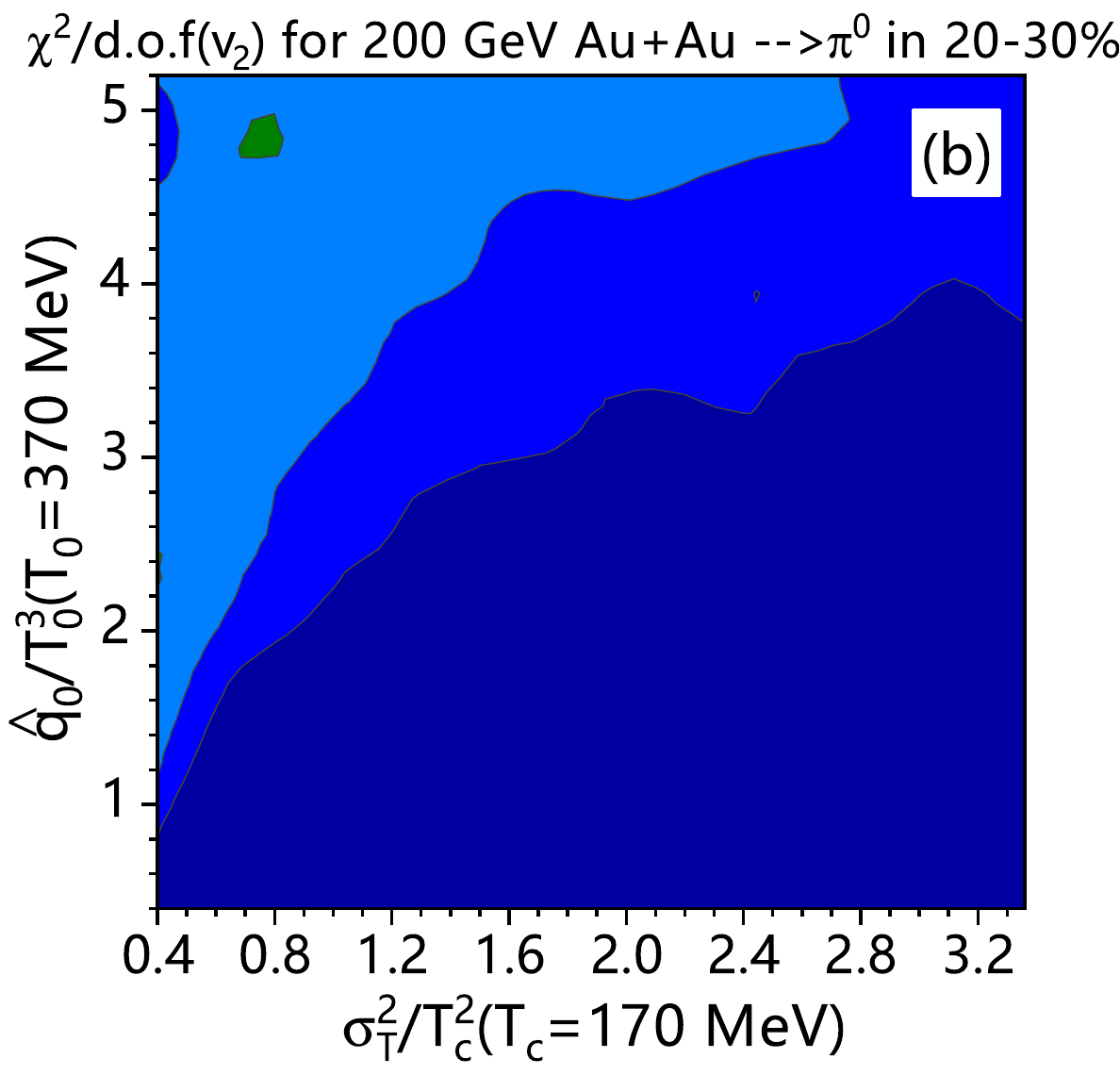}
\hspace{-3mm}
\includegraphics[width = 0.37\textwidth]{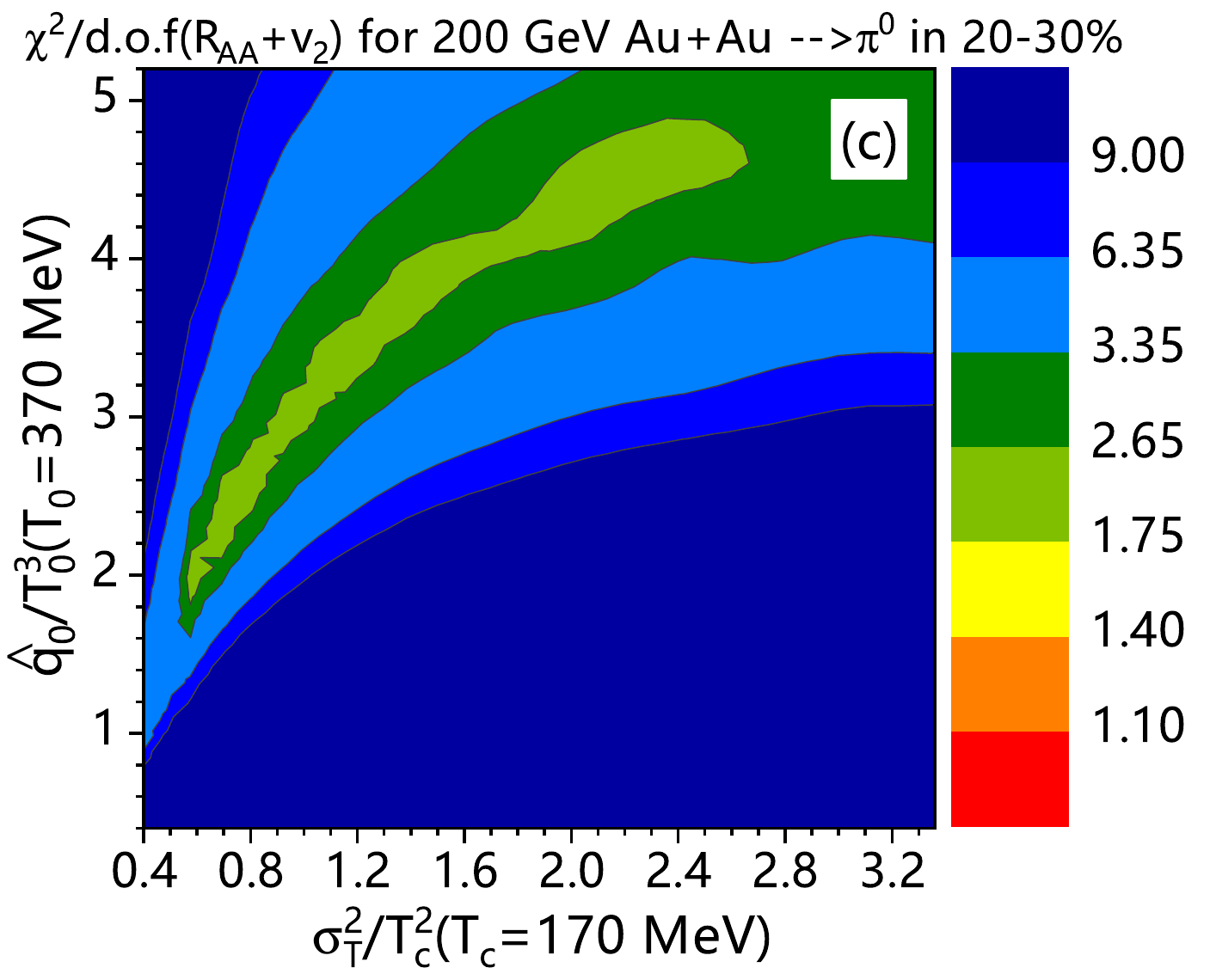}
\caption{(Gaussian-dependence) The $\chi^2/{\rm d.o.f}$ analyses of single hadron $R_{AA}(p_{\rm T})$ (panel (a)) and elliptic flow $v_{2}(p_{\rm T})$ (panel (b)) as a function of $\sigma_T^2/T_{\rm c}^2$ and $\hat{q}_0/T_0^3$ from fitting to experimental data \cite{PHENIX:2008saf,PHENIX:2012jha,PHENIX:2010nlr} in 20-30\% Au + Au collisions at $\sqrt{s_{\rm NN}}=200$~GeV. The global $\chi^2/{\rm d.o.f}$ fitting results for both $R_{AA}(p_{\rm T})$ and $v_{2}(p_{\rm T})$ are shown in the panel (c).}
\end{center}
\label{fig:G-RHIC-20-30-x2}
\end{figure*}

\vspace{12pt}

\section{Gaussian temperature dependence of $\hat{q}/T^3$ in QGP phase} \label{sec:gauss-QGP}

In the last section, the numerical results for the linear $T$-dependence assumption show that $\hat{q}/T^3$ goes down as the medium temperature increases. The going-down $T$ dependence of $\hat{q}/T^3$ stimulates an attempt to make a Gaussian assumption regarding the $T$ dependence of $\hat{q}/T^3$.
We assume that the apex of the Gaussian distribution is located at critical temperature, as shown in Eq.~($\ref{eq:qhat-gaus}$).
This assumption of Gaussian temperature dependence will be submitted into Eq. ($\ref{eq:qhat-f}$) with $\hat{q}_h = 0$ for the QGP phase only and Eq. ($\ref{eq:De}$) and ($\ref{eq:Ng}$) for the jet energy loss.

\subsection{Fit $R_{AA}$ and $v_2$ at RHIC}

In Eq. ($\ref{eq:qhat-gaus}$) for the assumption of Gaussian temperature dependence, the introduced $\hat{q}_0/T_0^3$ remains the scaled jet transport parameter at the initial time at the center of the medium, whereas $\sigma_T^2/T_{\rm c}^2$ is the squared Gaussian width. Starting with Au + Au collisions at $\sqrt{s_{\rm NN}}=200$~GeV, we choose $\sigma_T^2/T_{\rm c}^2 \in [0.35,3.5]$ with bin size 0.35, and $\hat{q}_0/T_0^3 \in [0.4,5.2]$ with bin size 0.4, and obtain 130 pairs of $(\sigma_T^2/T_{\rm c}^2,\hat{q}_0/T_0^3)$ for Eqs. ($\ref{eq:qhat-gaus}$) and ($\ref{eq:qhat-f}$), with $\hat{q}_h = 0$.
We first obtain $R_{AA}(p_{\rm T})$ in 0–5\% collisions and compare them with the data \cite{PHENIX:2008saf,PHENIX:2012jha} to get a 2-dimensional contour plot for $\chi^2/{\rm d.o.f}$, as shown in Fig. \ref{fig:G-RHIC-0-5} (a). According to the best-fitting region obtained for $(\sigma_T^2/T_{\rm c}^2,\hat{q}_0/T_0^3)$, we obtained $\hat{q}/T^3$ as a function of $T$ in Fig. \ref{fig:G-RHIC-0-5} (b). The red solid curve represents the downward $\hat{q}/T^3$-dependence with $(\sigma_T^2/T_{\rm c}^2,\hat{q}_0/T_0^3)=(1.4,3.6)$ whereas the blue dashed curve shows a constant dependence with $(\sigma_T^2/T_{\rm c}^2,\hat{q}_0/T_0^3)=(\infty,5.6)$. These two dependence forms yielded almost the same $R_{AA}(p_{\rm T})$, as shown in Fig. \ref{fig:G-RHIC-0-5} (c). $R_{AA}(p_{\rm T})$ doesn't ``care" whether $\hat{q}/T^3$ is of Gaussian temperature dependence or not at RHIC.

\begin{figure*}[tbh]
\begin{center}
\includegraphics[width = 0.30\textwidth,height=0.27\textwidth]{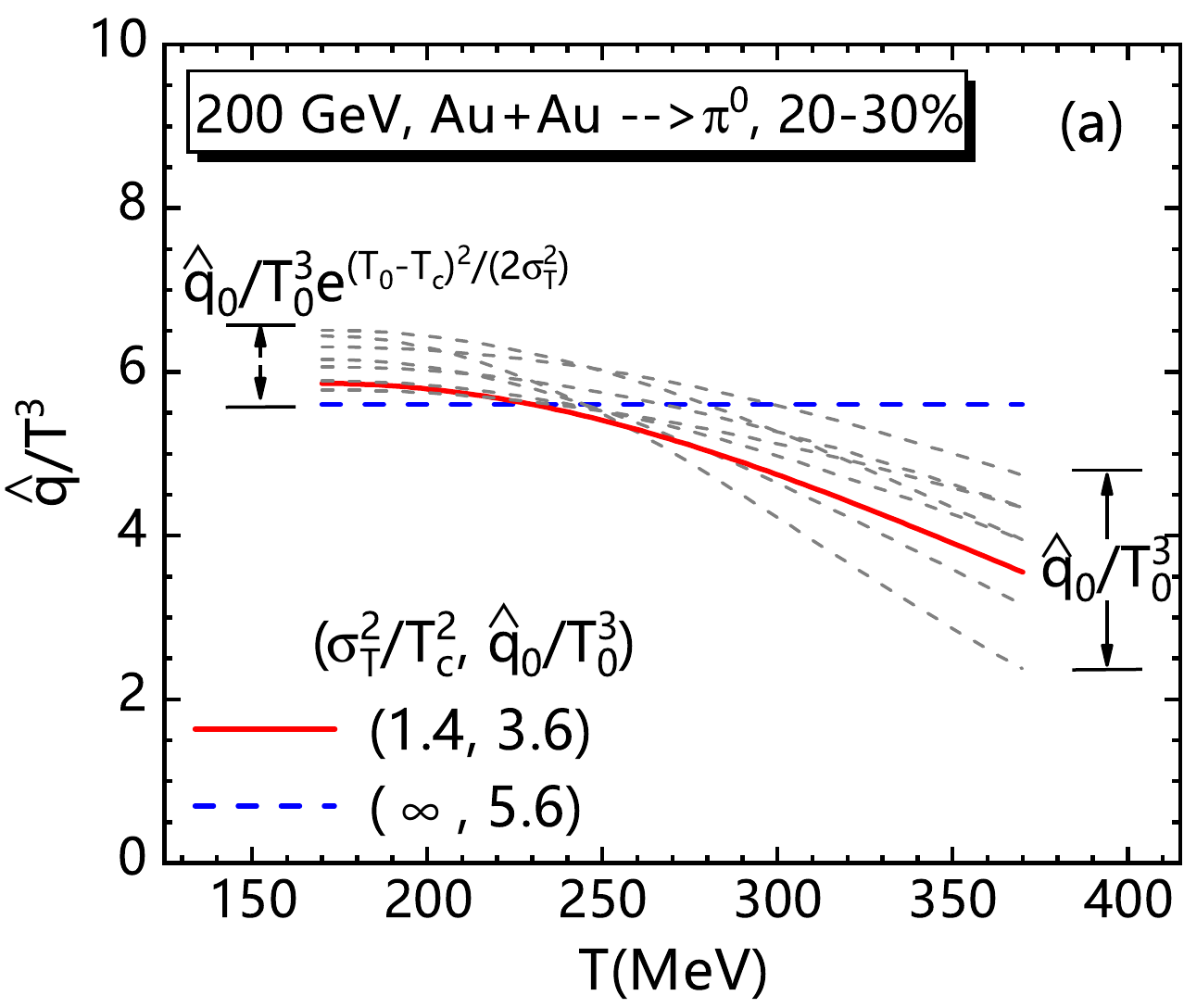}
\includegraphics[width = 0.32\textwidth,height=0.27\textwidth]{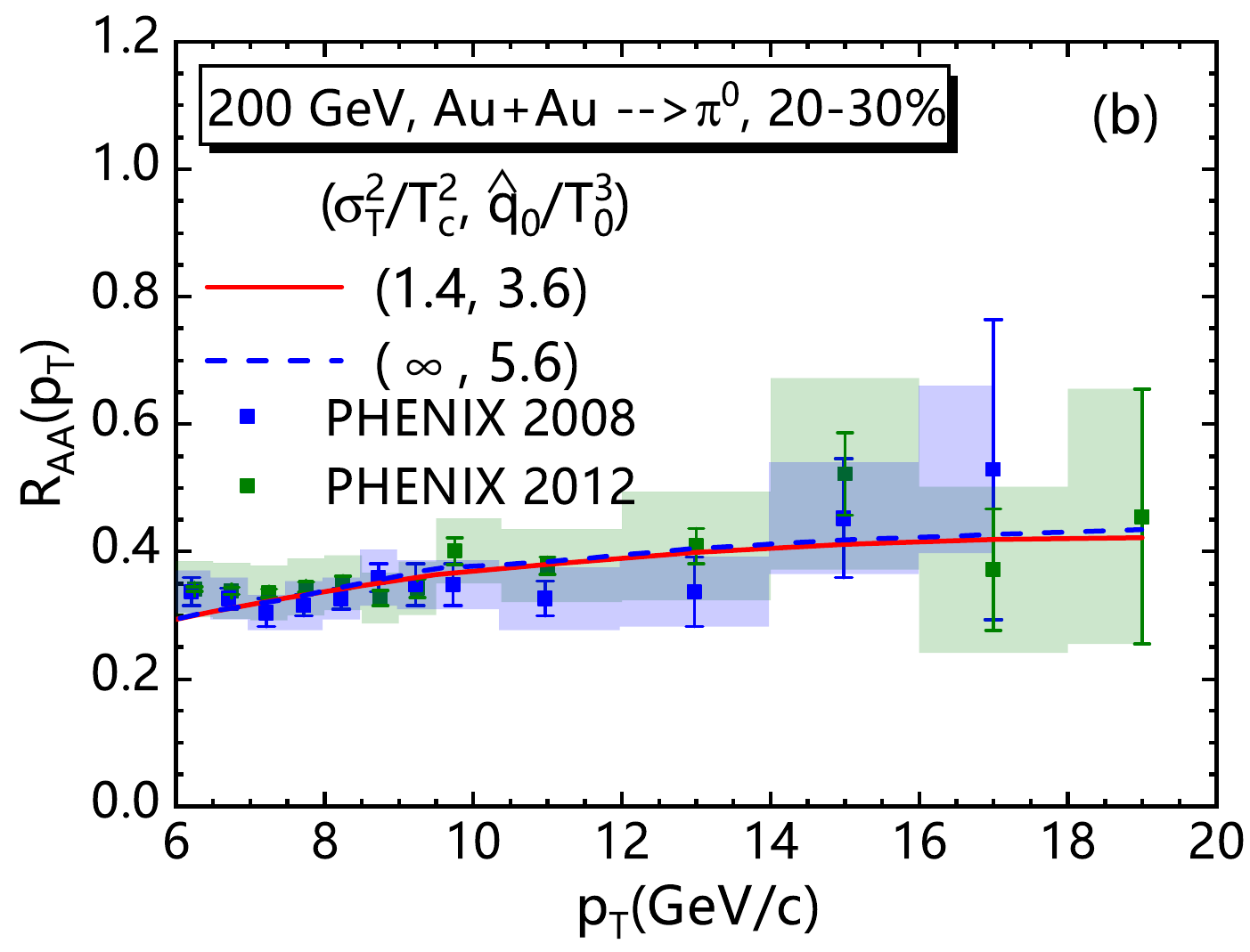}
\includegraphics[width = 0.32\textwidth,height=0.27\textwidth]{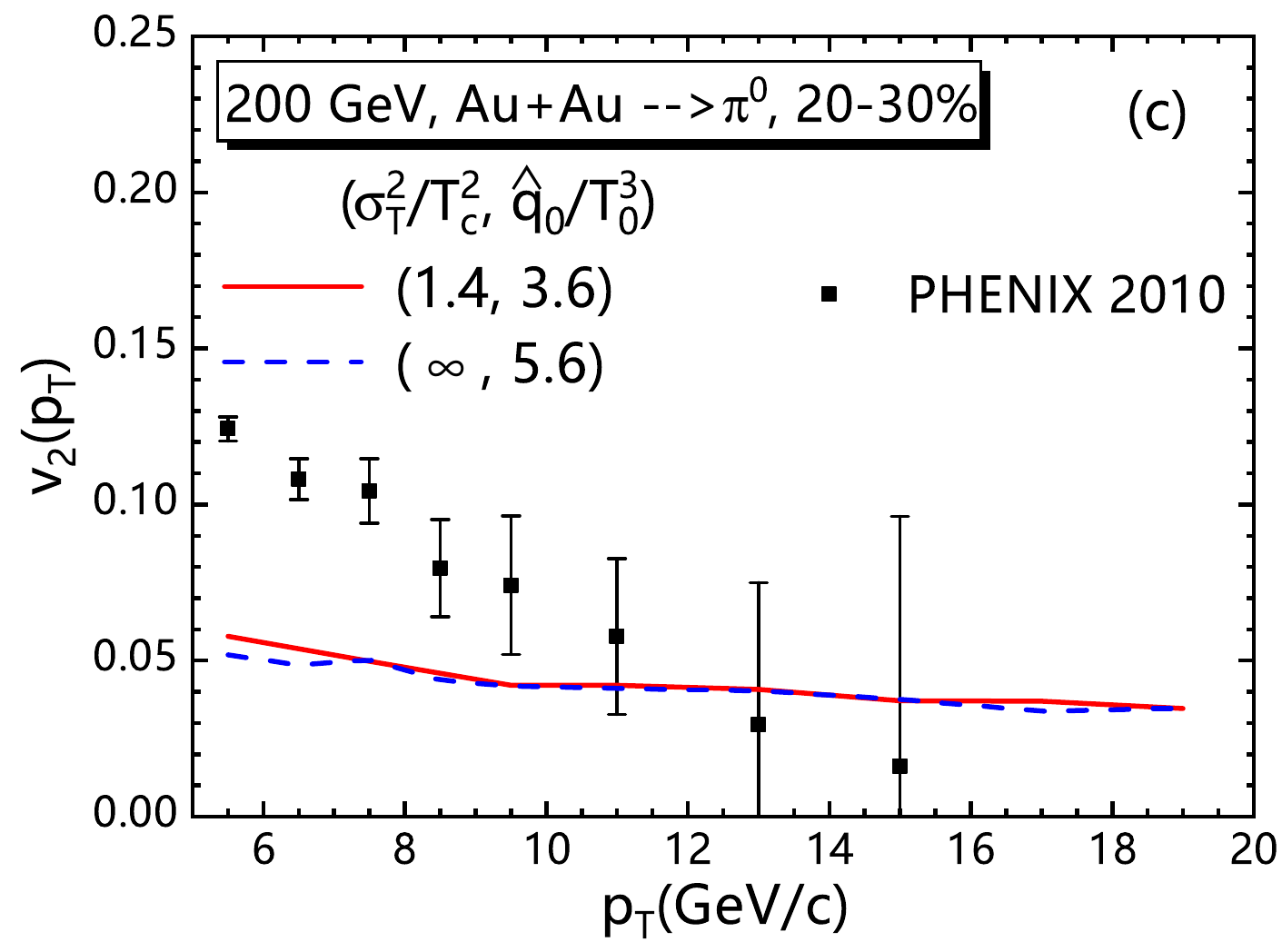}
\caption{(Gaussian-dependence) In 20-30\% Au + Au collisions at $\sqrt{s_{\rm NN}}=200$ GeV, the scaled dimensionless jet transport parameters $\hat{q}/T^3$ as a function of medium temperature $T$ from the best fitting region of global $\chi^2$ fits of Fig. \ref{fig:G-RHIC-20-30-x2} (c) are shown in panel (a). The single hadron suppression factors $R_{AA}(p_{\rm T})$ and elliptic flow parameter $v_{2}(p_{\rm T})$ are shown in panel (b) and (c), respectively, with couples of $(\sigma_T^2/T_{\rm c}^2,\hat{q}_0/T_0^3)=(1.4,3.6)$ (red solid curve) and $(\infty,5.6)$ (blue dashed curve) compared with PHENIX \cite{PHENIX:2008saf,PHENIX:2012jha,PHENIX:2010nlr} data.}
\end{center}
\label{fig:G-RHIC-RAA-v2-20-30}
\end{figure*}

\begin{figure*}[tbh]
\begin{center}

\includegraphics[width = 0.37\textwidth]{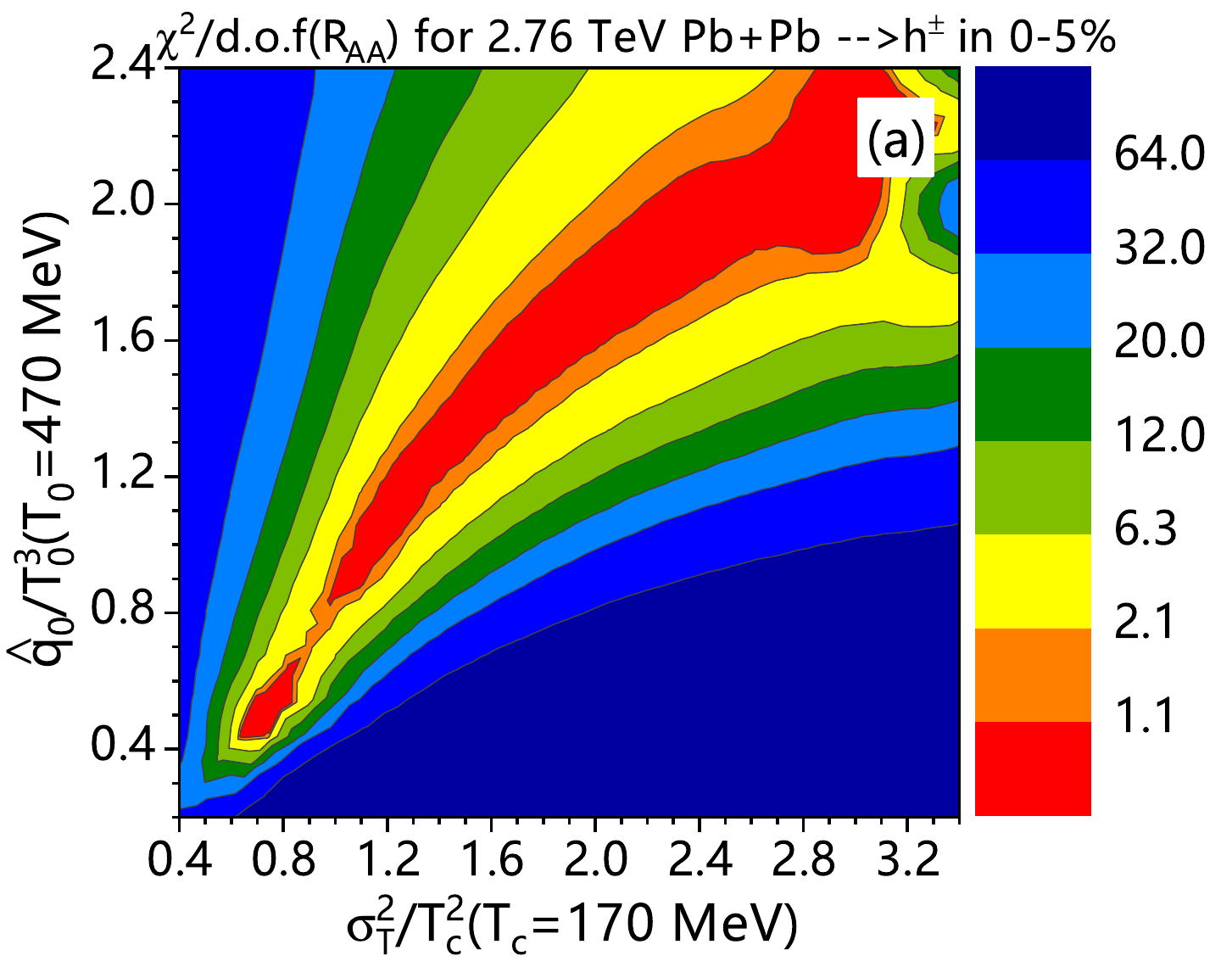}
\hspace{5mm}
\includegraphics[width = 0.324\textwidth,height = 0.28\textwidth]{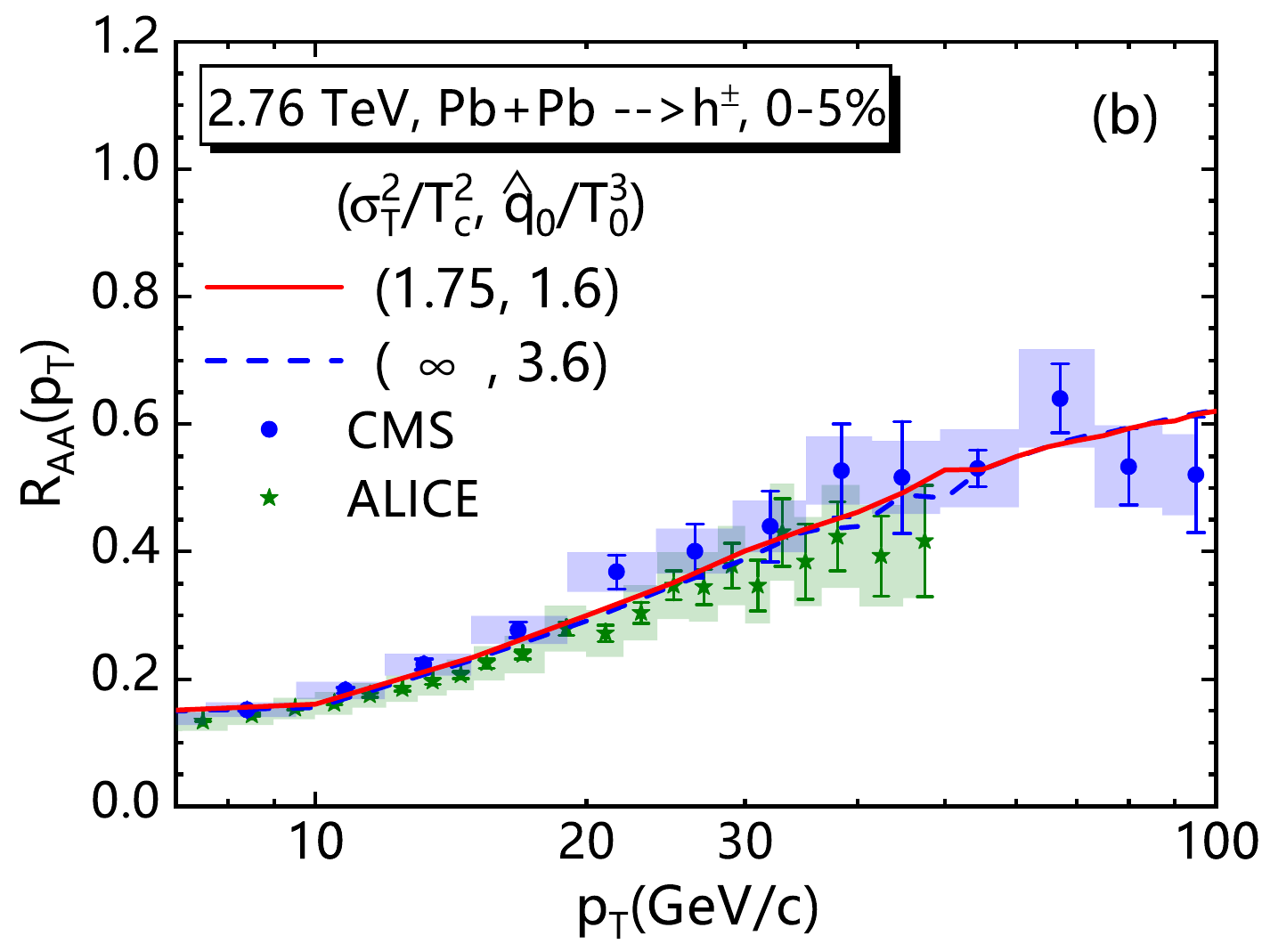}

\caption{(Gaussian-dependence) Panel (a): The $\chi^2/{\rm d.o.f}$ analyses for single hadron $R_{AA}(p_{\rm T})$ as a function of $\hat{q}_{\rm c}/T_{\rm c}^3$ and $\hat{q}_0/T_0^3$ from fitting to experimental data \cite{ALICE:2012aqc,CMS:2012aa} in the most central 0-5\% Pb + Pb collisions at $\sqrt{s_{\rm NN}}=2.76$~TeV. Panel (b): The single hadron suppression factors $R_{AA}(p_{\rm T})$ with couples of $(\sigma_T^2/T_{\rm c}^2, \hat{q}_0/T_0^3)=(1.75,1.6)$ (red solid curve) and $(\sigma_T^2/T_{\rm c}^2, \hat{q}_0/T_0^3)=(\infty,3.6)$ (blue dashed curve) compared with experimental data.}
\end{center}
\label{fig:G-LHC-0-5}
\end{figure*}

\begin{figure*}[tbh]
\begin{center}
\includegraphics[width = 0.32\linewidth]{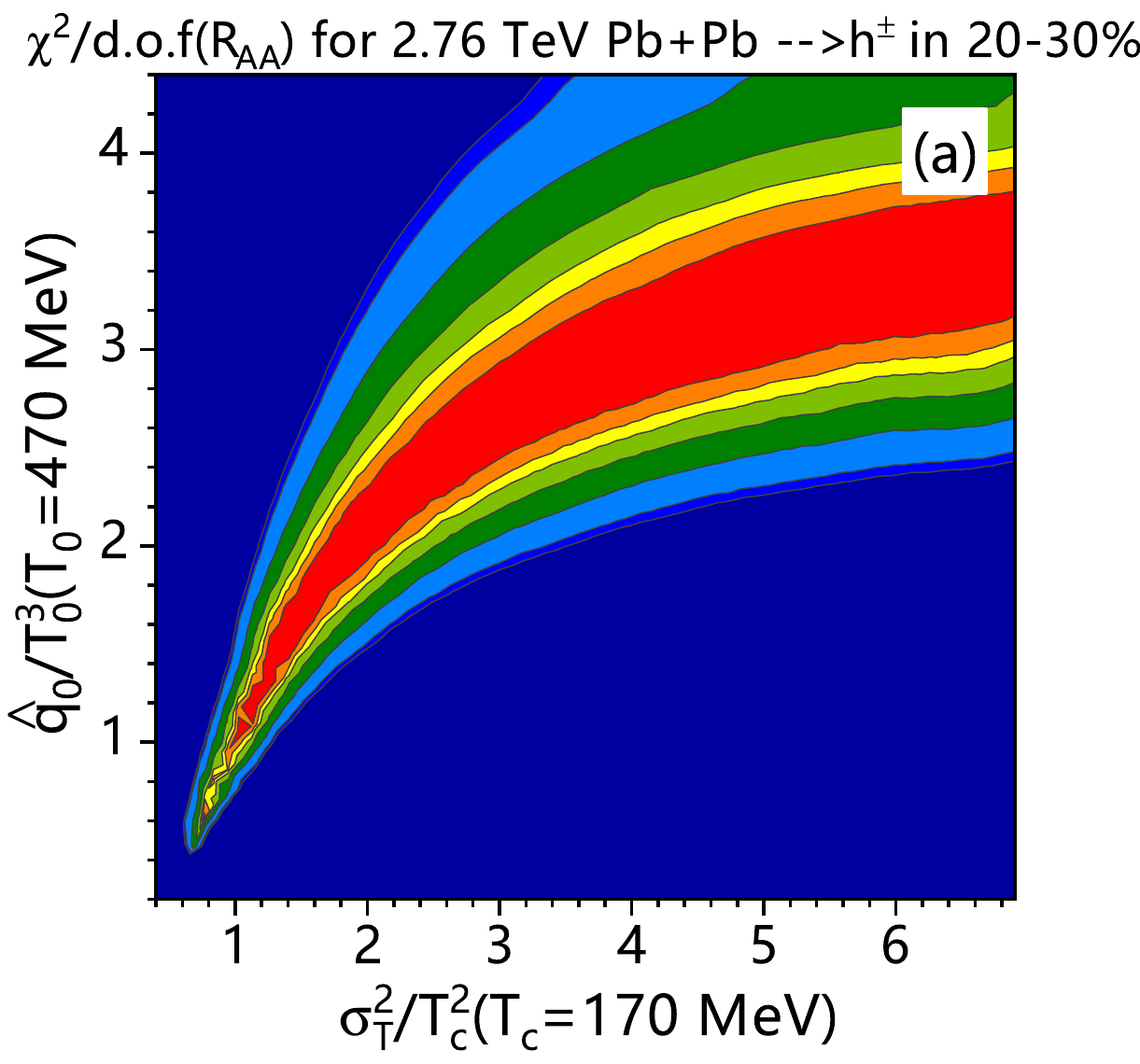}
\hspace{-3mm}
\includegraphics[width = 0.312\linewidth]{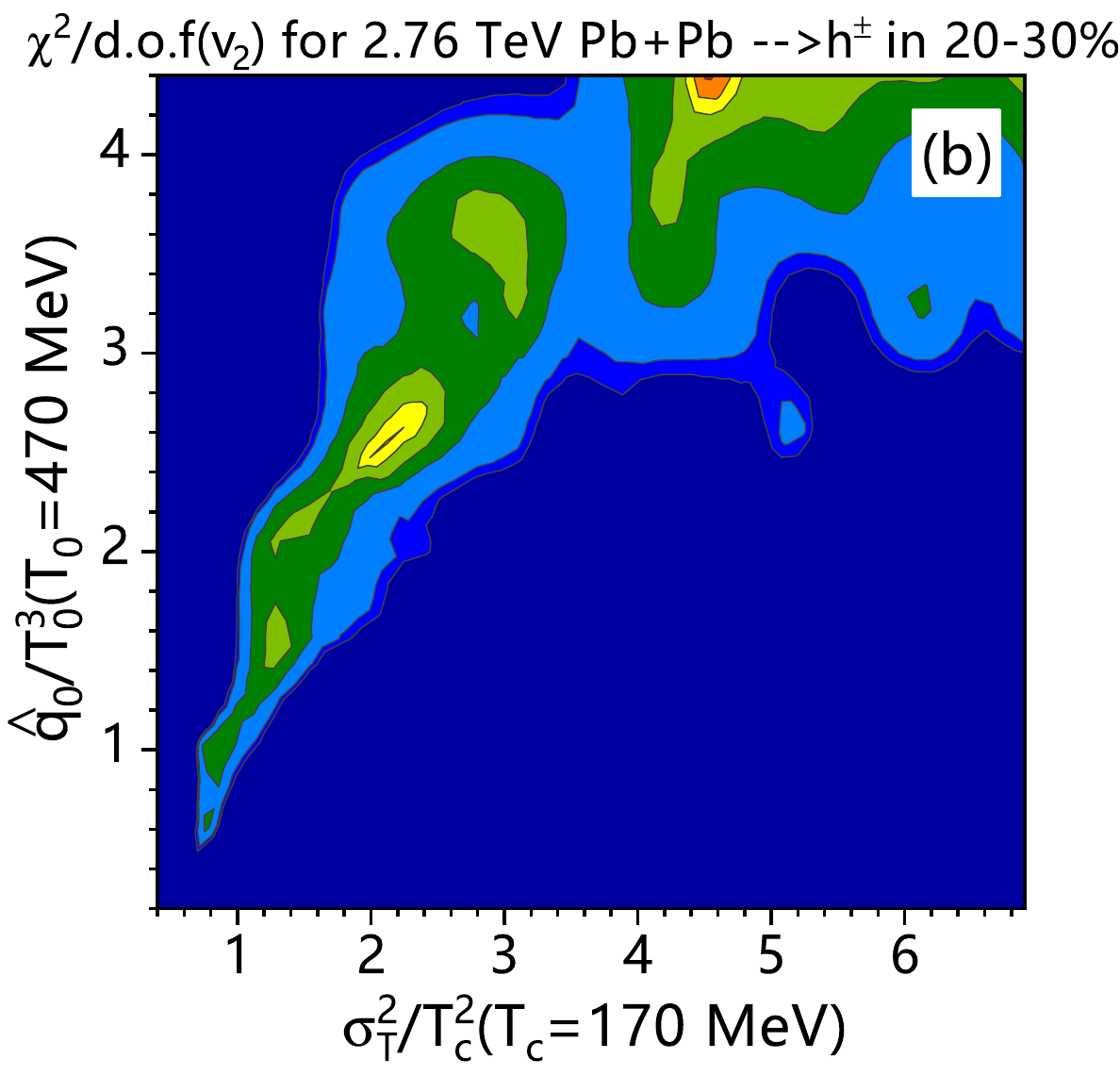}
\hspace{-3mm}
\includegraphics[width = 0.37\linewidth]{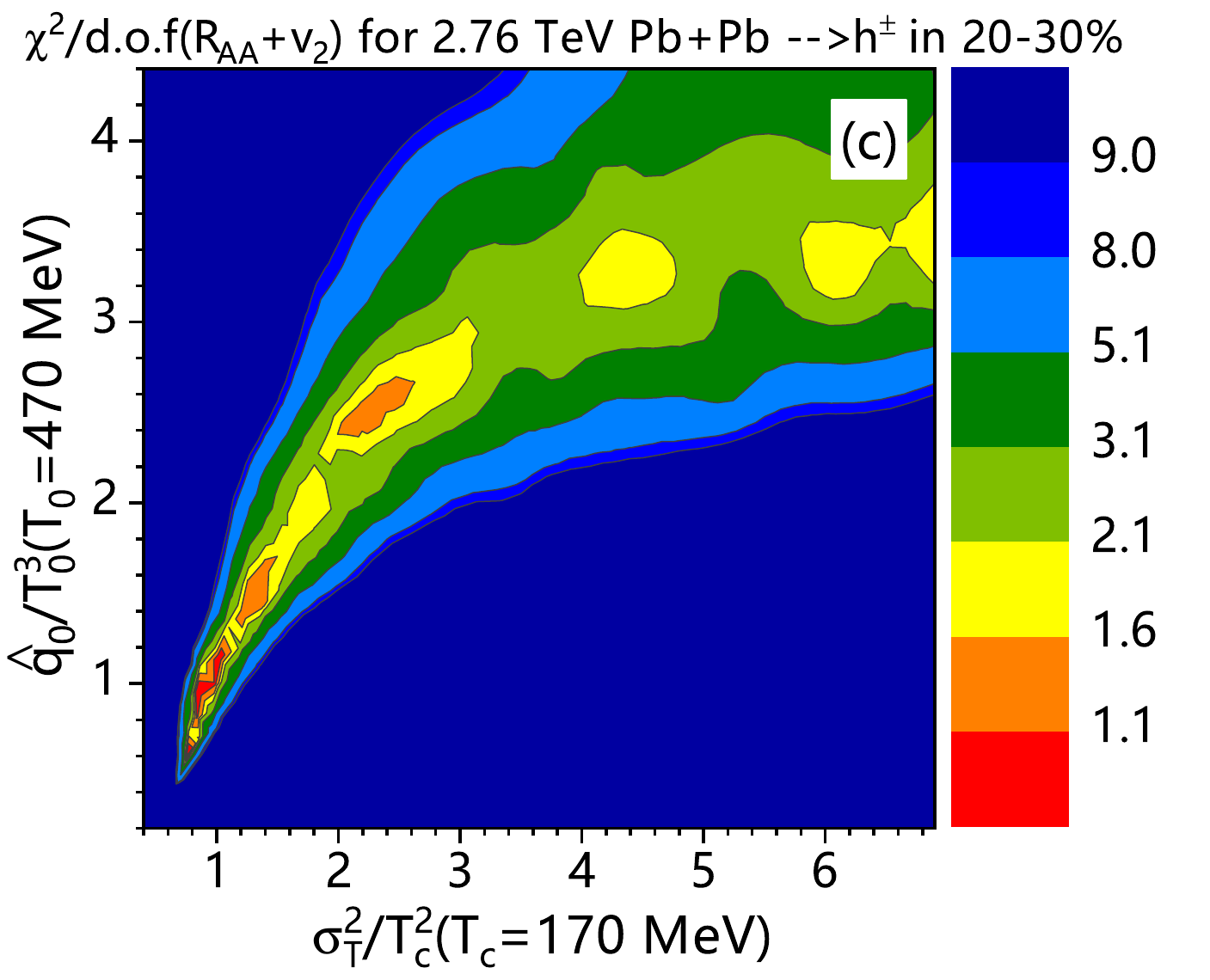}

\caption{(Gaussian-dependence)
The $\chi^2/{\rm d.o.f}$ analyses of single hadron $R_{AA}(p_{\rm T})$ (panel (a)) and elliptic flow $v_{2}(p_{\rm T})$ (panel (b)) as a function of $\sigma_T^2/T_{\rm c}^2$ and $\hat{q}_0/T_0^3$ from fitting to experimental data \cite{ALICE:2012aqc,CMS:2012aa,CMS:2012tqw,ALICE:2012vgf} in 20-30\% Pb + Pb collisions at $\sqrt{s_{\rm NN}}=2.76$~TeV. The global $\chi^2/{\rm d.o.f}$ fitting results for both $R_{AA}(p_{\rm T})$ and $v_{2}(p_{\rm T})$ are shown in panel (c).}
\end{center}
\label{fig:G-LHC-20-30-x2}
\end{figure*}

\begin{figure*}[tbh]
\begin{center}
\includegraphics[width = 0.30\textwidth,height=0.27\textwidth]{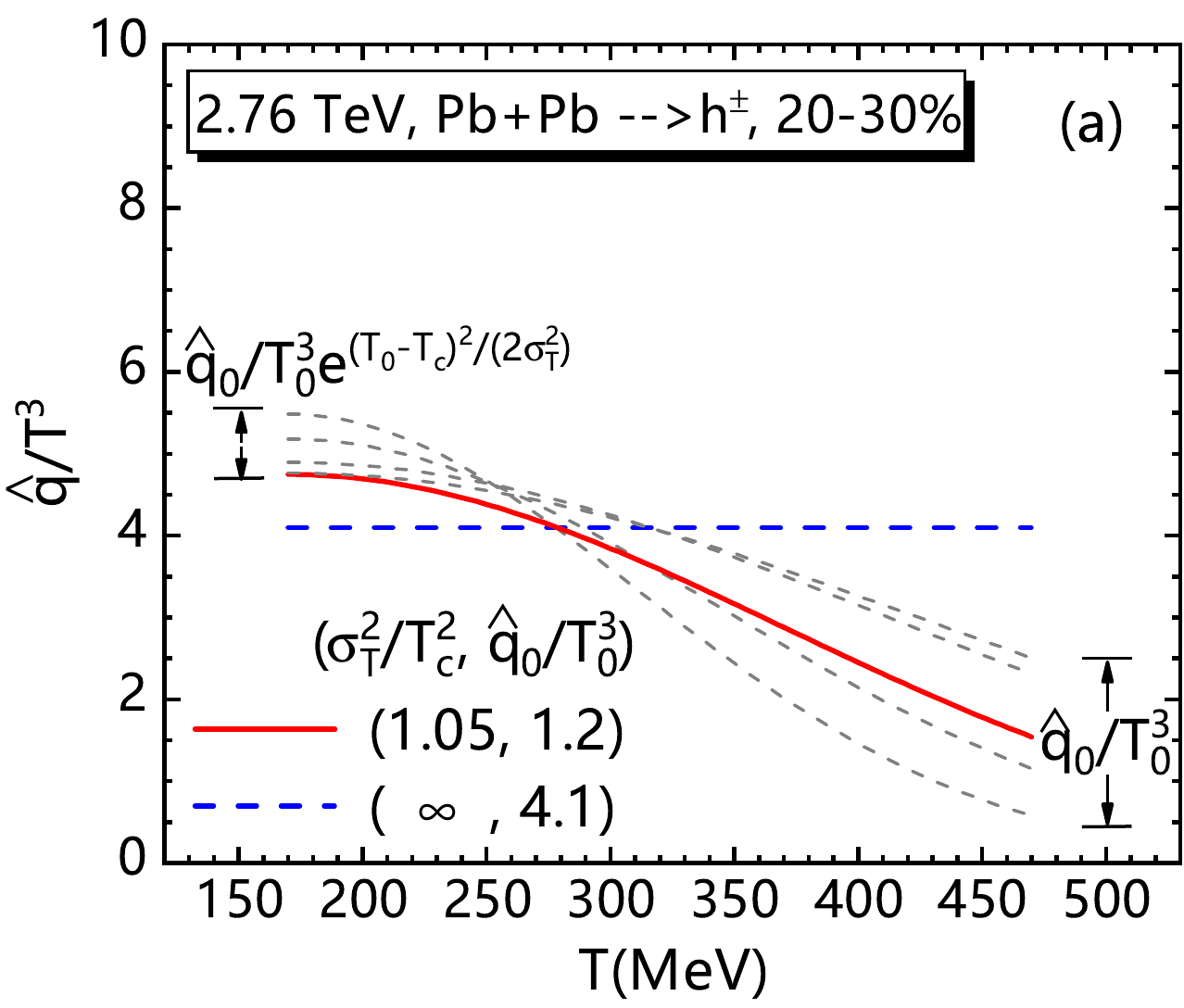}
\includegraphics[width = 0.32\textwidth,height=0.27\textwidth]{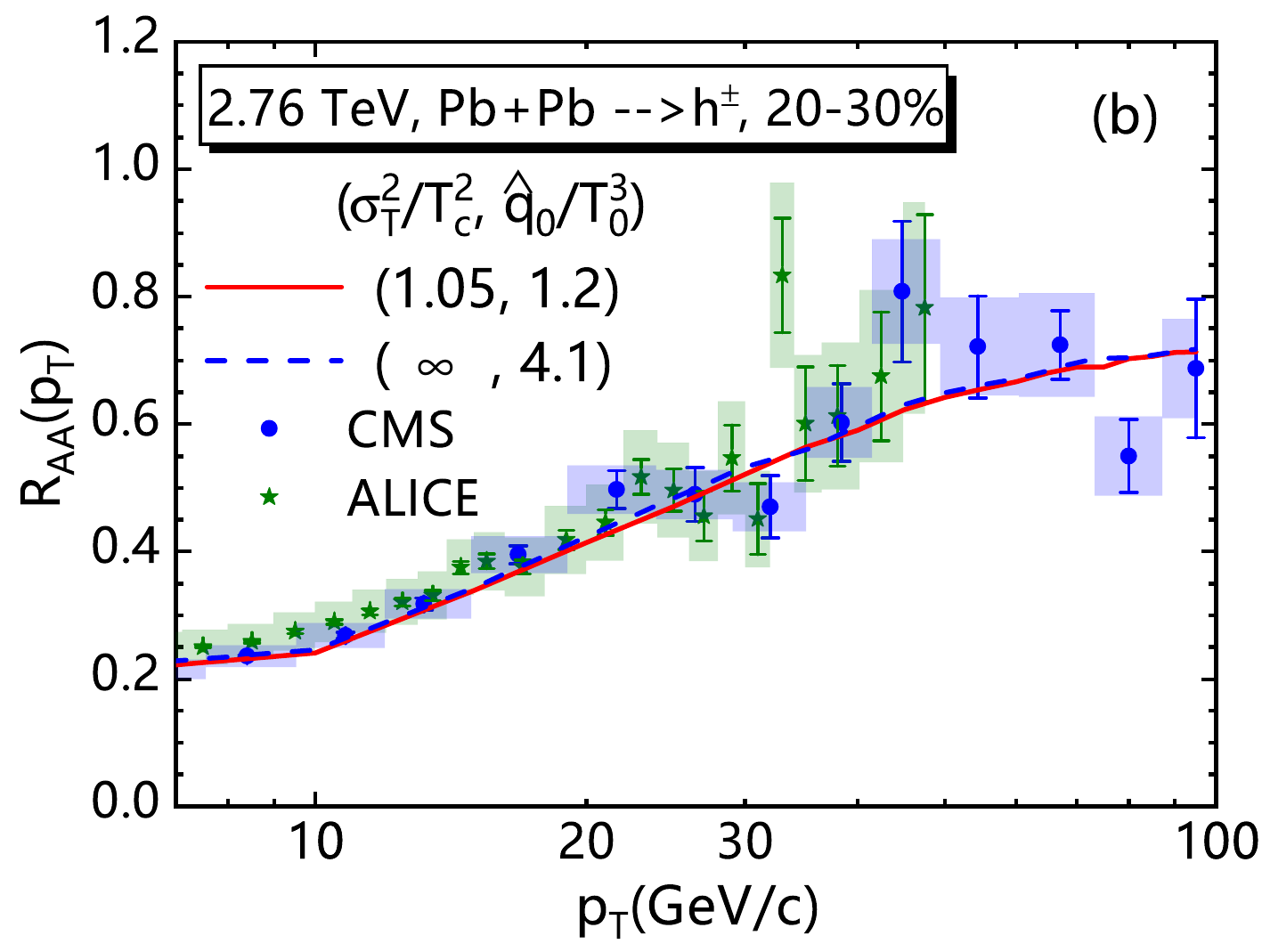}
\includegraphics[width = 0.32\textwidth,height=0.27\textwidth]{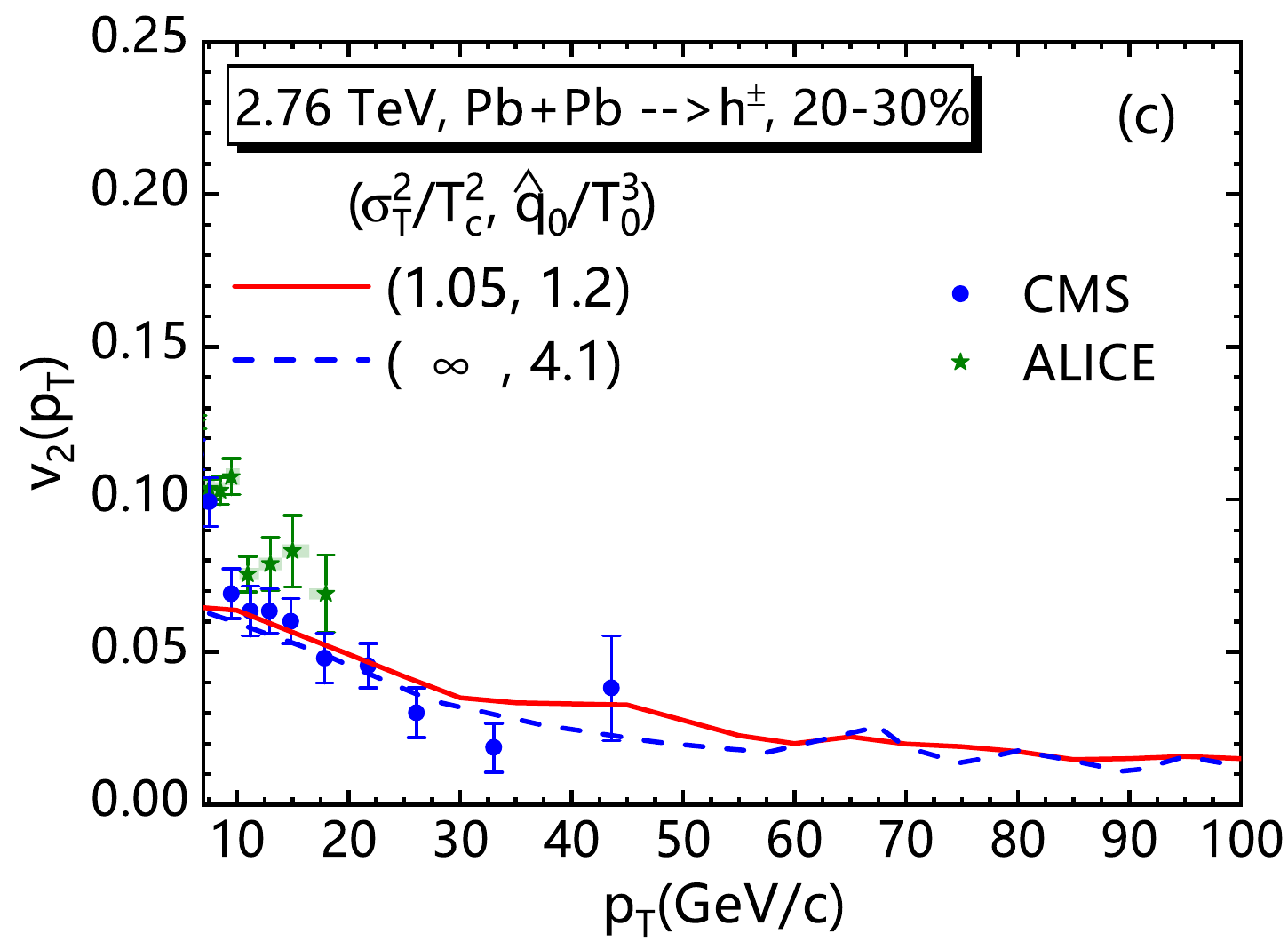}

\caption{(Gaussian-dependence) In 20-30\% Pb + Pb collisions at $\sqrt{s_{\rm NN}}=2.76$ TeV, the scaled dimensionless jet transport parameters $\hat{q}/T^3$ as a function of medium temperature $T$ from the best fitting region of global $\chi^2$ fits of Fig. \ref{fig:G-LHC-20-30-x2} (c) are shown in the panel (a). The single hadron suppression factors $R_{AA}(p_{\rm T})$ and elliptic flow $v_{2}(p_{\rm T})$ are shown in panels (b) and (c), respectively, with couples of $(\sigma_T^2/T_{\rm c}^2, \hat{q}_0/T_0^3)=(1.05,1.2)$ (red solid curve) and $(\infty,4.1)$ (blue dashed curve) compared with experimental data \cite{ALICE:2012aqc,CMS:2012aa,CMS:2012tqw,ALICE:2012vgf}.}
\end{center}
\label{fig:G-LHC-RAA-v2-20-30}
\end{figure*}

\begin{table*}
\begin{center}
\begin{tabular}{c|cc|cc|cc}
\hline \hline & \multicolumn{2}{|c|}{Constant} & \multicolumn{2}{c|}{ Linear $T$ dependence} & \multicolumn{2}{c}{ Gaussian $T$ dependence } \\
\cline { 2 - 7 } & $\hat{q}_{\rm c}/T_{\rm c}^3=\hat{q}_0/T_0^3$ & $\chi^2 /${\rm d.o.f} & $(\hat{q}_{\rm c}/T_{\rm c}^3,\hat{q}_0/T_0^3)$ & $\chi^2 /${\rm d.o.f} & $(\sigma_T^2/T_{\rm c}^2,\hat{q}_0/T_0^3)$ & $\chi^2 /${\rm d.o.f} \\
\hline \hline \multicolumn{7}{c}{20-30\% Au + Au collisions at 200 GeV} \\
\hline$R_{A A}$ & (5.8,5.8) & 0.17 & (6.9,1.1) & 0.17 & (2.8,4.4) & 0.48 \\
$v_2$ & -- & -- & (8.7,0.5) & 2.39 & (0.4,2.4) & 2.98 \\
\hline $(R_{AA}+v_2)$ & --  & -- & (7.5,0.2) & 1.51 & (0.7,2.4) & 2.45 \\
\hline \hline \multicolumn{7}{c}{20-30\% Pb + Pb collisions at 2.76 TeV} \\
\hline$R_{A A}$ & (4.0,4.0) & 0.59 & (5.0,3.6) & 0.54 & (4.8,3.2) & 	0.56 \\
$v_2$ & -- & -- & (5.4,0.2)  & 1.52 & (2.1,2.4) &	1.23 \\
\hline $(R_{AA}+v_2)$ & --  & -- & (5.1,0.2) & 0.94 & (1.4,1.6) & 1.23 \\
\hline \hline
\end{tabular}
\caption{Optimal parameter and corresponding $\chi^2/{\rm d.o.f}$ for different data sets in different $\hat{q}/T^3$ form. Note that for the constant form of $\hat{q}/T^3$, the $v_2$ data were not utilized.}
\label{tab:chi2}
\end{center}
\end{table*}

Shown in Fig. \ref{fig:G-RHIC-20-30-x2} (a) and (b) are
the $\chi^2/{\rm d.o.f}$ analyses of single hadron $R_{AA}(p_{\rm T})$ and elliptic flow $v_{2}(p_{\rm T})$ as functions of $\sigma_T^2/T_{\rm c}^2$ and $\hat{q}_0/T_0^3$ from fitting to experimental data \cite{PHENIX:2008saf,PHENIX:2012jha,PHENIX:2010nlr} in 20-30\% Au + Au collisions at $\sqrt{s_{\rm NN}}=200$~GeV, respectively. The global $\chi^2/{\rm d.o.f}$ fitting results for both $R_{AA}(p_{\rm T})$ and $v_{2}(p_{\rm T})$ are shown in Fig. \ref{fig:G-RHIC-20-30-x2} (c). Although $\chi^2/{\rm d.o.f} (R_{AA})$ performs inactively for temperature dependence in 20–30\%, as well as in 0–5\% centrality, $\chi^2/{\rm d.o.f} (v_{2})$ expresses a great favor in the small Gaussian width, which gives $\hat{q}/T^3$ going down more rapidly with $T$. Consequently, global $\chi^2/{\rm d.o.f}$ fits for both $R_{AA}(p_{\rm T})$ and $v_{2}(p_{\rm T})$ provide an explicit constraint on the introduced parameters $(\sigma_T^2/T_{\rm c}^2,\hat{q}_0/T_0^3)$.

With the best global fitting values for $(\sigma_T^2/T_{\rm c}^2,\hat{q}_0/T_0^3)$, we show the Gaussian $T$ dependence of $\hat{q}/T^3$ in Fig.~\ref{fig:G-RHIC-RAA-v2-20-30} (a). Choosing the constant dependence with $(\sigma_T^2/T_{\rm c}^2,\hat{q}_0/T_0^3)=(\infty,5.6)$ (blue dashed curve) and a Gaussian $T$ dependence with $(\sigma_T^2/T_{\rm c}^2,\hat{q}_0/T_0^3)=(1.4,3.6)$ (red solid curve) for $\hat{q}/T^3$, we again get the almost same $R_{AA}(p_{\rm T})$ as shown in Fig. \ref{fig:G-RHIC-RAA-v2-20-30} (b), and $v_{2}(p_{\rm T})$ with difference less than 5\% in Fig. \ref{fig:G-RHIC-RAA-v2-20-30} (c). Compared with the case of a linearly decreasing $T$ dependence, the two $v_{2}(p_{\rm T})$ in Fig. \ref{fig:G-RHIC-RAA-v2-20-30} (c) are closer to each other because the difference between the Gaussian $T$ dependence and the constant dependence is smaller, which leads to an almost invisible change in $v_{2}(p_{\rm T})$.

The use of a Gaussian form for $\hat{q}/T^3$ was intended to provide a more flexible temperature dependence by narrowing the Gaussian width compared with the linear form. Nevertheless, although $v_2$ data favor a higher Gaussian peak, $R_{AA}$ fitting imposes constraints. The global $\chi^2$-fitting results for the Gaussian temperature-dependence hypothesis presented in Fig. \ref{fig:G-RHIC-RAA-v2-20-30}(a) did not manifest a steeper decline with increasing temperature. A comparison of the red curves shown in Fig. \ref{fig:G-RHIC-RAA-v2-20-30}(a) with that in Fig. \ref{fig:L-RHIC-RAA-v2-20-30}(a), it is apparent that the linear temperature dependence results in a higher $\hat{q}_{\rm c}/T_{\rm c}^3$ at the critical temperature. Moreover, as stated previously, $v_2$ was more sensitive to energy loss near the critical temperature. Therefore, the performance of the Gaussian shape is not significantly better than that of the linear shape.

\subsection{Fit $R_{AA}$ and $v_2$ at the LHC}

The same process was performed for the Pb + Pb collisions at $\sqrt{s_{\rm NN}}=2.76$~TeV.
For the most central collisions, we choose $\sigma_T^2/T_{\rm c}^2 \in [0.35,3.5]$ with bin size 0.35 and $\hat{q}_0/T_0^3 \in [0.2,2.4]$ with bin size 0.2 and get 120 couples of $(\sigma_T^2/T_{\rm c}^2, \hat{q}_0/T_0^3)$ to get $R_{AA}(p_{\rm T})$ and the 2-dimensional $\chi^2/{\rm d.o.f}$-fitted contour plot, as shown in Fig. \ref{fig:G-LHC-0-5} (a). With a constant dependence with $(\sigma_T^2/T_{\rm c}^2, \hat{q}_0/T_0^3)=(\infty,3.6)$ and a Gaussian $T$ dependence with $(\sigma_T^2/T_{\rm c}^2, \hat{q}_0/T_0^3)=(1.75, 1.6)$, we obtain the same $R_{AA}(p_{\rm T})$ to fit the data well, as shown in Fig.~\ref{fig:G-LHC-0-5} (b).

For 20–30\% collisions, we choose $\sigma_T^2/T_{\rm c}^2 \in [0.35,7.0]$ and $\hat{q}_0/T_0^3 \in [0.2,4.4]$ and separately obtain 440 groups of $R_{AA}(p_{\rm T})$ and $v_2(p_{\rm T})$ to perform $\chi^2/{\rm d.o.f}$ fitting, as shown in Fig.~\ref{fig:G-LHC-20-30-x2} (a) and (b), respectively. The global fits for both $R_{AA}(p_{\rm T})$ and $v_2(p_{\rm T})$ are presented in Fig.~\ref{fig:G-LHC-20-30-x2} (c) where the constraints for the introduced parameters $( \sigma_T^2/T_{\rm c}^2,\hat{q}_0/T_0^3)$ are obtained. With these constraints, the Gaussian $T$-dependence of $\hat{q}/T^3$ is shown in Fig.~\ref{fig:G-LHC-RAA-v2-20-30} (a).
Choosing a Gaussian $T$ dependence with $(\sigma_T^2/T_{\rm c}^2, \hat{q}_0/T_0^3)=(1.05,1.2)$ (red solid curve) and the constant dependence with $(\sigma_T^2/T_{\rm c}^2, \hat{q}_0/T_0^3)=(\infty,4.1)$ (blue dashed curve) for $\hat{q}/T^3$, we again get almost the same $R_{AA}(p_{\rm T})$ shown in Fig.~\ref{fig:G-LHC-RAA-v2-20-30} (b), and different $v_{2}(p_{\rm T})$ in Fig.~\ref{fig:G-LHC-RAA-v2-20-30} (c). Compared with the constant case, the decreasing Gaussian $T$-dependence of $\hat{q}/T^3$ gives $v_{2}(p_{\rm T})$ an enhancement of approximately 10\%.

At both RHIC and the LHC, numerical results for simultaneously fitting $R_{AA}(p_{\rm T})$ and $v_{2}(p_{\rm T})$ show that the Gaussian $T$ dependence of $\hat{q}/T^3$ is smoothly going down with $T$ and similar to the linearly decreasing $T$ dependence of $\hat{q}/T^3$. Compared with the constant $\hat{q}/T^3$, the going-down $T$-dependence of $\hat{q}/T^3$ enhances the hadron azimuthal anisotropy by approximately 5\%-10\% to improve $v_{2}(p_{\rm T})$ to fit the data.

Thus far, we have demonstrated the constraining power of the experimental data on three temperature-dependent forms of $\hat{q}/T^3$. To more clearly distinguish the separate constraining effects of $R_{AA}$ and $v_2$, we list the best-fit parameters and corresponding minimum $\chi^2/{\rm d.o.f}$ values for each scenario in Table \ref{tab:chi2}. These values correspond to the results shown in Figs. \ref{fig:L-RHIC-20-30-x2}, \ref{fig:L-LHC-20-30-x2},\ref{fig:G-RHIC-20-30-x2}, and \ref{fig:G-LHC-20-30-x2}.  For the constant form of $\hat{q}/T^3$, $v_2$ data are not utilized.

\section{Linear temperature dependence of $\hat{q}/T^3$ in QGP and hadron phases} \label{sec:QGP-hadron}

For the complete study of the linear temperature dependence of $\hat{q}/T^3$, we use Eqs. ($\ref{eq:qhat-linear}$), ($\ref{eq:qhat-f}$), and ($\ref{eq:qhat-had}$) to include the contributions from both QGP and hadron phases in 20–30\% A + A collisions.

\begin{figure*}[htb]
\begin{center}
\includegraphics[width=0.32\textwidth]{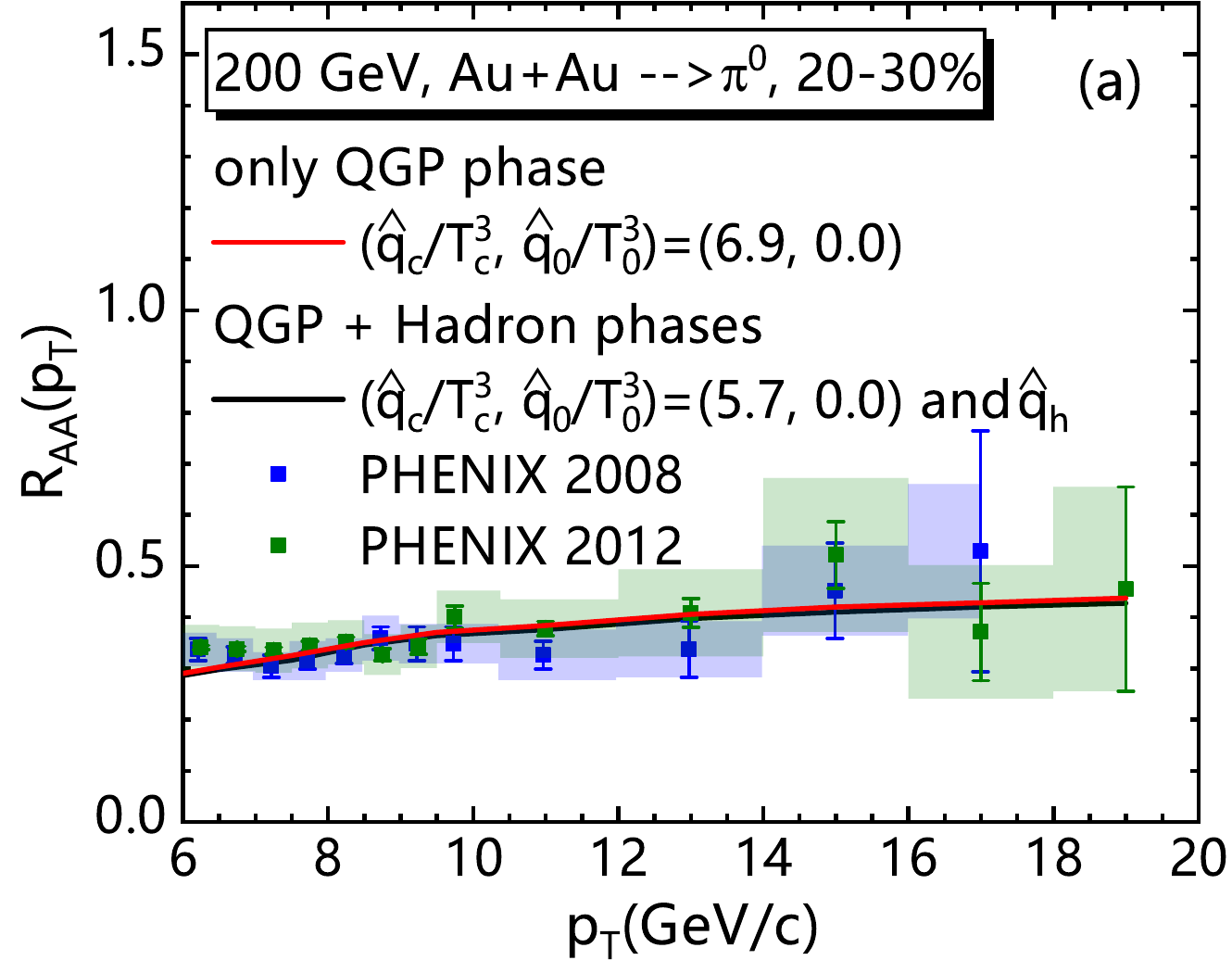}
\hspace{5mm}
\includegraphics[width=0.32\textwidth]{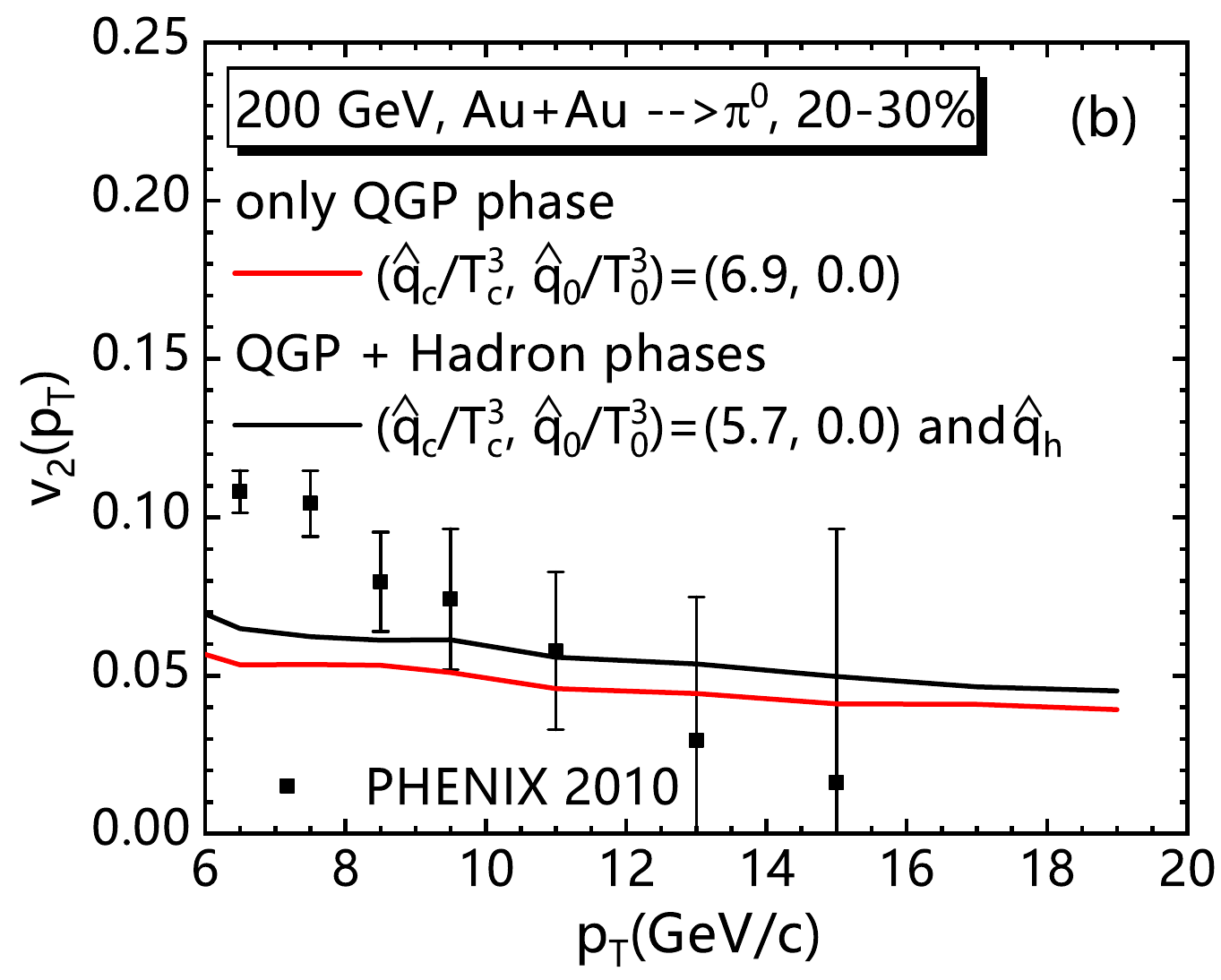}
\caption{Single hadron suppression factors $R_{AA}(p_{\rm T})$ (panel (a)) and elliptic flow $v_2(p_{\rm T})$ (panel (b)) with jet energy loss of QGP + Hadron phases (black solid curves with $(\hat{q}_{\rm c}/T_{\rm c}^3,\hat{q}_0/T_0^3)=(5.7,0.0)$ and $\hat{q}_h$) and of only QGP phase (red solid curves with $(\hat{q}_{\rm c}/T_{\rm c}^3,\hat{q}_0/T_0^3)=(6.9,0.0)$ from Fig.~\ref{fig:L-RHIC-RAA-v2-20-30} ) for 20-30\% Au + Au collisions at $\sqrt{s_{\rm NN}}=200$ GeV.}
\end{center}
\label{fig:H-RHIC-20-30}
\end{figure*}

\begin{figure*}[htb]
\begin{center}
\includegraphics[width=0.32\textwidth]{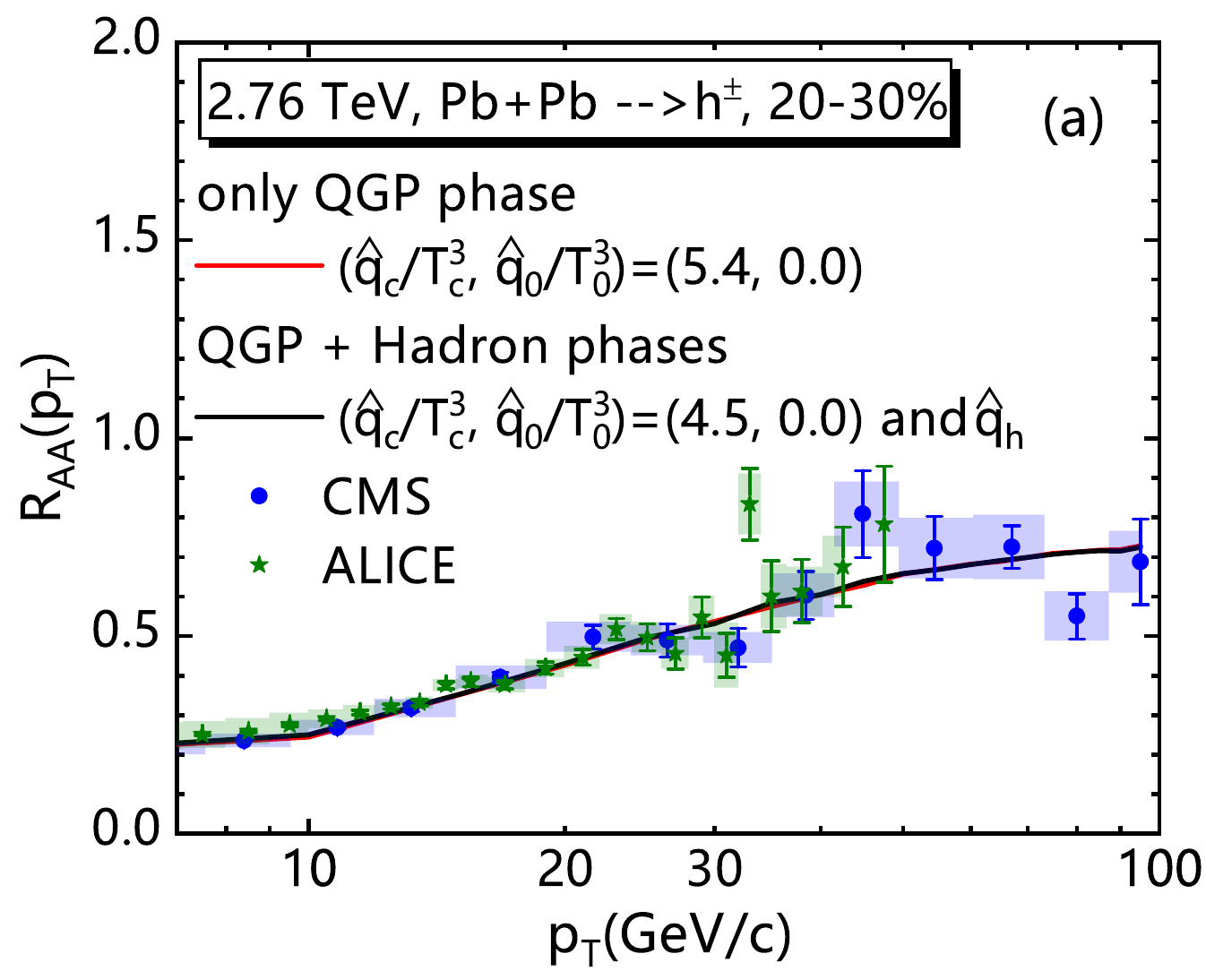}
\hspace{5mm}
\includegraphics[width=0.32\textwidth]{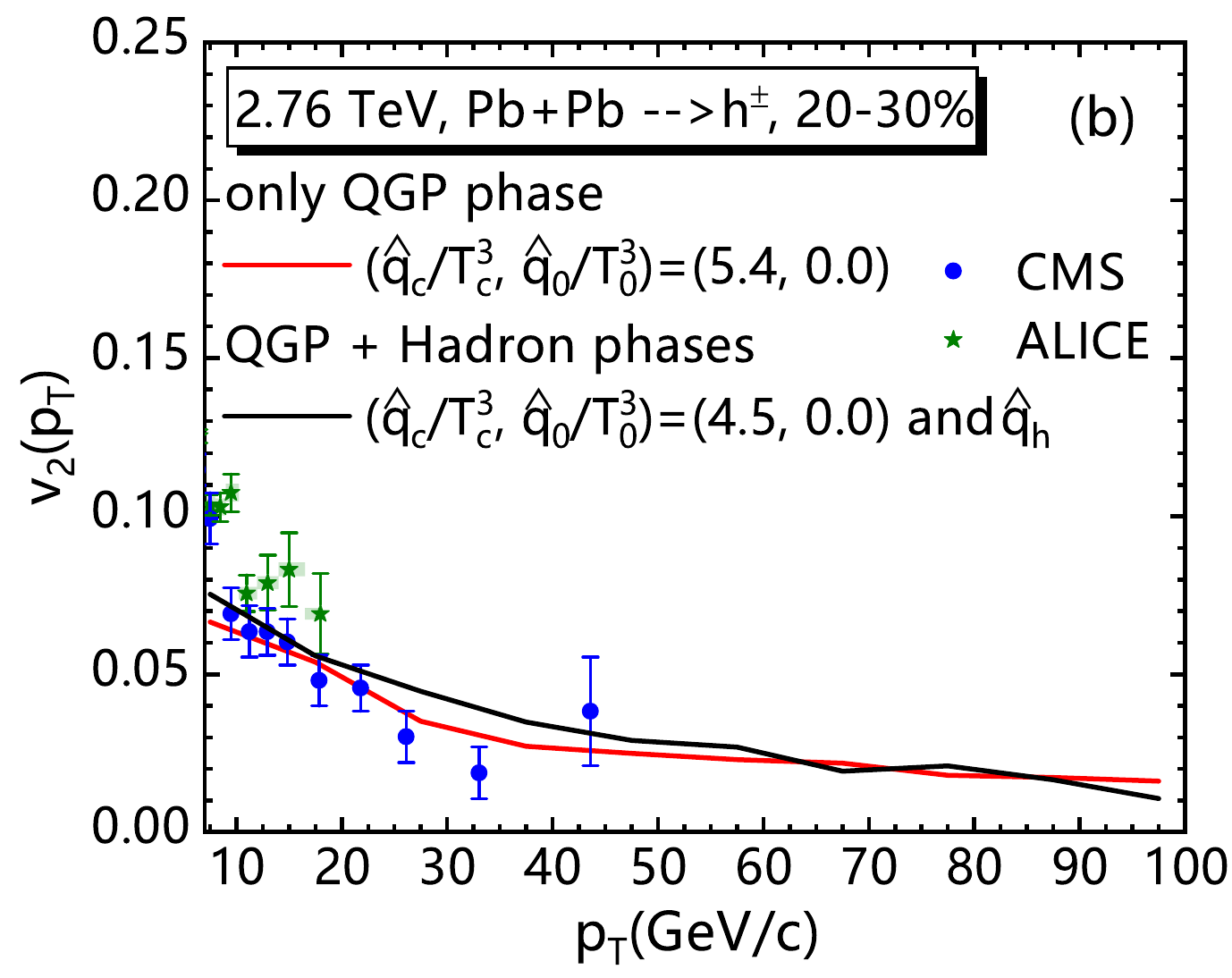}
\caption{Single hadron suppression factors $R_{AA}(p_{\rm T})$ (panel (a)) and elliptic flow $v_2(p_{\rm T})$ (panel (b)) with jet energy loss of both QGP + Hadron phases (black solid curves with $(\hat{q}_{\rm c}/T_{\rm c}^3,\hat{q}_0/T_0^3)=(4.5,0.0)$ and $\hat{q}_h$) and of only QGP phase (red solid curves with $(\hat{q}_{\rm c}/T_{\rm c}^3,\hat{q}_0/T_0^3)=(5.4,0.0)$ from Fig. \ref{fig:L-LHC-RAA-v2-20-30}) for 20-30\% Pb + Pb collisions at $\sqrt{s_{\rm NN}}=2.76$ TeV.}
\end{center}
\label{fig:H-LHC-20-30}
\end{figure*}

\begin{figure*}[htbp]
\begin{center}
\includegraphics[width=0.32\textwidth]{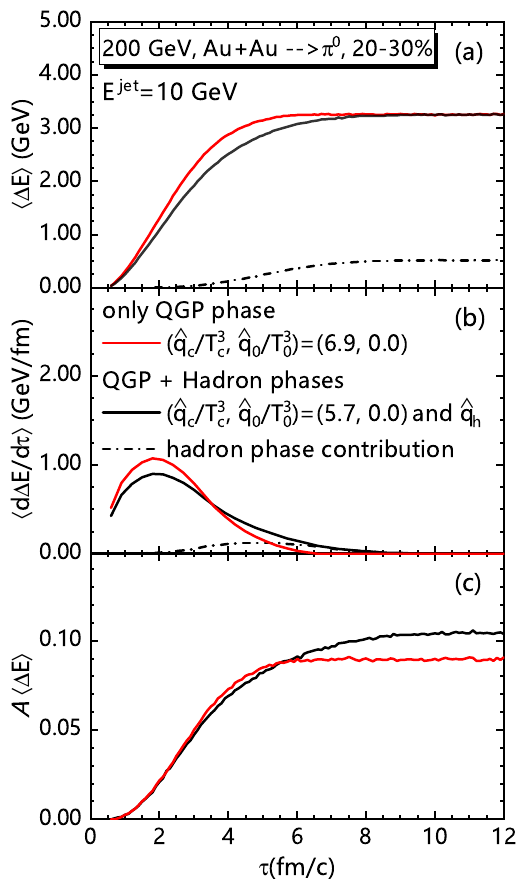}
\hspace{5mm}
\includegraphics[width=0.32\textwidth]{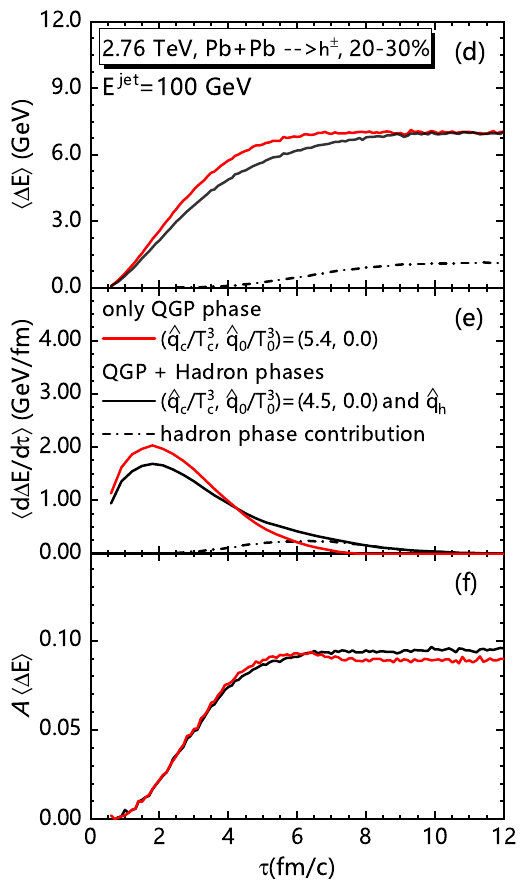}
\caption{Comparisons of the average jet energy loss distribution between QGP + Hadron phases (black solid curves) and only QGP phase (red solid curves from Fig. \ref{fig:L-de-tau} and \ref{fig:L-deltaE}) in noncentral A + A collisions at RHIC (left panels) and the LHC (right panels), respectively. The black dot-dashed curves are for hadron phase contribution in the case of QGP + Hadron phases.
From top to bottom are the average accumulative energy loss, differential energy loss, and energy loss asymmetry, respectively.}
\end{center}
\label{fig:H-de-tau}
\end{figure*}

\subsection{Best fits for $R_{AA}$ and $v_2$ due to energy loss in QGP and hadron phases}

We begin with Au + Au collisions in 20–30\% centrality at $\sqrt{s_{\rm NN}}=200$ GeV. Shown in Fig.~\ref{fig:H-RHIC-20-30} are hadron suppression factors $R_{AA}(p_{\rm T})$ (panel (a)) and elliptic flow $v_2(p_{\rm T})$ (panel (b)) with jet energy loss of both QGP and hadronic phases (black solid curves) and of only QGP phase (red solid curves from Fig. \ref{fig:L-RHIC-RAA-v2-20-30}) for 20-30\% Au + Au collisions at $\sqrt{s_{\rm NN}}=200$ GeV, respectively.
With the same ${\hat{q}_0/T_0^3=0.2}$ shown in Fig. \ref{fig:L-RHIC-RAA-v2-20-30}, we have to decrease the value of $\hat{q}_{\rm c}/T_{\rm c}^3$ from 6.9 to 5.7 due to the included hadronic phase to get the same $R_{AA}(p_{\rm T})$.
The 20\% reduction in $\hat{q}/T^3$ is consistent with Ref. \cite{Chen:2010te}.
The jet energy loss in the hadron phase gives an enhancement of 10\% for the elliptic flow $v_2(p_{\rm T})$, which means that the jet energy loss of the hadronic phase has an important and non-negligible contribution to $v_2(p_{\rm T})$.

As shown in Fig.~\ref{fig:H-LHC-20-30} is for the case of 20-30\% Pb + Pb collisions at $\sqrt{s_{\rm NN}}=2.76$ TeV.
For consistency, $R_{AA}(p_{\rm T})$ in Fig.~\ref{fig:H-LHC-20-30} (a), the $v_2(p_{\rm T})$ (black solid curve) with {$(\hat{q}_{\rm c}/T_{\rm c}^3,\hat{q}_0/T_0^3)=(4.5,0.0)$} of QGP phase and $\hat{q}_h$ of hadronic phase has an additional enhancement by 5\% compared with {$(\hat{q}_{\rm c}/T_{\rm c}^3,\hat{q}_0/T_0^3)=(5.4,0.0)$} of only QGP phase (red solid curve from Fig. \ref{fig:L-LHC-RAA-v2-20-30}) in Fig. \ref{fig:H-LHC-20-30} (b).
This enhancement was less than that at RHIC because the fraction of hadron phase contribution to the jet energy loss at the LHC was less than that at RHIC, as shown in Fig. \ref{fig:H-de-tau} (a) and (d).

It is worth noting that the description of $p_{\rm T}$ dependence of $v_2$ still deserves further improvement. This study considers only the $T$ dependence of $\hat{q}/T^3$. Taking into account the dependence of $\hat{q}/T^3$ on the parton energy $E$, which is more directly related to the $p_{\rm T}$ distribution of the jet energy loss \cite{JETSCAPE:2021ehl, JETSCAPE:2022jer, He:2018gks}, is likely to aid in improving the $p_{\rm T}$ dependence of $v_2$.
Moreover, considering the dependence of $\hat{q}/T^3$ on the collision energy, $\sqrt{s_{\rm NN}}$ can also contribute to a more accurate description \cite{Zhu:2022dlc}.
Additionally, elastic energy loss and jet-induced medium responses have a significant impact on hadron production at intermediate transverse momentum, which could enhance $v_2$ in this $p_{\rm T}$ region \cite{Qin:2015nma, Zhao:2021vmu}. In the future, we will incorporate these considerations into our model to further improve its descriptive power for experimental data.

\subsection{Jet energy loss distribution in QGP and hadron phases}

Similarly to Figs.~\ref{fig:L-de-tau} and~\ref{fig:L-deltaE}, we show in Fig.~\ref{fig:H-de-tau} the comparisons of the average jet energy loss distribution between QGP + Hadron phases (black solid curves) and only QGP phase (red solid curves) in noncentral A + A collisions at RHIC (left panels) and the LHC (right panels), respectively. The black dot-dashed curves represent the hadron phase contribution in the case of QGP + Hadron phases. From top to bottom are the average accumulated energy loss, differential energy loss, and energy loss asymmetry, respectively.

The hadron phase contribution to the total energy loss of QGP + Hadron phases was approximately 17\% at RHIC in Fig.~\ref{fig:H-de-tau} (a) while about 14\% at the LHC in Fig. \ref{fig:H-de-tau} (d). Because of the first-order phase transition in the current model shown in Fig.~\ref{fig:L-de-tau} (c) and (f), the hadron phase contribution happens mainly in the $T_{\rm c}$ nearby shown in Fig.~\ref{fig:H-de-tau} (b) and (e), which strengthens the azimuthal anisotropy of the system and then enhances the elliptic flow parameter. This is similar to the peak of the energy loss rate pushed to $T_{\rm c}$ owing to the linear $T$ dependence of $\hat{q}/T^3$ in Fig.~\ref{fig:L-de-tau} (b) and (e). Such enhancements in the azimuthal anisotropy are also exhibited by the energy loss asymmetry shown in Fig.~\ref{fig:H-de-tau} (c) and (f).

\subsection{Temperature dependence of $\hat{q}/T^3$ in QGP and hadron phases}

\begin{figure}[htbp]
\begin{center}
\includegraphics[width=0.35\textwidth]{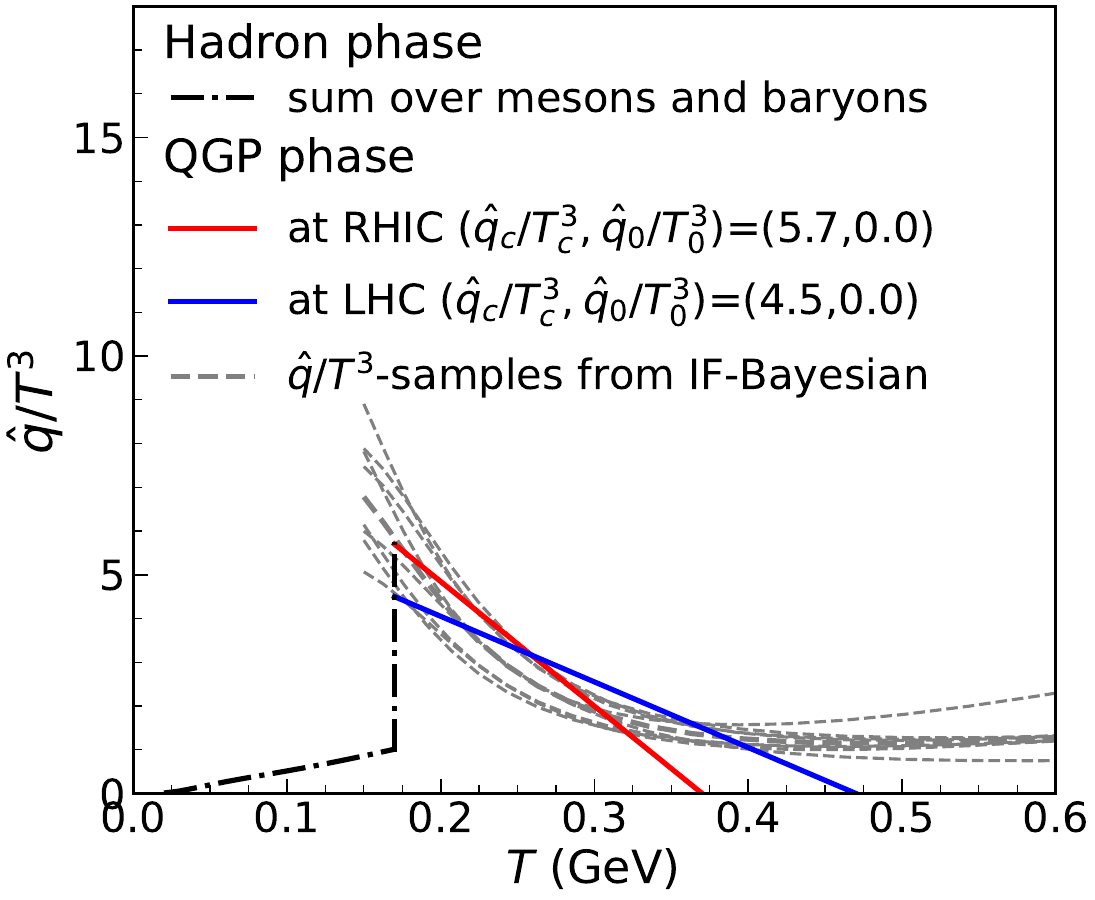}
\caption{The scaled jet transport coefficient $\hat{q}/T^3$ as a function of medium temperature $T$. $T<$ 0.17 GeV for the hadronic phase (dot-dashed curve), and $T>$ 0.17 GeV for the QGP phase (solid curves). The $\hat{q}/T^3$ of QGP phase for Au + Au collisions at $\sqrt{s_{\rm NN}}=200$ GeV is denoted in red, and for Pb + Pb collisions at $\sqrt{s_{\rm NN}}=2.76$ TeV is denoted in blue. As comparisons, the $\hat{q}/T^3$ posterior samples from single hadron, dihadron, and $\gamma$-hadron production at RHIC and the LHC energies with IF-Bayesian analysis \cite{Xie:2022ght} are also shown in grey curves.}
\end{center}
\label{fig:qhad}
\end{figure}

Shown in Fig.~\ref{fig:qhad} is the $\hat{q}/T^3$ of the hadronic phase (dot-dashed line) and QGP phase (solid lines) as a function of medium temperature $T$. The hadronic phase $\hat{q}/T^3$ is given by Eq.~($\ref{eq:qhat-had}$). The values of $\hat{q}/T^3$ for the QGP phase with linear temperature dependence are given by Eq.~(\ref{eq:qhat-linear}) with $(\hat{q}_{\rm c}/T_{\rm c}^3,\hat{q}_0/T_0^3)=(5.7,0.0)$ at RHIC (denoted by the red curve) and $(4.5,0.0)$ at the LHC (denoted by the blue curve), respectively. The contributions to $\hat{q}/T^3$ of the QGP and hadron phases combined using Eq.~(\ref{eq:qhat-f}) were applied to $R_{AA}(p_{\rm T})$ and $v_{2}(p_{\rm T})$ at RHIC/LHC, as shown in Fig.~\ref{fig:H-RHIC-20-30} and Fig.~\ref{fig:H-LHC-20-30}. For comparison, representative samples of $\hat{q}/T^3$ extracted from single hadron, dihadron, and $\gamma$-hadron production at RHIC and LHC energies with information field (IF) Bayesian analysis \cite{Xie:2022ght} are also shown in Fig.~\ref{fig:qhad}, which are denoted by grey curves. Ref.~\cite{Xie:2022ght} does not consider the jet energy loss in the hadronic phase with a pseduocritical temperature $ T_{\rm c}= 0.165$ GeV. When considering the experimental data covering a wider temperature range, $\hat{q}/T^3$ still indicates a large value at the critical temperature. Our numerical results for temperature-dependent $\hat{q}/T^3$ are consistent with those obtained using the IF-Bayesian method.

After adding the hadron phase contribution to the jet energy loss, one should decrease the QGP phase contribution so as to obtain a total energy loss equal to that of the QGP phase alone. The decreased QGP-phase energy loss makes $v_{2}(p_{\rm T})$ smaller, whereas the added hadron-phase energy loss makes $v_{2}(p_{\rm T})$ larger.
The numerical results show that the competition between them gives $v_{2}(p_{\rm T})$ a larger value than in the case of only the QGP phase because of the energy loss contribution of the hadronic phase concentrated near the critical temperature. Regardless of the added hadron phase or the linear going-down $T$-dependence of $\hat{q}/T^3$ in the QGP phase, as shown in Fig. \ref{fig:qhad}, the jet is made to lose much more energy near the critical temperature, which results in a larger $v_{2}(p_{\rm T})$ for the large $p_{\rm T}$ to fit data better.
This numerical result of stronger jet quenching in the near-$T_{\rm c}$ region is consistent with that of a previous theoretical study \cite{Liao:2008dk}.

One may notice that we extracted $\hat{q}/T^3$ at RHIC and the LHC with different parameter ranges. This is mainly because, in this work, we observed that extracting $\hat{q}/T^3$ at RHIC and LHC separately could provide a better $\chi^2$ result than the simultaneous fit at both collision energies. 
In this study, we only considered the dependence of $\hat{q}$ on the temperature. As mentioned previously, if the dependencies of $\hat{q}$ on parton energy $E$ and $\sqrt{s_{\rm NN}}$, as well as the jet-induced medium responses and elastic energy loss, were all considered, the value of $\hat{q}$ could be constrained more accurately from $R_{AA}$ and $v_2$ at both RHIC and LHC energies simultaneously.
Nevertheless, the $\hat{q}/T^3$ obtained in this study is also consistent with the $\hat{q}/T^3$ extracted by the IF-Bayesian approach \cite{Xie:2022ght} and JETSCAPE \cite{JETSCAPE:2021ehl,JETSCAPE:2022jer}, which constrained $\hat{q}/T^3$ from RHIC and the LHC simultaneously, as shown in Fig.~\ref{fig:qhad}. We hope that in the future, by updating the model and enriching the information on $\hat{q}$, we can better describe $R_{AA}$ and $v_2$  simultaneously for both RHIC and the LHC energies.

\section{SUMMARY} \label{sec:summary}

In this study, within a next-to-leading-order perturbative QCD model, the medium-temperature dependence of jet energy loss was studied via the nuclear modification factor $R_{AA}(p_{\rm T})$ and elliptic flow parameter $v_2(p_{\rm T})$ of large transverse momentum hadrons.
Owing to the jet quenching, medium-modified fragmentation functions based on the higher-twist energy-loss formalism were used in the numerical calculations.
We assumed that the scaled jet transport coefficient $\hat{q}/T^3$ depends on the medium temperature in linear or Gaussian form, with which we calculated the single hadron suppression factor $R_{AA}(p_{\rm T})$ and elliptic flow parameter $v_2(p_{\rm T})$ and compared them with experimental data. To constrain the $\hat{q}/T^3$ temperature dependence forms, a global $\chi^2/{\rm d.o.f}$ fitting was performed on the the experimental data. 
Finally, the jet energy loss in the hadronic phase was also considered.

With the linear $T$ dependence of $\hat{q}/T^3$ for only the QGP phase, the global $\chi^2/{\rm d.o.f}$ fitting for both $R_{AA}(p_{\rm T})$ and $v_2(p_{\rm T})$ shows that $\hat{q}_{\rm c}/T_{\rm c}^3=$ 6.0-8.0 at RHIC and 4.0-6.0 at the LHC {while $\hat{q}_0/T_0^3=$ 0.0-4.2 at both RHIC and the LHC}, as shown in Fig. \ref{fig:L-RHIC-RAA-v2-20-30} (a) and Fig. \ref{fig:L-LHC-RAA-v2-20-30} (a). The numerical results indicate that $R_{AA}(p_{\rm T})$ and $v_2(p_{\rm T})$ are both more sensitive to the value of $\hat{q}/T^3$ near the critical temperature $T_{\rm c}$ than near the initial highest temperature $T_0$.
Furthermore, the fitting results show a decreasing trend of $\hat{q}/T^3$ depending on the medium temperature, which is also supported by the Gaussian $T$ dependence of $\hat{q}/T^3$ for only the QGP phase.

Compared with the case of constant $\hat{q}/T^3$, the going-down $T$ dependence of $\hat{q}/T^3$ causes a hard parton jet to lose more energy near the critical temperature $T_{\rm c}$ and therefore strengthens the azimuthal anisotropy for large $p_{\rm T}$ hadron production.
As a result, the elliptic flow parameter $v_2(p_{\rm T})$ for large $p_{\rm T}$ hadrons was enhanced by approximately 10\% to better fit the data at RHIC/LHC. Considering the first-order phase transition from QGP to hadron and the hadron phase contribution to the jet energy loss, $v_2(p_{\rm T})$ is again enhanced by 5\%-10\% at RHIC/LHC.

\section*{ACKNOWLEDGMENTS}

We thank Xin-Nian Wang, Bowen Xiao and Guang-You Qin for their exciting discussions.

\end{document}